# Design and Construction of the BESIII Detector

M. Ablikim, Z.H. An, J.Z. Bai, Niklaus Berger, J.M. Bian, X. Cai, G.F. Cao, X.X. Cao, J.F. Chang, C. Chen, G. Chen, H.C. Chen, H.X. Chen, J. Chen, J.C. Chen, L.P. Chen, P. Chen. X.H. Chen, Y.B. Chen, M.L. Chen, Y.P. Chu,X.Z. Cui, H.L. Dai, Z.Y. Deng, M.Y. Dong, S.X. Du, Z.Z. Du, J. Fang, C.D. Fu, C.S. Gao, M.Y. Gong, W.X. Gong, S.D. Gu, B.J. Guan, J. Guan, Y.N. Guo, J.F. Han, K.L. He, M. He, X. He, Y.K. Heng, Z.L. Hou, H.M. Hu, T. Hu, B. Huang, J. Huang, S.K. Huang, Y.P. Huang, Q. Ji, X.B. Ji, X.L. Ji, L.K. Jia, L.L. Jiang, X.S. Jiang, D.P. Jin, S. Jin, Y. Jin, Y.F. Lai, G.K. Lei, F. Li, G. Li, H.B. Li, H.S. Li, J. Li, J. Li, J.C. Li, J.Q. Li, L. Li, L. Li, R.B. Li, R.Y. Li, W.D. Li, W.G. Li, X.N. Li, X.P. Li, X.R. Li, Y.R. Li, W. Li, D.X. Lin, B.J. Liu, C.X. Liu, F. Liu, G.M. Liu, H. Liu, H.M. Liu, H.W. Liu, J.B. Liu, L.F. Liu, Q. Liu, Q.G. Liu, S.D. Liu, W.J. Liu, X. Liu, X.Z. Liu, Y. Liu, Y.J. Liu, Z.A. Liu, Z.Q. Liu, Z.X. Liu, J.G. Lu, T. Lu, Y.P. Lu, X.L. Luo, H.L. Ma, Q.M. Ma, X. Ma, X.Y. Ma, Z.P. Mao, J. Min, X.H. Mo, J. Nie, Z.D. Nie, R.G. Ping, S. Qian, Q. Qiao, G. Qin, Z.H. Qin, J.F. Qiu, R.G. Liu, Z.Y. Ren, G. Rong, L. Shang, D.L. Shen, X.Y. Shen, H.Y. Sheng, Y.F. Shi, L.W. Song, W.Y. Song, D.H. Sun, G.X. Sun, H.S. Sun, L.J. Sun, S.S. Sun, X.D. Sun, Y.Z. Sun, Z.J. Sun, J.P. Tan, S.Q. Tang, X. Tang, N. Tao, H.L. Tian, Y.R. Tian, X. Wan, D.Y. Wang, J.K. Wang, J.Z. Wang, K. Wang, K.X. Wang, L. Wang, L. Wang, L.J. Wang, L.L. Wang, L.S. Wang, M. Wang, N. Wang, P. Wang, P.L. Wang, Q. Wang, Y.F. Wang, Z. Wang, Z. Wang, Z.G. Wang, Z.Y. Wang, C.L. Wei, S.J. Wei, S.P. Wen, J.J. Wu, L.H. Wu, N. Wu, Y.H. Wu, Y.M. Wu, Z. Wu, M.H. Xu, X.M. Xia, H.S. Xiang, G. Xie, X.X. Xie, Y.G. Xie, G.F. Xu, H. Xu, Q.J. Xu, J.D. Xue, L. Xue, L. Yan, G.A. Yang, H. Yang, H.X. Yang, S.M. Yang, M. Ye, B.X. Yu, C. Yuan, C.Z. Yuan, Y. Yuan, S.L. Zang, B.X. Zhang, B.Y. Zhang, C.C. Zhang, C.C. Zhang, D.H. Zhang, H.Y. Zhang, J. Zhang, J.W. Zhang, J.Y. Zhang, L.S. Zhang, M. Zhang, Q.X. Zhang, W. Zhang, X.M. Zhang, Y. Zhang, Y.H. Zhang, Y.Y. Zhang, Z.X. Zhang, S.H. Zhang, D.X. Zhao, D.X. Zhao, H.S. Zhao, J.B. Zhao, J.W. Zhao, J.Z. Zhao, L. Zhao, P.P. Zhao, Y.B. Zhao, Y.D. Zhao, B. Zheng, J.P. Zheng, L.S. Zheng, Z.P. Zheng, B.Q. Zhou, G.M. Zhou, J. Zhou, L. Zhou, Z.L. Zhou, H.T. Zhu, K. Zhu, K.J. Zhu, Q.M. Zhu, X.W. Zhu, Y.S. Zhu, Z.A. Zhu, B.A. Zhuang, J.H. Zou, X. Zou, J.X. Zuo

*Institute of High Energy Physics, Beijing 100049, China*

M. H. Ye

*China Center of Advanced Science and Technology, Beijing 100080, China*

Yangheng Zheng, Cong-Feng Qiao, Xiaorui Lu, Hongbang Liu, Jifeng Hu

*Graduate University of Chinese Academy of Sciences, Beijing 100049, China*

Y. T. Gu, X. D. Ruan

*Guangxi University, 13, Xiuling Road,Nanning, Guangxi 530005, China*

G. M. Huang

*Huazhong Normal University, Wuhan 430079, China*

Yun Zeng, Yonghong Yan

*Hunan University, Changsha 410082, China*

Igor Boyko, Vladimir Bytev, Dmitry Dedovich, Sergey Grishin, Yuri Nefedov, Alexey Zhemchugov, Andrei Sarantsev

*Joint Institute for Nuclear Research (JINR), 141980 Dubna, Moscow Region, Russia*

Zhenjun Xiao, Jialun Ping, Libo Guo Chenglin Luo

*Nanjing Normal University, Nanjing 210097, China*

Shenjian Chen, Ming Qi, Xiaowei Hu, Lei Zhang

*Nanjing University, Nanjing 210093, China*

Yajun Mao, Zhengyun You, Yutie Liang

*Peking University, Beijing 100871, China*

Xueyao Zhang, Xingtao Huang, Jianbin Jiao, Jiaheng Zou

*Shandong University, Jinan 250100, China*

Mai-Ying Duan, Fu-Hu Liu, Qi-Wen Lv, Fei-Peng Ning, Xiao-Dong Wang

*Shanxi University, Taiyuan 030006, China*

Yongfei Liang, Changjian Tang, Yiyun Zhang

*Sichuan University, Chengdu, Sichuan 610065, People's Republic of China*

Y.N. Gao, H. Gong, B.B. Shao, Y.R. Tian, S.M. Yang

*Tsinghua University, Beijing 100084, China*

F. A. Harris, J. W. Kennedy, Q. Liu, X. Nguyen, S. L. Olsen, M. Rosen, C. P. Shen, and G. S. Varner
*University of Hawaii, Honolulu, Hawaii 96822, USA*

X. Yu, Y. Zhou, H. Liang, Y. Chen, J. Xue, Q. Liu, B. Liu, Z. Cheng, L. Zhou, H. Yang, H.F. Chen, Cheng Li, M. Shao, Y.J. Sun, J. Yan, Z.B. Tang, X. Li, L. Zhao, L. Jiang, Z.P. Zhang, J. Wu, Z.Z. Xu, Q. Shan, Z. Xue, X.L. Wang, Q. An, S.B. Liu, J.H. Guo, L. Zhao, C.Q. Feng, X.Z. Liu, H. Li, W. Zheng, H. Yan, Z. Cao, X.H. Liu
*University of Science and Technology of China, Hefei 230026, China*

Sachio Komamiya, Tomoyuki Sanuki, Taiki Yamamura.
*University of Tokyo, Hongo, Bunkyo-ku, Tokyo 113-0033, Japan*

T. Zhao
*University of Washington, Seattle, WA 98195, USA*

Mingxing Luo
*Zhejiang University, Hangzhou 310028, China*

**Abstract**

This paper will discuss the design and construction of BESIII [1], which is designed to study physics in the $\tau$-charm energy region utilizing the new high luminosity BEPCII double ring $e^+e^-$ collider [2]. The expected performance will be given based on Monte Carlo simulations and results of cosmic ray and beam tests. In BESIII, tracking and momentum measurements for charged particles are made by a cylindrical multilayer drift chamber in a 1 T superconducting solenoid. Charged particles are identified with a time-of-flight system based on plastic scintillators in conjunction with *dE/dx* (energy loss per unit pathlength) measurements in the drift chamber. Energies of electromagnetic showers are measured by a CsI(Tl) crystal calorimeter located inside the solenoid magnet. Muons are identified by arrays of resistive plate chambers in the steel magnetic flux return. The level 1 trigger system, Data Acquisition system and the event filter system based on networked computers will also be described.

**Contents**

# 1. Introduction

*1.1. Overview*

The original Beijing electron-positron collider BEPC [3], designed to operate in the τ-charm energy region, and its detectors, the Beijing Spectrometer (BES) [4] and the upgraded BESII [5], were operated from 1989 to 2004 at the Institute of High Energy Physics (IHEP) of the Chinese Academy of Sciences in Beijing. The original BES detector was upgraded to BESII in 1996, and the BEPC was also improved over the years. A variety of important τ-charm results, involving $J/\psi$, $\psi(2S)$, $\tau$, $D$ and $D_S$ particles produced at or near the threshold were obtained during its 15 year operation. BEPC was a single ring electron-positron storage ring operated in single bunch mode, and the instantaneous luminosity reached approximately $10^{31}$ cm$^{-2}$s$^{-1}$ before it was shut down.

The BEPCII and BESIII [6] program began in 2003 after formal approval by the Chinese government. The new BEPCII collider, installed in the same tunnel as the BEPC, is a double-ring multi-bunch collider with a design luminosity of $\sim 1 \times 10^{33}$ cm$^{-2}$s$^{-1}$ optimized at a center-of-mass energy of $2 \times 1.89$ GeV, a factor of approximately one hundred increase over its predecessor. In addition to the τ-charm studies, BEPCII will also be used as a high flux synchrotron radiation light source.

A new detector BESIII, that fully utilizes advanced detector technologies developed over the past two decades, has been completed and recently begun its operation. The advanced design of BESIII will allow it to take advantage of the high luminosity delivered by BEPCII and to collect large data samples so that τ-charm physics can be studied with high precision.

The surprising discoveries of the narrow $D_{sj}$ mesons [7], several hidden charm resonances in the 4 GeV region, and the $X(1835)$ at BESII during the past few years have considerably enhanced the interest in the spectroscopy of hadrons with and without open charm. At the boundary between the perturbative and non-perturbative regimes of QCD, the energy region $\sqrt{s} = 2 - 4.6$ GeV offers vast and diverse physics opportunities. Results from BESIII are expected to play an important role in the understanding of the Standard Model and will also provide important calibrations for the Lattice Gauge community. The rich physics program of the BESIII experiment includes:

- Tests of electroweak interactions with very high precision in both the quark and lepton sectors.
- High statistics studies of light hadron spectroscopy and decay properties.
- Studies of the production and decay properties of $J/\psi$、$\psi$ (2S) and $\psi$ (3770) states with large data samples and search for glueballs, quark-hybrids, , multi-quark states and other exotic states via charmonium hadronic and radiative decays.
- Studies of τ-physics.
- Studies of charm physics, including the decay properties of $D$ and $D_s$ and charmed baryons.
- Precision measurements of QCD parameters and CKM parameters.
- Search for new physics by studying rare and forbidden decays, oscillations, and CP violations in *c-hadron* and *τ-lepton* sectors.

*1.2. BEPCII storage ring*

The engineering run of BEPCII in collision mode was successfully completed in July of 2008, and physics data taking was started in March of 2009. The instantaneous luminosity reached $0.32 \times 10^{33}$ cm$^{-2}$s$^{-1}$ at a center-of-mass energy of $2 \times 1.89$ GeV. The design parameters of BEPCII are summarized and compared with those of BEPC in Table 1.

Table 1
BEPCII design parameters compared with those of BEPC

| Parameters | BEPCII | BEPC |
|---|---|---|
| Center of mass Energy (GeV) | 2 - 4.6 | 2 - 5 |
| Circumference (m) | 237.5 | 240.4 |
| Number of rings | 2 | 1 |
| RF frequency $f_{rf}$ (MHz) | 499.8 | 199.5 |
| Peak luminosity at $2 \times 1.89$ GeV (cm$^{-2}$s$^{-1}$) | $\sim 10^{33}$ | $\sim 10^{31}$ |
| Number of bunches | $2 \times 93$ | $2 \times 1$ |
| Beam current (A) | $2 \times 0.91$ | $2 \times 0.035$ |
| Bunch spacing (m/ns) | 2.4/8 | - |
| Bunch length ($\sigma_z$) cm | 1.5 | ~5 |
| Bunch width ($\sigma_x$) μm | ~380 | ~840 |
| Bunch height ($\sigma_y$) μm | ~5.7 | ~37 |
| Relative energy spread | $5 \times 10^{-4}$ | $5 \times 10^{-4}$ |
| Crossing angle (mrad) | ±11 | 0 |

BEPCII is designed to collide $e^+e^-$ beams in the energy range of $\sqrt{s} = (2 - 4.6)$ GeV, and its luminosity is optimized at $2 \times 1.89$ GeV. Housed in the existing tunnel, the collider was essentially rebuilt. The LINAC has new klystrons, a new electron gun, and a new positron source. In order to increase the average luminosity, the "top-off" injection scheme up to 1.89 GeV is adopted. This allows filling without dumping the beam remaining and requires that the injection and collision optics be almost the same. The electron injection rate has reached 200 mA per minute, and the positron injection rate 50 mA per minute. All major components of the collider are new, including injection kickers, beam control optics, superconducting RF cavities, a superconducting micro-β final focusing system, beam pipes, the vacuum system, magnets and power supplies, and the control system. The old dipoles were modified and are used in the outer ring. BEPCII will be operated in multi-bunch mode with 93 bunches stored in each ring spaced by 8 ns or 2.4 meters. Electrons and positrons will collide at the interaction point with a horizontal crossing-angle of ±11 mrad. The single beam current is designed to be 0.91 A in the collider mode. BEPCII will also be used as a synchrotron radiation facility with a 250 mA electron beam at 2.5 GeV energy. The design luminosity of BEPCII is expected to reach $10^{33}$ cm$^{-2}$s$^{-1}$ at 3.78 GeV, roughly 100 times higher than the luminosity of BEPC. The luminosity of BEPCII is expected to be lowered to $6\times10^{32}$ cm$^{-2}$s$^{-1}$ at 3.0 and 4.2 GeV. High beam currents and the large number of closely spaced bunches are required to achieve the high luminosity. Another important factor that helps to improve the luminosity is that the vertical beam size is compressed by the micro-beta technique using superconducting quadrupole magnets placed very close to the interaction point.

*1.3 Event rates and final states*

The expected data samples in each calendar year of the BESIII operated in the the τ-charm energy region are summarized in Table 2. The integrated luminosity used to calculate these rates was based on one half of the design luminosity, and the total running time was assumed to be $10^7$sec/year. Many technical choices for the BESIII detector were made based on the high data rates, the types of secondary particles, their energy spectra, event topology and multiplicity. In typical hadranic final states, the most probable momentum of charged particles produced is approximately 0.3 GeV/c, and an overwhelming majority of particles have momentum below 1 GeV/c, and the most probable energy of photons is about 100 MeV. The average multiplicity is on the order of four for charged particles and photons in final states [1].

Table 2
Expected BESIII data samples in a calendar year

| States | Energy (GeV) | Peak luminosity ($10^{33}$ cm$^{-2}$s$^{-1}$) | Physics cross-section (nb) | Events/year |
|---|---|---|---|---|
| $J/\psi$ | 3.097 | 0.6 | 3,400 | $1\times10^{10}$ |
| $\psi(2S)$ | 3.686 | 1.0 | 640 | $3\times10^9$ |
| $\tau^+\tau^-$ | 3.670 | 1.0 | 2.4 | $1.2\times10^7$ |
| $D^0\bar{D}^0$ | 3.770 | 1.0 | 3.6 | $1.8\times10^6$ |
| $D^+D^-$ | 3.770 | 1.0 | 2.8 | $1.4\times10^6$ |
| $D_sD_s$ | 4.030 | 0.6 | 0.32 | $1\times10^6$ |
| $D_sD_s$ | 4.170 | 0.6 | 1.0 | $2\times10^6$ |

*1.4. BESIII Components*

BESIII is configured around a 1 T superconducting solenoid (SSM), considered to be optimum for precise momentum measurements for charged tracks in the τ-charm energy region, and the steel structure of its flux return. The 1 T superconducting solenoid replaces the 0.4 T conventional magnet of the original BES detector. The coil of the superconducting magnet is located outside of the electromagnetic calorimeter and has a mean radius 1.482 m and a length of 3.52 m.

Fig. 1 shows the configuration of BESIII, the main spectrometer components are indicated. The multilayer drift chamber (MDC) surrounds the beryllium beam pipe. Two superconducting quadrupoles (SCQs) are inserted in the conical shaped MDC end caps as close as possible to the interaction point. The time-of-flight (TOF) system consisting of two layers of plastic scintillator counters is located outside of the main drift chamber. The CsI(Tl) electromagnetic calorimeter (EMC) is placed outside of the TOF system and inside the SSM. The muon identifier (MU) consists of layers of resistive plate chambers (RPCs) inserted in gaps between steel plates of the flux return yoke. The polar angle coverage of the spectrometer is 21°< θ < 159°, and the solid angle coverage is ΔΩ/4π = 0.93. The main parameters and expected performance of BESIII are listed in Table 3, along with parameters of the BESII detector, which are given as references.

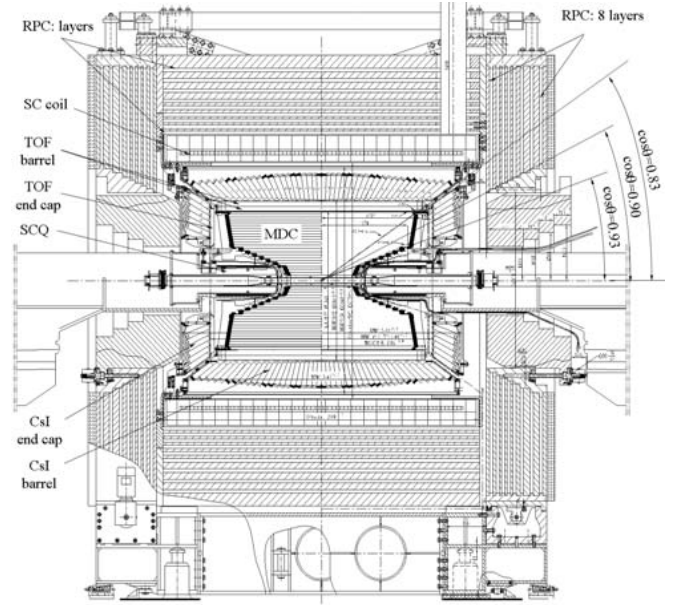

Fig. 1. Schematic drawing of the BESIII detector.

Table 3
Detector parameters and performance comparison between BESIII and BESII

| Sub-system | | | BESIII | BESII |
|---|---|---|---|---|
| MDC | Single wire $\sigma_{r\phi}$ (μm) | | 130 | 250 |
| | $\sigma_p/p$ (1 GeV/c) | | 0.5% | 2.4% |
| | $\sigma$ (dE/dx) | | 6 % | 8.5% |
| EMC | $\sigma_E/E$ (1GeV) | | 2.5% | 20% |
| | Position resolution (1 GeV) | | 0.6 cm | 3 cm |
| TOF | $\sigma_T$ (ps) | Barrel | 100 | 180 |
| | | End cap | 110 | 350 |
| Muon | No. of layers (barrel/end cap) | | 9/8 | 3 |
| | cut-off momentum (MeV/c) | | 0.4 | 0.5 |
| | Solenoid magnet Field (T) | | 1.0 | 0.4 |
| | ΔΩ/4π | | 93% | 80% (used) |

Tagging the *D* decays just above their production threshold by secondary vertices was not considered since the decay length of *D*'s is only about 30 μm. Since the momentum of charged particles from *D* decays is low, the multiple scattering makes precision determination of the secondary vertices by a vertex chamber impractical.

*1.4.1 Multilayer Drift chamber*

The multilayer drift chamber design was optimized for effectively tracking the relatively low momentum particles produced in the τ-charm energy region with excellent momentum resolution and good *dE/dx* measurement capability. The MDC also produces signals for the level 1 triggers to select good physics events and reject various backgrounds. The MDC inner radius is 59 mm and the outer radius is 810 mm.

A multilayer small cell design was adopted for the MDC, and a helium based gas mixture He-$C_3H_8$ 60:40 was chosen to minimize the effect of multiple scattering while maintaining a reasonable *dE/dx* resolution. In order to place the final focusing quadrupole as close as possible to the interaction point (IP), the end plates of the inner chamber have a stepped conical shape. The solid angle coverage of the MDC is $\Delta\Omega/4\pi$ = 0.93. The single cell position resolution is expected to be better than 130 μm in the *r-φ* plane and the position resolution in the beam direction at the vertex is expected to be ~2 mm measured by stereo wire superlayers with stereo angles in the range of -3.4° to +3.9°. The transverse momentum resolution is expected to be 0.5% at 1 GeV/c in the 1 T solenoid field. A 3σ π/K separation is possible up to 770 MeV/c with the expected 6% *dE/dx* resolution for particles with incident angle of 90° .

*1.4.2 Time-of-flight system*

The time-of-flight system is based on plastic scintillator bars read out by fine mesh photomultiplier tubes directly attached to the two end faces of the bars. It consists of a barrel and two end caps. The barrel TOF has two layers of staggered scintillating bars mounted on the outer surface of the carbon fiber composite shell of the MDC. Each layer has 88 scintillator bars that are 5 cm thick with a trapezoidal cross-section. The two single layer end caps, each with 48 fan-shaped counters, are located outside of the MDC end caps. The expected ~100 ps time resolution allows 3σ π/K separation to reach approximately 900 MeV/c at 90°. The solid angle coverage of the barrel TOF is |cos(θ)| < 0.83, while that of the end cap is 0.85< cos(θ) < 0.95. TOF counters also play a critical role as fast triggers for charged particles. The Hamamatsu fine mesh PMTs are directly attached to the ends of the scintillator bars without the long Lucite light guides used in the TOF system of the original BES detector. This is the main factor that contributes to the significant improvements in time resolution.

*1.4.3 CsI(Tl) crystal calorimeter*

The electromagnetic calorimeter is designed to precisely measure the energies of photons above 20 MeV and to provide trigger signals. It has good e/π discrimination capability for momentum higher than 200 MeV/c. The EMC consists of 6,240 large CsI(Tl) crystals located outside of the TOF counters and inside the coil of the solenoid. Its performance is greatly enhanced compared to the sampling EM calorimeter of the BES/BESII that was made of streamer tubes sandwiched in between lead plates. The inner radius of the BESIII EMC is 94 cm. The length of the crystals is 28 cm or 15 radiation length ($X_0$) with a front face measuring nominally 5.2 cm × 5.2 cm. The depth and the segmentation of the EMC are chosen to optimize the detector performance and to control the cost of the detector. The total weight of crystals is approximately 24 tons. The design energy resolution of EM showers is $\sigma_E/E = 2.5\%\sqrt{E}$ and the position resolution is $\sigma = 0.6\,cm/\sqrt{E}$ (E in GeV) at 1 GeV. The angular coverage of the barrel EMC is 144.7° < θ < 33.5° (|cos(θ)| < 0.83). The two end caps cover 21.3° < θ < 34.5° and 145.4° < θ < 158.7° (0.85 < |cos(θ)| < 0.95).

*1.4.4 Muon identifier*

The muon identifier consists of resistive plate counters interspersed in the steel plates of the magnetic flux return yoke of the solenoid magnet. The main function of the muon identifier is to separate muons from charged pions, other hadrons and backgrounds based on their hit patterns in the instrumented flux return yoke. It is important that the detector can identify muons with momentum as low as possible. The barrel muon identifier has nine layers of steel plates with a total thickness of 41 cm and nine layers of RPCs with the first layer placed in front of the steel. There are eight layers of counters in the end caps with the first layer behind 4 cm thick steel. The original BES detector had only a total of three active muon detector layers.

Muons as minimum ionizing particles lose 0.16 MeV of energy in the 28 cm long CsI crystals. The minimum momentum of muons that can be identified is also affected by the bending in the solenoid magnetic field. The minimum muon momentum at which the muon identifier starts to become effective is approximately 0.4 GeV/c.

*1.4.5 Trigger system*

The trigger, data acquisition and online computing systems are designed to accommodate multi-beam bunches separated by 8 ns and high data rate. The entire readout electronics system is pipelined to achieve a nearly dead-time free operation. The spectrometer must work in a high event rate and high background rate environment with high reliability, and must process a large amount of data in real-time.

The trigger system consists of two-levels. Level 1 (L1) is a hardware trigger, and the next level (L3) is a software trigger using an online computer farm. At L1, sub-triggers generated by the TOF, MDC and EMC sub-detectors are processed by the global trigger logic. The entire trigger system operates at 41.65 MHz synchronized with the accelerator RF. This trigger clock is distributed to readout electronics crates and synchronizes the operation of the entire BESIII data acquisition. The trigger logic is largely implemented in FPGA based hardware. Optical links are used between the readout crates and the trigger VME crates to improve the trigger performance and eliminate ground loops. The maximum L1 rate is expected to be 4 kHz at 3.097 GeV.

*1.4.6. DAQ system and event filter*

The BESIII DAQ (Data AcQuisition) system is based on VME and an online computer farm designed to read out large amounts of data from the front-end electronics system and record valid data on permanent storage devices. It adopts multi-level buffering, parallel processing, high speed optical data communication and high speed Ethernet technologies.

After L1 triggers are received, the DAQ system reads event data stored in buffers for each sub-detector from the VME crates and transfers the data to the online computer farm where the data are assembled into complete events and filtered.

The event filtering system (Level 3 trigger) is based on a computer farm running event reconstruction and filtering software. Commercial servers (farm) are used to run custom software to further suppress background events, to categorize the events, and to control and monitor the DAQ operation. The maximum data rate written to permanent storage is expected to be about 40 MB/sec while running at the $J/\psi$ peak recording with ~2 kHz of physics and ~1 kHz of background events.

*1.4.7. Detector control system*

The detector control system (DCS) that monitors the environmental parameters and the performance of spectrometer and accelerator is critical for BESIII operation. The slow control system has six subsystems that monitor thousands of sensors distributed throughout the spectrometer and detector hall. Critical parameters such as environmental temperature, humidity and radiation levels are monitored in real time. The system also reads the monitoring outputs of thousands of hardware devices and reports the device status, voltages, current and other parameters. The DCS controls and monitors the detector high voltage (HV) system. The DCS also controls and monitors the gas systems and provides safety interlocking among detector systems and between detector and accelerator.

The DCS is organized into the front-end layer, local control layer and global control layer (GCL). A LabVIEW based software framework is used for the data collection control. Data of approximately 9,000 readouts are recorded once every 10 seconds. Network communication and web server technologies are used for the data collection and GCL.

*1.5. Luminosity Determination*

Accurately measuring the integrated luminosity that BESIII receives is critical for achieving the physics goals of BESIII. The luminosity will be determined based on the three main QED processes $e^+e^- \to e^+e^-$, $\mu^+\mu^-$ and $\gamma\gamma$ using the entire BESIII detector. The cross-sections of these processes are very large and accurately known. By measuring the rates of these QED processes, and correcting for spectrometer acceptances and efficiencies, the luminosity can be accurately determined.

The BESIII luminosity measurements are made in two stages. First a crude determination of the luminosity is made in the real time L3 event filtering process using mainly the end cap EMC. This online luminosity is mainly used for data quality monitoring. The more accurate luminosity measurements will be made by offline data analysis to further eliminate backgrounds and to apply more detailed corrections for detector efficiencies and other effects determined from Monte Carlo simulation.

All three QED processes can be used to calculate the luminosity in the continuum regions, while $e^+e^- \to \gamma\gamma$ can be used in both the continuum and the resonance regions. For $J/\psi$ and $\psi(2S)$ resonances, special algorithms are adopted to use $e^+e^- \to e^+e^-$ and $\mu^+\mu^-$ to calculate the luminosity. Uncertainties for luminosity measurements from the three final states are different, and the results of the three independent measurements can be combined.

The main sources of uncertainties for absolute luminosity measurements are from trigger efficiency determinations, radiative corrections, Monte Carlo simulations and from requirements to eliminate backgrounds. It is expected that an accuracy of (3 - 6)% can be achieved initially. Within a year of running, the integrated luminosity can be measured to a level of less than 1%.

The luminosity monitors installed near the IP made of fused silica blocks are used to make relative determinations of the luminosity bunch-by-bunch by measuring the rate of photons from radiative Bhabha scattering events. This luminosity monitor is mostly used for beam tuning purposes.

## 2. Interaction region

*2.1. Interaction region beam-line magnets*

The collision hall for BESIII and the RF hall occupy the two straight sections in the north and south sides of the rings. Unlike the original BEPC, BEPCII is a double ring machine. The existing 237.5 m circumference of the BEPC tunnel imposes a strong limitation on the available space for the spectrometer and machine components in the interaction region. The horizontal beam crossing angle between the electron and positron beams is chosen to be ±11 mrad, based partially on the limited space in the interaction region (-14 m to +14 m). The relatively large crossing angle allows the beam bunches to be closely packed without parasitic collisions. The locations and parameters of the six pairs of quadrupole magnets that focus the beam in the interaction region (IR) and a pair of beam bending dipoles (OWBL) are given in Table 4 [8].

Fig. 2 shows the arrangement details of the final beam focusing magnets. Two very compact superconducting quadrupole micro-β magnets, designed and fabricated by Brookhaven National Lab in the United States, provide the final vertical focusing for the two beams. These superconducting magnets are located inside the 1 T spectrometer solenoid, starting only 0.55 m from the interaction point. Each has a multi-function coil package consisting of a vertical focusing quadrupole (SCQ), a horizontal bending dipole (SCB), a vertical steering dipole (VDC), a skew quadrupole (SKQ) and three anti-solenoid (AS1, 2, 3) windings for compensating the coupling due to the spectrometer solenoid.

Table 4
Parameters of the IR magnets

| Name | Position (m) | Length (m) | Horizontal displacement (m) | Quad. focusing strength ($m^{-2}$) |
|---|---|---|---|---|
| SCQ | 1.096 | 0.407 | 0.0 | -2.5787 |
| Q1A | 3.55 | 0.2 | 0.07754 | 1.245 |
| Q1B | 4.05 | 0.4 | 0.09375 | 0.655 |
| Q2 | 5.552 | 0.5 | 0.13567 | -0.3732 |
| Q3 | 9.553 | 0.5 | 0.24368 | -0.2376 |
| Q4 | 12.554 | 0.4 | 0.32341 | 0.6536 |
| OWBL | 13.52 | 0.932 | - | - |

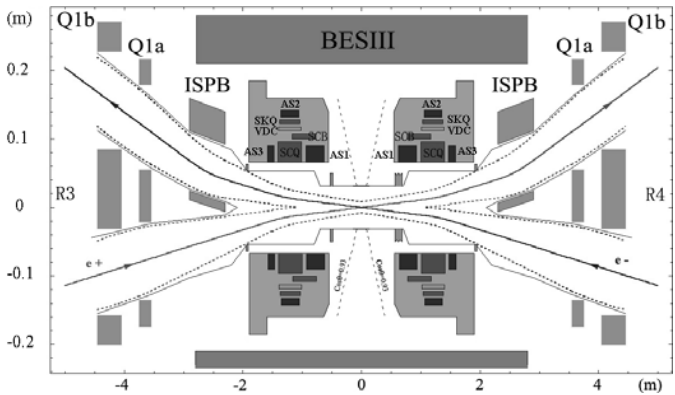

Fig. 2. Arrangement of the BEPCII beam focusing magnets.

Shared by two beams, the horizontal bending dipole coils in the two superconducting quadrupoles deflect the beams by ±26 mrad. Two special septum magnets labeled as ISPB provide further horizontal beam bending. These 0.6 m long, 0.4 T septum magnets enhance the separation between the electron and positron beams deflecting them in the inner ring by ±65.5 mrad. In order to minimize backgrounds from synchrotron radiation, the ISPB only acts on the outgoing beams and all the vacuum pipes up to ISPB are made of copper except for ±15 cm around the interaction point. On each side of the interaction region, a doublet of dual aperture room temperature quadrupoles labeled as Q1a and Q1b provide the horizontal focusing.

The BESIII detector solenoid has the maximum field strength of 1.0T over an effective length of ±1.8m around the IP. The longitudinal field inside the beam pipe is compensated by the three anti-solenoid (AS1, 2, 3) windings in each SCQ to avoid strong coupling of the betatron oscillations. AS1 cancels the integral longitudinal field from the IP to the SCQ, AS2 approximately cancels the field over the active length of SCQ and AS3 cancels the integral field in the fringe region beyond where the cold mass ends. A skew quadrupole SKQ included in the SCQ is used for fine tuning.

*2.2. Beam pipe*

The material of the beam pipe must be minimized to prevent secondary interactions and to minimize multiple scattering of particles produced in the $e^+e^-$ collisions before entering the sensitive volume of the spectrometer. The beam pipe walls must be very thin and be made from a material with low density and low atomic number. This requirement is especially important within the detector acceptance. The beam pipe walls must have good thermal conduction and good mechanical properties in order to withstand the high heat load, differential pressure and to maintain high vacuum that is very important for reducing the beam-gas backgrounds. The design dynamic vacuum pressure is $5\times10^{-10}$ torr in the IR. Its electric conductivity must be high in order to shield the RF radiation from the beam bunches. Beryllium is an ideal material that can meet all these requirements. Since beryllium is an expensive material that is difficult to machine and weld, fabricating a precise, thin wall beryllium beam pipe that is the first beryllium pipe of its kind designed and fabricated in China [9] imposed many technical challenges.

The cross-section of the BESIII beam pipe mounted on the end plates of the drift chamber is shown in Fig. 3. It is 1000 mm long with an inner diameter of 63 mm and an outer diameter of 114 mm. The beam pipe consists of two parts: the 29.6 cm long beryllium central pipe and two copper extension pipes joined together with the beryllium central pipe by welding. At the full beam current, the beam pipe is subject to a substantial heat load. The expected maximum heat load on the inner wall of the beam pipe is about 700 W, and it must be actively cooled.

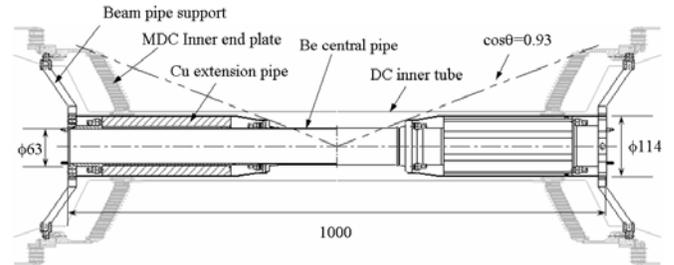

Fig. 3. Cross-sectional view of the beam pipe.

The thickness of the inner wall that maintains the ultra high vacuum in the beam pipe is 0.8 mm and the outer beryllium wall is 0.6 mm. A 0.8 mm channel between the two walls is used for flowing cooling fluid (high purity mineral oil). The beryllium central pipe is welded on to two double wall copper extension pipes that are cooled by water flowing in cooling channels.

Cooling is an important consideration for the beam pipe design. The beam pipe heating is mainly caused by the high order mode heating due to the trapped wake fields of the multi-bunch beam in the beam pipe wall. Interactions with synchrotron radiation photons, particles from beam-gas scattering and other beam losses also contribute significantly to the beam pipe heating. The heating power for the inner wall of the beam pipe is expected to be less than 700 W and the heating power for the 29.6 cm long central beryllium section can not exceed 200 W. Due to the proximity to the drift chamber, the temperature of the outer surface of the beam pipe must be maintained to within ±1°C of the ambient temperature. The beam pipe must be actively cooled.

The central beryllium beam pipe is terminated by two beryllium flanges. Four coolant connectors are mounted on each flange. The coolant used for the central beryllium pipe is a special paraffin oil or synthetic mineral oil (SMO). This oil was developed as an electron discharge machining fluid (EDM-1) with high flash point and low viscosity. Water cannot be used as a coolant because it is corrosive to beryllium. The two copper extension tubes welded on to the beryllium central tube are made of copper and cooled by water flowing in the channels between the thick outer tube and thin inner tube. Two copper flanges terminate the 1 m long beam pipe. Sixteen temperature sensors are attached on the surface of the beam pipe for monitoring and control purposes.

The inner surface of the beryllium beam pipe is coated with a 14.6 μm gold in order to attenuate the low energy synchrotron radiation photons and to reduce the hit rate of the MDC, in particular the inner layers.

The total thickness that particles produced at the IP must penetrate corresponds to 1.04% (Be: 0.4%, gold: 0.44%, SMO: 0.2%) of a radiation length at normal incidence.

*2.3. IP radiation monitoring*

The radiation background level and its fluctuations must be monitored in order to prevent the fragile beam pipe from accidental damage due to unusual beam conditions and also to ensure the radiation backgrounds in the spectrometer are within an acceptable limit. The temperature of the beam pipe must also be monitored to prevent damage due to high heat load, and the monitoring system must be able to react quickly. Twelve radiation monitors, six on each end, are mounted in the small gaps between the central beryllium beam pipe and the copper extension pipe (z = ± 200, r = 75 mm), and between the SMO cooling tubes. Each radiation monitor unit contains a PIN diode, two temperature sensors and their associated electronics, including amplifiers and ADCs.

XRB100s-CB380 PIN silicon diodes with a 380 μm thick intrinsic layer and non-transparent window are chosen as radiation sensors [10]. Beam particles and high energy photons that convert into electrons and positrons in the intrinsic layer of the reversely biased photodiode generate electric current proportional to the instantaneous radiation dose. Alarm or beam abort signals are generated within 10 ms once the instantaneous dose rate exceeds predetermined limits.

The sensitivity of the XRB100s-CB380 diode is approximately 30 nA/mGy·s$^{-1}$. The reverse current of the diode has a radiation damage rate of 3 nA/kGy. The dose measurement sensitivity is determined by the accuracy of measuring diode reverse current. The PIN diodes are sensitive to radiation dose rate at a level of approximately 20 μA/Gy·s$^{-1}$, corresponding to about 1 nA current increase at a dose rate of 0.05 mGy/s.

The temperature coefficient of the diodes, however, is relatively high (11%/°C) and it is necessary to compensate for changes in temperature. Glass-encapsulated thermistors YSI55036 [11] with 10 kΩ nominal resistance were chosen to measure the temperature of the PIN diodes in order to compensate the temperature effect. The design accuracy of the temperature measurement is 0.02 °C within the working temperature range of 10 °C to 40 °C. The resolution of the current measurement circuit is 0.1 nA with a 15 bit ADC.

*2.4. Luminosity monitor*

The very tight space near the interaction region does not allow accurate calorimetry based real time small angle luminosity monitors to be installed. The luminosity monitors discussed here have an accuracy level of ~15% and are mainly used to provide an instantaneous feedback for accelerator tuning. Two luminosity monitors [12], one on each side of the IP, are installed in the gaps between two parasitic sections of the vacuum chamber and between the ISPB septum for horizontal beam bending and the horizontal focusing quadrupoles Q1a. Luminosity monitors are situated ±3.1 m from the IP facing the incoming electron and positron beams.

The instantaneous luminosity computation is based on radiative Bhabha scattering ( $e^+e^- \to e^+e^-\gamma$ ) in which a photon is produced. Most of the photons are in the very forward angle within about 1 mrad of the beam. Radiative Bhabha photons escape the copper beam pipe wall through 3 mm thick windows and can be detected by the luminosity monitors.

The available space for the luminosity monitor is 180 mm long with a front width of 56 mm and rear width of 69 mm. The design of the luminosity monitor must be very compact. The top view of the luminosity monitor is shown in Fig. 4. A 3.5 mm thick tungsten target converts radiative Bhabha photons into electrons and positrons in front of a block of fused silica that has a cross-section of 40 mm × 45 mm and is 66 mm long. Two PMTs (Hamamatsu R7400U) are attached through a light guide to the back face of the silica block to read out the Cherenkov light generated by the electrons and positrons in the fused silica block.

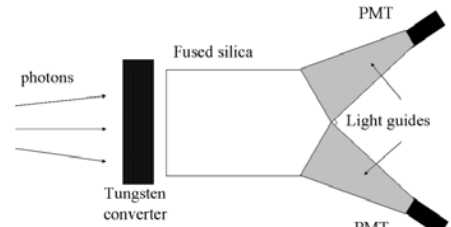

Fig. 4. Top view of the luminosity monitor.

This luminosity monitor was tested with cosmic rays. The number of photoelectrons produced by a minimum ionizing particle was measured to be about seven, and the detection efficiency better than 98%. The time resolution is 0.25 ns. The high speed electronics design fulfills the requirement to detect and record events within 4ns, sufficient to separate beam bunches that are spaced 8 ns apart. In order to minimize the influence from the very noisy environment of the collider, methods like discrimination, coincidence, and anti-coincidence techniques are adopted. The collision status and the bunch by bunch luminosity can be monitored based on instantaneous counts of these two monitors.

By steering the positron beam while observing how the counting rate of the luminosity monitor changes, the horizontal and vertical sizes of the electron and positron beam bunches were estimated to be 0.591 ± 0.018 mm and 0.032 ± 0.001 mm. The BEPCII collider was running at very low beam currents during these measurements.

*2.5. Beam related backgrounds*

*2.5.1. Beam background sources and protection*

During regular colliding beam operation, the background particles hitting the spectrometer are mostly produced in the vicinity of the interaction region where the six pairs of quadrupole magnets are located. Background sources, including synchrotron radiation, lost beam particles due to the beam-gas elastic Coulomb scattering, the inelastic bremsstrahlung and Touschek scattering, were studied using BEPCI/BESII and were simulated in BEPCII/BESIII by various computer programs in order to understand effects of backgrounds on detector safety and performance and to optimize the design of the collimator and mask systems.

Synchrotron radiation (SR) photons, mostly soft x-rays, are generated when beam particles undergo acceleration in magnetic fields of dipoles and quadrupoles. Photons are emitted in the direction near the tangent of the instantaneous trajectories. A thin layer of gold coating on the surface of the beryllium beam pipe can effectively attenuate the synchrotron radiation photons and greatly reduce their chances of entering the MDC.

A system of collimators and masks was placed at various locations around the collider ring and in the interaction region to reduce the chances that particles associated with lost beam particles affect the spectrometer.

*2.5.2. Beam background studies*

Background hit rates, instantaneous and accumulated radiation doses received by various spectrometer components were simulated by computer programs. Based on these studies, collider parameters, including the lattice, beam orbits and vacuum pipe apertures, were optimized, and collimators, shielding and masks were designed to minimize the effect of background sources.

The synchrotron radiation backgrounds are mainly produced when beam particles are focused or bent in the apertures of the final focusing magnets SCQ, Q1A, Q1B, Q2, Q3, Q4 and the weak dipole (OWBL). The space distribution of the synchrotron radiation photons and their interactions were simulated by three computer programs, SRGEN, SRSIM and EGS[13]. SRGEN followed positions and lateral motions of electrons and positrons in the magnet apertures, and generated synchrotron phones. The SRSIM and EGS were used to simulate photon interactions within the IR and detector components. The heating of the beam pipe and hit rates of MDC wires due to synchrotron radiation were calculated.

The synchrotron radiation power is quite high at BEPCII. The calculated radiation power distribution in the vicinity of the interaction point generated in the apertures of all beam line magnets is shown in Fig. 5. These low energy X-ray photons mainly affect inner layers of the MDC. The 14.6 μm thick gold film coated on the beryllium beam pipe is sufficient to limit the background hit rate in the MDC due to synchrotron radiation to an acceptable level.

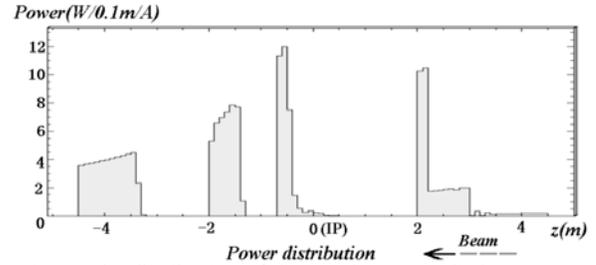

Fig. 5. The distribution of the synchrotron radiation power.

The beam particles that circulate in the collider rings interact with the molecules of residual gas in the beam pipe. Beam-gas backgrounds are mainly from elastic Coulomb scattering and inelastic bremsstrahlung. The deflected beam particles that escape from the beam pipe can directly enter the spectrometer or create electromagnetic showers when they are scattered from surfaces of beam line and spectrometer components in the IR. The Touschek scattering in which the beam particles in the same bunch interact is another source of beam related background. Effects of these background sources to the MDC, TOF and EMC were studied [14] using the computer simulation packages DECAY TURTLE [15] and GEANT 3 [16]. Eleven masks of various apertures around the collider were designed based on these studies in order to control these backgrounds. The accelerator parameters and vacuum assumed in these studies are given in Table 5.

Table 5
Accelerator parameters used in background simulations

| Accelerator parameters | | | | |
|---|---|---|---|---|
| Energy (GeV) | Current (mA) | Emittance eX (mm·mrad) | Emittance coupling eY/eX | Energy spread $\sigma_E$ |
| 1.89 | 900 | 0.144 | 1.5% | $5.16 \times 10^{-4}$ |
| Vacuum at various locations ($10^{-9}$ torr of 20% CO + 80% $H_2$) | | | | |
| (0, 224.1) m | (-13.42,-5) m | (-5, -2) m | (-2, +2) m | (+2, +5) m |
| 5.00 | 0.79 | 4.20 | 17.0 | 4.20 |

Fig. 6 shows the simulated hit rates of the first layer wires of the MDC due to beam-gas interactions and Touschek scatterings. The total hit rates and contributions from different processes are given. The wire labeling is such that wire number 1 and 40 are near the horizontal plane.

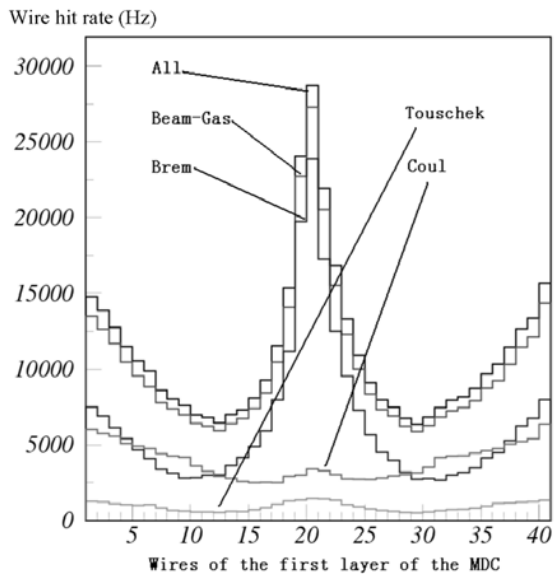

Fig. 6. Hit rates of the first layer wires of the MDC.

Fig. 7 shows the simulated hit rates of the 21st layer wires in the middle of the MDC due to beam-gas interactions and Touschek scatterings. The contribution from Touschek scatterings is larger than that from the beam-gas interactions in this case. Here wire number 1 and 160 are near the horizontal plane.

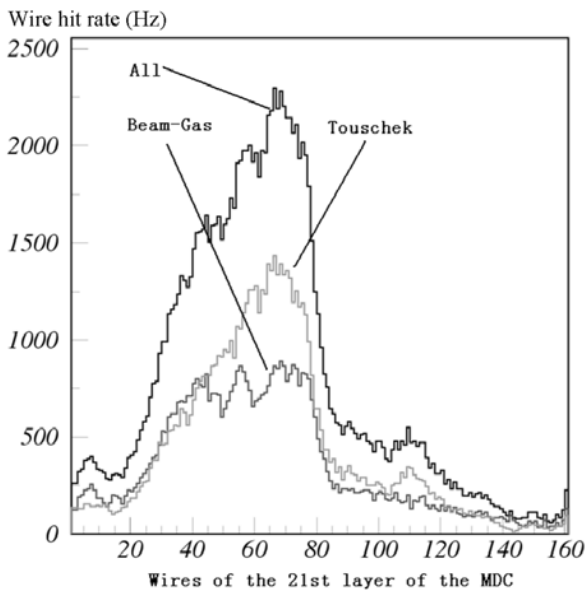

Fig. 7. Wire hit rates of the 21st layer of the MDC.

Hit rates in the first two layers of MDC wires, the accumulated radiation doses of the EMC crystals and the background hit rates in the TOF scintillation counters are given in Table 6. The threshold of the TOF hits was assumed to be 500 keV equivalent.

Table 6
Background hit rates and accumulated doses

|  | MDC wire hit rates (kHz) | | EMC Dose (rad/year) | | TOF counter hit rates (kHz) | |
| --- | --- | --- | --- | --- | --- | --- |
|  | Layer 1 | Layer 2 | Barrel | End cap | Barrel | End cap |
| Maximum | 31.60 | 18.40 | 0.77 | 5.27 | 3.93 | 3.14 |
| Average | 9.06 | 6.53 | 0.14 | 0.93 | 2.88 | 2.23 |

### 3. Solenoid magnet and flux return

The 1.0 T superconducting solenoid magnet (SSM) [17] of BESIII enables accurate momentum measurements of charged particles produced in the $e^+e^-$ collisions. Its instrumented steel flux return yoke serves as the hadron absorber for hadron/muon separation and provides the overall structure and support for the spectrometer components. The forward sections of the flux return are each split horizontally and are mounted on rollers that can be opened to allow access to the detector. The total weight of the flux return is approximately 498 metric tons.

*3.1 Parameters of the superconducting solenoid magnet*

The main parameters of the solenoid are summarized in Table 7. The superconducting solenoid magnet is designed to provide a uniform 1.0 T axial field for the 1.62 m diameter and 2.58 m long drift chamber. The inner diameter of the SSM is determined by the CsI crystal calorimeter that is located inside the superconducting coil.

Table 7
Main parameters of the solenoid coil

| Items | Parameters |
| --- | --- |
| Cryostat radius: outer/inner | 1.700 m/1.375 m |
| Cryostat length | 3.91 m |
| Coil mean radius | 1.482 m |
| Coil length | 3.52 m |
| SC core wire material ratio | Nb-Ti/Cu/Al (1/0.9/28.2) |
| Stabilizer | 99.998% aluminum |
| Cable dimension | 3.7 mm × 20 mm |
| Number of turns | 848 |
| Nominal current | 3,369 A |
| Central field | 1.0 T |
| Total weight | 15 t |
| Effective cold mass | 3.6 t |
| Inductance | 1.7 H |
| Stored energy | 9.8 MJ |
| Radiation thickness | 2.24 $X_0$ |
| Coolant | Liquid He at 4.5 K |

*3.2 Construction of the superconducting solenoid*

The BESIII superconducting magnet was constructed at IHEP with assistance from Wang NMR Incorporated [18]. It is the first superconducting magnet of this type built in China. The vacuum vessel of the superconducting coil, the chimney and valve box are shown in Fig. 8. The cross-sectional view of the superconducting cable, its backing cylinder and liquid helium cooling tube are also shown. The single layer coil was wound inside an aluminum support cylinder.

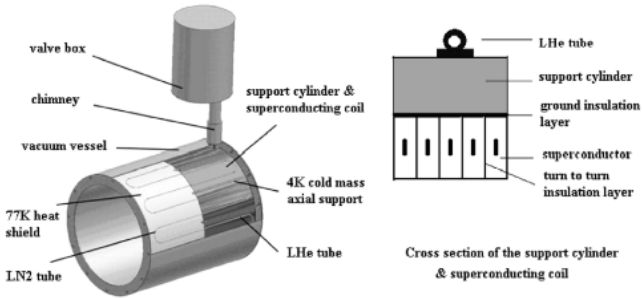

Fig. 8. BESIII superconducting solenoid and the cross-section of the superconducting cable.

The superconducting cable [19] was supplied by Hitachi Cable Ltd. The 0.7 mm diameter NbTi/Cu superconductor strands were formed into the 12 strand Rutherford cable measuring 1.26 mm × 4.2 mm. The Rutherford cable was imbedded in the center of a stabilizer made of high purity (99.998%, RRR 500) aluminum with outside dimensions of 3.7 mm × 20.0 mm. The conductor joints were made using helium inert gas welding. A quench protection system was designed for the BESIII SSM [20].

The thickness of the turn-to-turn conductor insulation, designed to withstand 100 V, is 0.075 mm consisting of 0.025 mm thick polyimide, 0.040 mm thick fiberglass reinforced epoxy and 0.010 mm thick epoxy. The total thickness of the coil to ground insulation, required to withstand 2,000 V, is 0.60 mm. It composes of two insulation films, each consists of two fiberglass reinforced epoxy layers and a polyimide layer, bond together by epoxy films.

Fifteen layers of super insulation films separate the liquid nitrogen thermal shield and the liquid helium cooled cold mass. The thermal shield is isolated from the vacuum vessel by another fifty layers of super insulation. The superconducting magnet is indirectly cooled to an operating temperature of 4.5 K by forced flow of two phase helium through cooling tubes wound on the outside surface of the support cylinder.

*3.3. Flux return yoke*

*3.3.1. Yoke design*

The magnetic flux of the solenoid is returned by two pole tips, the two forward yokes and the octagonal shaped barrel yoke. The 50 ton weight of the spectrometer inner components is supported by the steel structure. The shape and locations of the pole tips are designed to ensure the field uniformity of the solenoid. The barrel and forward yokes are segmented into layers, and are instrumented with resistive plate counters (RPCs) for muon/hadron separation. The yoke steel (barrel: 56 cm thick, end: 43 cm thick) is divided into nine layers and the thicknesses of the layers were chosen to optimize the performance of the muon identification as discussed later in Section 7.

An end yoke consists of two halves that are mounted on separate rollers. The two halves can be opened independently by driving motors. The valve box for the SSM is located on the top of the barrel yoke. There is a hole on top of the east side barrel yoke as shown in Fig. 1 for the 320 mm diameter chimney to pass through.

*3.3.2. Yoke construction*

The material chosen for the yoke is No. 10 low carbon steel (Chinese standard) that has sufficient strength and acceptable magnetic properties. The low cost steel has 0.1% carbon content on average with a possible range of variation from 0.07% to 0.13%. The magnetic properties of the steel plates used in the yoke construction were closely monitored during the construction to ensure that their quality was within specifications. The nine layers of steel plates are connected by screws to form an octant, and octants are joined together by connecting brackets. The design of the yoke structure minimizes the dead space at the boundary of octants.

*3.3.3. Magnetic field optimization*

The uniformity of the axial magnetic field in the tracking volume, especially in the region near the two ends of the solenoid, is affected by the design of the flux return yoke. A steel shielding is added outside the gap between the barrel and end yokes to reduce the fringe field. Calculations show that the non-uniformity of the magnetic field in the tracking volume is less than 2% not including the superconducting quadruple magnet (SCQ) and rises to 13% once the SCQ is taken into account. The effect of SCQs on the SSM field uniformity is particularly strong near SCQs.

*3.4. Magnetic field map*

The uniformity of the SSM is affected by the circumferential asymmetry of the return yoke, the final focusing quadrupoles and the SCQ. As discussed in Section 2, three anti-solenoid coils are included in the SCQ for magnetic field compensation. For measuring the momentum of charged particles, the three components of the magnetic field vectors in the tracking volume of the drift chamber must be precisely known. The goal of the magnetic field mapping, and subsequent field computations and fitting was to determine the magnetic field to a precision of better than 1 mT with a position accuracy of better than 1 mm in the axial and radial directions. The angular accuracy should be better than 0.1°. A set of polynomials, a few trigonometric-Bessel terms and several dipole terms were used to fit the mapping data. The fitted parameters will be used for particle momentum calculations and for accelerator beam tuning.

*3.4.1. Mapping procedure*

The magnetic field in a cylindrical volume of 2.6 m in diameter and 3.5 m in length inside the SSM was mapped by a specially designed, automated field mapping device that can move precisely in the *r* and *z* directions driven by ultrasonic motors, and in the φ direction by a rotating motor. Three precise Hall magnetic field sensors that measure the three components of the magnetic field $B_z$, $B_r$ and $B_\phi$ are encapsulated in the field probe. The precision of sensors is better than 0.1% (0.001 T in nominal field of 1.0 T). Three such 3D probes are positioned along a radial line having the same φ-coordinate. They covered the radial ranges of 0 < r < 450 mm, 200 < r < 650 mm and 450 < r < 900 mm.

The step size was optimized based on the expected non-uniformity of the magnetic field. In the central region where the magnetic field was more uniform, a step size of 5 cm was used. In the region near the two ends of the solenoid where the field changed more rapidly, a 2.5 cm step size was used. The step size in φ was chosen to be 11.25°. As an example of the field mapping results, the distribution of measured $B_z$ for points on a cylindrical surface at R = 450 mm and Z from -1100 mm to 1100 mm is shown in Fig. 9 for case 2 listed in Table 8.

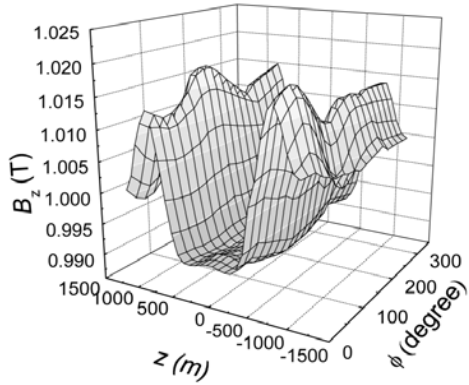

Fig. 9. Distribution of $B_z$ in a portion of the mapping volume.

The mapping was done for seven combinations of field values in the SSM and SCQs. The excitation currents of the SCQs will be set to different values when the machine is operated at different beam energies. The SCQ current values corresponding to zero, 1.55, 1.89 and 2.1 GeV beam energy were used. The nominal field values of the SSM were zero, 0.9 T and 1 T. The seven data sets corresponding to the seven field combinations, as listed in Table 8, are sufficient to allow field computations for different operating conditions that might be used. The field mapping was controlled by a computer and completed in 20 days.

Table 8
Field mapping combinations

| Case No. | SCQ current settings (GeV) | SSM field (T) |
|---|---|---|
| 1 | 0 | 1.0 |
| 2 | 1.55 | 1.0 |
| 3 | 1.89 | 1.0 |
| 4 | 2.1 | 1.0 |
| 5 | 1.89 | 0 |
| 6 | 1.55 | 0.9 |
| 7 | 1.89 | 0.9 |

*3.4.2. Linearity of the SCQ magnetic field and current*

The magnetic fields of the SCQs can significantly alter the magnetic field distribution in the tracking volume. Since the accelerator will run at energies different from the three energies listed in Table 8, the linearity of the SCQ magnetic fields as a function of the driving currents was studied. Three mapping runs were taken when the SCQ current settings were adjusted for beam energies of 1.55 GeV (case 2), 1.89 GeV (case 3) and 2.1 GeV (case 4) while SSM (1.0 T) and AS coils (1.0 T) settings were kept the same. Bx, By and Bz, for all measured data points within the MDC tracking volume were recorded. The values of Bx3, By3 and Bz3 in case 3 were calculated by using the measured field values in case 2 and case 4 assuming a linear relation exists between the magnetic fields and the SCQ currents. The standard deviations of differences between measured and the extrapolated values of case 3 magnetic field components Bx3, By3 and Bz3 were 2.65 G, 2.64 G, 2.85 G, respectively. This study shows that for other beam energies, the SCQ currents can be set based on a linear relation between the SCQ magnetic fields and driving currents, and the effect of nonlinear saturation of the return yoke is insignificant.

## 4. Multilayer drift chamber

### 4.1. MDC design considerations

#### 4.1.1. Overview

The main functions of the BESIII MDC are:
- to reconstruct charged tracks in 3D space.
- to determine the momentum of charged particles produced in the $e^+e^-$ collisions.
- to measure *dE/dx* for identifying charged particles.
- to reconstruct long lifetime hadrons (mainly $K_S^0$ and Λ) that decay in the MDC volume.
- to form the L1 trigger based on reconstructed charged tracks and reject background tracks.
- to provide extrapolated track positions at the outer detector components.

The diameter of the MDC is constrained by the limited space and the total cost of the spectrometer. The outer radius of the MDC is chosen to be 810 mm and the inner radius is 59 mm, 2 mm from the beam pipe. The maximum length of the outer chamber is 2,582 mm. The polar angle coverage of the outermost and innermost wire layers is $|\cos\theta| \leq 0.83$ and $|\cos\theta| \leq 0.93$, respectively. The MDC consists of an outer chamber and an inner chamber, which are joined together at the end plates, sharing a common gas volume. The inner chamber can be replaced in case of radiation damage. The end plate of the inner chamber has a stepped conical shape to accommodate the intrusion of the micro-β focusing magnets.

The overall design of the BESIII MDC [21] is similar to the design of the CLEO-III drift chamber [22]. A small cell design with 43 sense wire layers was chosen and the shape of a drift cell is almost square. In the 1 T solenoid field, once the radius and the number of layers are fixed, the momentum measurement precision is determined mainly by two factors: multiple scattering of the charged tracks by materials inside the drift chamber volume and the single sense wire position resolution. Within the center of mass energy range of 2 to 4.6 GeV, the momentum of most secondary charged particles generated in the $e^+e^-$ collisions is well below 1 GeV/c. Multiple Coulomb scattering in the MDC gas volume plays a critical role in the momentum determination, so the materials in the drift chamber must be minimized. For this reason, thin aluminum field wires and a helium based gas mixture were used. The drift chamber gas mixture was chosen to be $He/C_3H_8$ 60:40. This helium based gas with sufficient hydrocarbon gas has a long radiation length of 550 m and can ensure the stable operation of the drift chamber. The specific primary ionization is adequate for *dE/dx* measurements.

*4.1.2 Drift cells*

The BESIII MDC adopted the small cell drift chamber design. A sense wire is surrounded by 8 field wires forming a drift cell. The cell height and width are identical for cells in different layers in order to keep the gas gain the same under the same high voltage. The cell width of the inner chamber (layers 1 to 8) is approximately 12 mm and the cell width of the outer chamber is 16.2 mm on average. The single wire resolution is dominated by electron diffusion. Disadvantages of choosing a cell size that is too small include increased readout channels and the multiple scattering caused by additional wires [23].

The 25 μm diameter sense wires are gold plated tungsten with 3% rhenium. In order to minimize the material, field wires are made of 110 μm diameter gold plated aluminum. The aluminum is a partially tempered 5056 aluminum alloy. The drift field lines in a 1 T solenoid field, the distance-time (*r-t*) relations of drift cells and position resolution of the MDC were simulated by Garfield [24]. The average single wire position resolution is expected to be ≤ 130 μm. Main factors that contribute to the single wire resolution are electron diffusion estimated to be about 60 μm and the time measurement uncertainties due to electronics estimated to be less than 20 μm. The statistical fluctuations of the primary ionizations along tracks and uncertainties of sense wire positions also contribute.

*4.1.3 Layer arrangement*

The drift chamber contains 43 sense wire layers arranged as 11 superlayers. There are 4 layers of sense wires in each of superlayers 1 to 10 and 3 layers in the last superlayer. Layers 1 to 8 and 21 to 36 are small angle stereo layers. Layers 9 to 20 and 37 to 43 are axial. The field wires are shared between neighboring cells within a superlayer.

Z-positions of tracks along sense wires are determined by stereo and axial wires. For the stereo angles chosen, the z-resolution of stereo wires is expected to be in the range of 3 mm to 4 mm. A layer of stereo wires with identical radius on the end plates forms a rotating hyperboloid surface. The maximum inward radial shift in the middle is 2.4 mm. Effects of such a shift on the track position measurements must be considered carefully. Additional field wires are added between superlayers at the boundary of axial and stereo layers. These additional field wires are necessary to compensate electric field variations along the stereo wires [25]. Track segments reconstructed in superlayers are linked and used in the L1 trigger. Details of the layer arrangement are given in Table 9.

Table 9
Geometric parameters of the wire layers

| Items | Parameters |
| --- | --- |
| Radius of inner most field wire | 73 mm |
| Radius of outer most field wire | 789 mm |
| θ-coverage: inner most layer | \|cosθ\| ≤ 0.93 |
| θ-coverage: outer most layer | \|cosθ\| ≤ 0.83 |

| Layer No. | Superlayer No. | Tilted angle |
| --- | --- | --- |
| 1 - 4 | 1 | U: -(3.0° – 3.3°) |
| 5 - 8 | 2 | V: +(3.4° –3.9°) |
| 9 -20 | 3 - 5 | A: 0° |
| 21-24 | 6 | U: -(2.4° – 2.8°) |
| 25- 28 | 7 | V: +(2.8° – 3.1°) |
| 29 - 32 | 8 | U: -(3.1° – 3.4°) |
| 33 - 36 | 9 | V: +(3.4° – 3.6°) |
| 37 -43 | 10 - 11 | A: 0° |

*4.1.4. Expected momentum resolution*

In a multilayer tracking chamber with equally spaced wire layers along the particle trajectories in a uniform axial magnetic field, a simple model can be used to estimate the transverse momentum resolution. In this model, the resolution $\sigma_{p_t}$ of the transverse momentum can be approximately expressed as

$$\frac{\sigma_{p_t}}{p_t} = \sqrt{\left(\frac{\sigma_{p_t}^{wire}}{p_t}\right)^2 + \left(\frac{\sigma_{p_t}^{ms}}{p_t}\right)^2},$$

where $p_t$ is the transverse momentum of particles, $\sigma_{p_t}^{wire}$ is the momentum resolution resulted from the uncertainties of position measurements of individual wires and $\sigma_{p_t}^{ms}$ is the momentum resolution contribution due to multiple scattering of tracks inside the tracking chamber. The first term on the right of the above equation is given by

$$\frac{\sigma_{p_t}^{wire}}{p_t} = \frac{3.3 \times 10^2 \times \sigma_x}{B \times L^2} \times p_t \times \sqrt{\frac{720}{n+5}}$$

where $\sigma_x$ is the position resolution of a single wire in meters, *B* is the magnetic field in Tesla, *L* is the track length in meters, $p_t$ is the transverse momentum in GeV/c and *n* is the total number of sense wire layers. The multiple scattering term can be expressed as

$$\frac{\sigma_{p_t}^{ms}}{p_t} = \frac{0.05}{B \times L} \times \sqrt{1.43 \frac{L}{X_0}} \left(1 + 0.038 \ln \frac{L}{X_0}\right)$$

where $X_0$ is the radiation length in meters of the material assumed to be uniformly distributed inside the tracking volume. Here we assume particles have β = 1. In the 43 layer MDC, the effective length L of tracks at a 90° polar angle is approximately 70 cm. By assuming $p_t$ = 1 GeV/c, $B$ = 1 T and $\sigma_x$ = 130 μm, we have $\sigma_{p_t}^{wire}$ = 0.32% and $\sigma_{p_t}^{ms}$ = 0.35%. The expected momentum resolution of the MDC for 1 GeV track at 90° is

$$\sigma_{p_t} = \sqrt{\sigma_{p_t}^{wire} + \sigma_{p_t}^{ms}} = \sqrt{0.32\% + 0.35\%} = 0.47\%.$$

This result agrees well with more elaborate Monte Carlo calculations [25]. Based on these studies, the momentum resolution of the MDC is expected to be better than 0.5% at $p_t$ = 1 GeV/c.

*4.1.5. Expected dE/dx performance*

Factors contributing to the *dE/dx* resolution include the fluctuations of the number of primary ionizations along the track, the recombination loss of electron-ion pairs at the corners of the drift cell where the electric field is low and fluctuations in the avalanche process. The density of the helium based gas is relatively low, and the most probable value of the number of primary ionization pairs is about 50 per centimeter for minimum ionizing particle tracks. Monte Carlo simulations show that the dE/dx resolution of the MDC is about 6% allowing 3σ π/K separation up to momenta of ~770 MeV/c.

*4.2. Prototype studies*

*4.2.1. Test setup*

The single wire resolution and *dE/dx* performance of the MDC drift cell design was studied with a full length prototype. Extensive cosmic ray and beam tests were conducted [26].

The prototype consists of 78 drift cells, arranged in 12 layers of 6 to 9 cells to model different regions of the cylindrical chamber. Among the layers, 5 layers have the same geometry as the inner chamber layers, and 7 layers have the outer chamber cell geometry. All wires are axial. The wires and gas mixtures used in the prototype are exactly the same as used in the final MDC.

Beam tests of the prototype were first performed using secondary beams produced at the 12 GeV/c pion beam line at KEK. The prototype chamber was placed in a 1 T magnetic field. Both positive (p and π$^+$) and negative (e$^-$ and π$^-$) beams were used. The momentum of the beams was adjusted from 0.6 GeV to 4 GeV. Later more tests were done at the IHEP test beam line using electrons and secondary particles including electrons, pions and protons produced by the beam of the 1.89 GeV electron LINAC hitting copper or beryllium targets. This test beam was extensively used during the BESIII construction. Its main parameters are summarized in Table 10.

The sense wire readout electronics was the same as the final MDC electronics as discussed in Section 4.4 below. Signals were amplified by trans-impedance preamplifiers. The preamplifier output signals were sent to the main amplifier and split into two branches. One branch is discriminated and fed to a TDC for timing measurements. The other was amplified, shaped and fed to an ADC for charge measurements.

Table 10
IHEP E3 test beam parameters

| Parameters | Momentum range |
|---|---|
| e$^{\pm}$ | 200 – 1,100 MeV/c |
| π$^{\pm}$ | 400 - 900 MeV/c |
| Protons | 500 - 1000 MeV/c |
| Momentum uncertainty Δp/p | ≤ 1% |
| Position uncertainty | 0.2 to 0.4 mm |
| Rate | 3 to 4 Hz |

*4.2.2. Optimizing drift cell operation conditions*

The effect of magnetic field on the *r-t* relation in the drift cells was studied in detail. Fig. 10 shows the *r-t* scatter plot of a drift cell in the 1 T magnetic field. The effect of the magnetic field on the performance of the prototype was found to be insignificant except that the boundary layers performed slightly worse than the inside layers. This is expected since the Lorentz angle of He/C$_3$H$_8$ 60:40 gas mixture is rather small. Based on test results, a high voltage of +2,200 V on the sense wires was found to be optimum for MDC operation. The field wires were at ground potential, except those at the boundary where positive compensating voltages of 100 V to 200 V were applied to reduce the left–right asymmetry caused by the magnetic field. The effect of different incident angles was also studied. The *r-t* relations for tracks of different incident angles and different types of cells were parameterized by fitting to a fifth order polynomial plus a first order polynomial to accommodate regions near the cell edge close to the field wire. The parameterization database is used in the data analysis for track reconstruction.

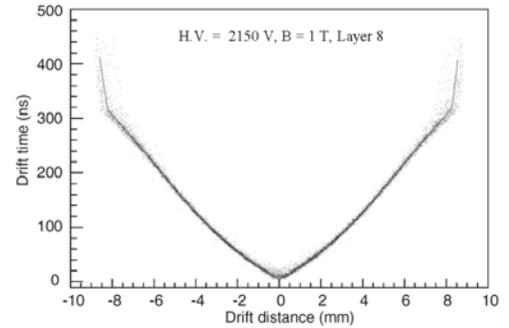

Fig. 10. The measured r-t relation of a drift cell and fitted curve in 1 T magnetic field.

*4.2.3. Single wire resolution*

The single wire resolution of a typical drift cell as a function of track distances to the sense wire is plotted in Fig. 11. A slight left-to-right asymmetry produced by the effect of the 1 T magnetic field is observed. After track fitting, the residual distribution was fitted to two Gaussian functions, and the standard deviation σ was calculated by

$$\sigma = \sqrt{\frac{N_1\sigma_1^2 + N_2\sigma_2^2}{N_2 + N_2}}$$

where $N_1$ and $N_2$ are the amplitudes and $\sigma_1$ and $\sigma_2$ are the standard deviations of the two Gaussian functions. The average position resolutions of drift cells in all 12 layers of the prototype are plotted in Fig. 12, and the average position resolution of all drift cells was found to be 112 μm.

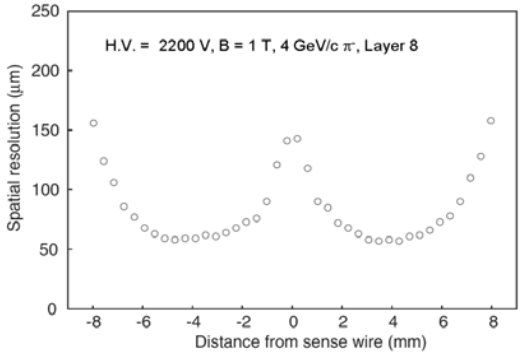

Fig. 11. Position resolution as a function of track impact parameter.

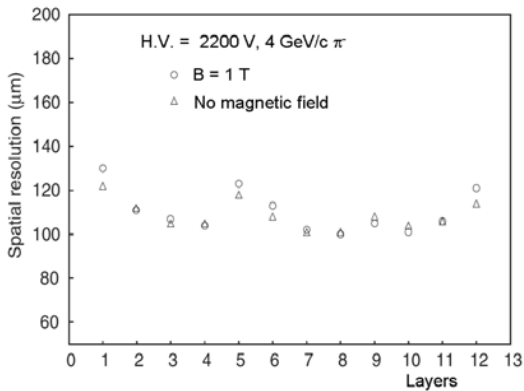

Fig. 12. Average position resolutions of all 12 layers.

*4.2.4. dE/dx resolution*

The data from the seven outer chamber layers were used for studying the *dE/dx* performance. The method of the truncated means that converts the Landau spectrum to a Gaussian-like spectrum was used to reduce the effect of the Landau tail fluctuations. Measured high *dE/dx* values are truncated and the remaining data are used to compute the *dE/dx* resolution. Beam tests were conducted using a 4 GeV/c $\pi^-$ beam at KEK. In Fig. 13, *dE/dx* resolutions as a function of fractions of the 35 *dE/dx* data samples along tracks are plotted. It was found that the best *dE/dx* resolution was 4.5% when 80% of the data sample were accepted. We expect the *dE/dx* resolution of the BESIII MDC under real experimental conditions will be somewhat worse than the result obtained in the beam test.

*4.3 Mechanical design and constructions*

*4.3.1. Structural components*

The cross-sectional view of one quarter of the MDC mechanical structure is shown in Fig. 14, where dimensions of the structure are given. The MDC consist of two parts, an outer chamber and an inner chamber, and they are joined together at the end plates. The inner chamber can be replaced in case it is damaged by radiation. The outside shell of the MDC outer chamber, made of 11.5 mm thick carbon fiber composite, provides the mechanical support to bear the total tension load of 3.8 metric tons. Eight large observation windows were opened in the shell. The inner chamber has only a 1.2 mm thick carbon fiber composite inner shell that also provides some mechanical strength. The end plates of the outer chamber consist of two sections, the conical section and the stepped section. Two thick aluminum rings at the end of the outer carbon fiber barrel connect the barrel to the two slightly conical outer end plates that are 18 mm thick and made of 7050 aluminum alloy. The inner section of the outer chamber end plate has a stepped design allowing micro-β SCQs to be placed as close as possible to the interaction point and also helps to minimize the deformation of the end plates under the tensioning load. Two cylindrical masks that are 22 mm thick and 679 mm long made of depleted uranium surround the front portion of SCQs and protect the spectrometer from background photons and electromagnetic debris created by lost beam particles.

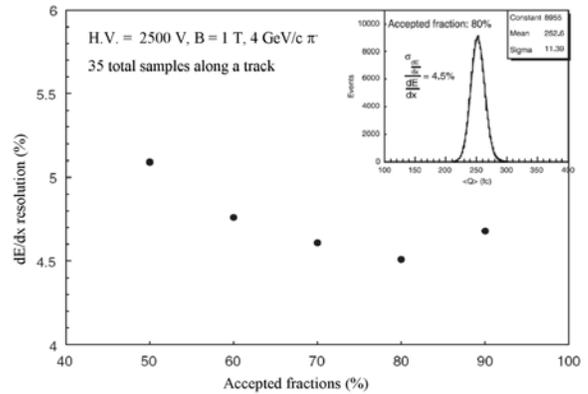

Fig. 13. Measured *dE/dx* resolution as a function of the fractions of 35 measurements used.

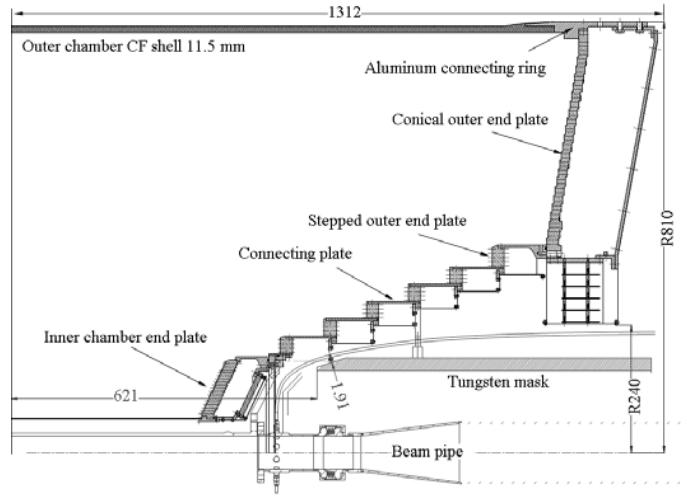

Fig. 14. The MDC mechanical structure.

A total of approximately $2 \times 28,640$ precision holes for the wire feed-throughs were drilled by a large CNC machine. The diameter of the drilled holes is $3^{+0.025}_{-0}$ mm. Drilling the deep holes accurately on the complex shaped end plates made of a high strength aluminum alloy was difficult even with modern multi-axial CNC machines. The completed end plate components were measured at the factory on large computer controlled measuring machines (CMMs) to ensure hole positions were within tolerances. The achieved average precision was better than 25 µm rms. A database of the sense wire hole positions was established based on the CMM measurements.

The inner section of the outer chamber end plate consists of six circular sections joined together by 3 mm thick aluminum connecting plates. Each step of the inner section holds two rows of axial sense wires (superlayers 3 to 5). Joining the precision machined disk sections and rings accurately required elaborate tooling and alignment procedures aided by four precision optic surveying instruments. The achieved accuracies of the end plate assembly is < 0.05 mm, and the non-concentricity is < 0.1 mm. The misalignment between the two end plates is < 0.1 mm, and their non-parallelism is < 0.12 mm. After the wiring of the outer chamber was completed, the axial deformation of the outer cylinder was less than 0.05 mm under the tension load of 3.8 tons.

The inner chamber has two 25 mm thick conical shaped end plates, holding 8 layers of stereo sense wires. They were joined together only by the 1.2 mm thick (0.45% $X_0$) 126 mm diameter inner shell. The axial deformation of the inner cylinder was within 50 µm at 100 kg wire tension load. The two endplates are required to be parallel within 200 µm and their concentricity better than 100 µm.

After the wiring of the inner chamber was completed, the inner chamber and the outer chamber were joined together by screws. The achieved accuracy of the parallel alignment of the inner chamber and the outer chamber was better than 120 µm, and the uncertainty of lateral positioning between inner and outer chamber was better than 200 µm.

*4.3.2. Wire feed-throughs*

Wire feed-throughs are a critical component in a precision drift chamber. Feed-throughs must provide accurate position registrations through the 18 mm to 25 mm thick aluminum end plates and must be able to hold the 2.2 kV anode high voltage with minimum leakage currents. A feed-through consists of three parts: a 3.18 mm diameter plastic outer bushing for high voltage insulation; a 1.56 mm diameter copper tube inserted in the bushing for high voltage connection and a 0.8 mm diameter wire crimp pin inserted in the central hole of the copper tube. Wire crimps pins for aluminum field wires are made of an aluminum alloy that is softer than the field wire so that mechanical crimping does not damage the wires. Crimp pins for the sense wires are made of soft copper.

The radius tolerances are ± 0.013 mm for the copper tubes and ± 0.025 mm for the insulating bushing. The tolerance on their concentricity is ± 0.025 mm. The material of the insulating bushing is Vectra A130, a liquid crystal polyester with 30% glass filling that has good mechanical stability, high bulk resistivity and high dielectric strength. Vigorous tests were conducted in order to ensure the design and material choices of the feed-throughs were correct.

*4.3.3. Assembly and stringing*

The wire stringing of the inner and outer chamber was completed within less than six months in a clean room using a specially designed wiring machine. Wires were crimped with handheld tools. The parameters related to the MDC wires are summarized in Table 11. Accurately controlling the wire tension was an important consideration. Since lengths of wires vary in the chamber, wire tensions must be adjusted in order to keep the gravity sags of wires to within the design value of 50 µm for the inner chamber, 100 µm for the stepped sections, and 120 µm for other layers. The effect of the overall wire sags to the position resolution will be corrected offline. The effect of the end plate deformation under the tensioning forces was calculated and compensated for during the wiring process.

Table 11
Parameters related to the MDC wires

| Items | Parameters |
|---|---|
| No. of sense wires | 6,796 |
| No. of field wires | 21,844 |
| Sense wire material | Gold plated W/Rh |
| Sense wire diameter | 25 µm |
| Sense wire tension | |
| Inner chamber | (15 to 18) g ± 5% |
| Outer chamber | (14 to 52) g ± 5% |
| Field wire material | Gold plated 5056 Al alloy, partially tempered |
| Field wire diameter | 110 µm |
| Field wire tension | |
| Inner chamber | (47 to 59) g ± 5% |
| Outer chamber | (47 to 173) g ± 5% |

In order to determine the final tension of wires, a long term tension study was conducted for a group of 110 µm aluminum field wires on a test bench. These wires were initially tensioned at 160 g. Long term wire creeping is expected to be important for aluminum field wires, and their tension can be reduced by up to 15% of the initial tension after 6 months due to creeping. The rate of creeping is high initially and becomes much slower after a few months under tension.

The wire stringing for the inner and outer chambers was done separately. Special tooling and a wiring machine that controls the wire feeding and tensioning were developed. For the outer chamber, the two end plates were attached to two strengthening rings that were connected by 252 tension rods without stressing the outer carbon fiber cylinder. The chamber structure was suspended vertically in a clean room during the wire stringing. Under computer control, the chamber structure can be rotated with its axis tilted at a variable angle from the vertical and can be flipped over. Rotations were necessary while stringing stereo wires and conducting cosmic ray tests.

Wire tensions were subject to change during the stringing process as more and more wires are added and the amount of end plate deformation increased. A finite element analysis was preformed to simulate this process in order to ensure the initial wire tension was set correctly so that the final tension would reach the designed value. The initial wire tension was determined based on considerations that included the wire length, gravitational sag, relaxations of aluminum wires due to creeping and the deformation of the endplates. Parameters of the initial wire tension, pre-stressing and releasing, wire string rate and the wire tension values at different time were considered and used as guidance for wire stringing. Measured tensions for all wires were in good agreements with the simulation three months after the last wire was strung. Calculated and measured tensions of field wires and sense wires are shown in Figs. 15 and 16, respectively. The wire lengths in the chamber were in the range from 0.8 m to 2.2 m. Tensions of wires with different lengths were different in order to match gravitational sags.

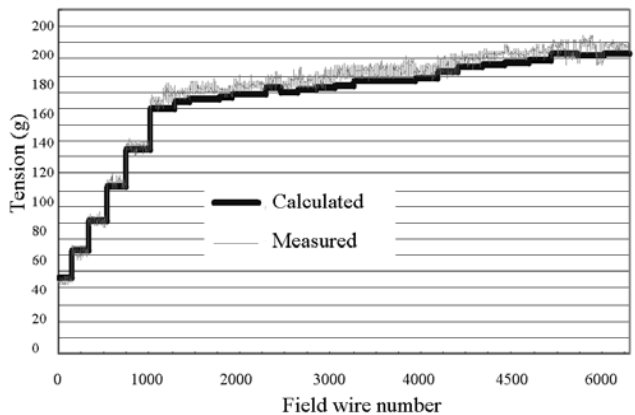

Fig. 15. Calculated and measured tensions of all field wires in the outer chamber.

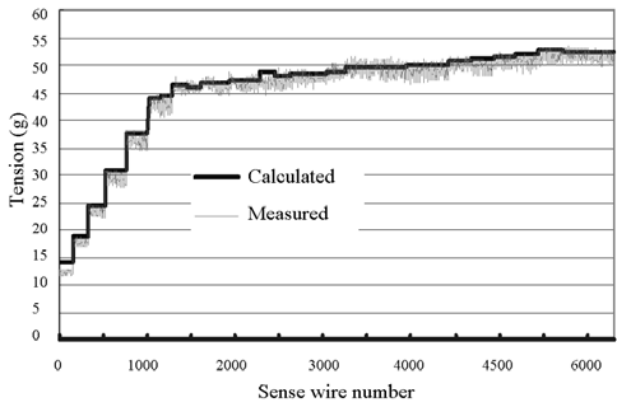

Fig. 16. Calculated and measured tensions of all sense wire in the outer chamber.

Ensuring the wire tension uniformity is a very important task for the wiring quality control. Limits for accepted variations of initial tensions were set to within ±6% of the average value for all field wires and ±4% for all sense wires. After wire tensions became stable, the tension variations were found to be within ±8% for all field wires and ±5% for sense wires. Wire tensions were measured by a 32 channel automatic tension meter. Wires to be measured were excited through capacitive coupling by varying the electromagnetic frequency applied to their neighboring wires, and resonant frequencies of 32 wires were measured simultaneously. The uncertainty of tension measurements was ≤ 1%. A wire would be replaced if its tension was outside the acceptable range. Different types of epoxy were used to seal the wire crimp pins, feed-through holes, screws, potential gas leak paths between the end plate sections and connecting plates and other joints.

### 4.3.4. Wiring quality assurance

Vigorous Q&A procedures were followed during the MDC wiring. The tension of every wire was measured. Approximately 1.4% of wires were replaced because their tensions were out of tolerance. Leakage current in air under the working high voltage for every wire was required to be less than 5 nA after a short period of training. No wires were replaced because leakage currents were too large. After the inner and outer chambers were connected and sealed, the gas tightness was tested. The helium leak rate was measured to be 20 mL/min. corresponding to about 1% of the planned gas flow rate of the MDC. This leakage rate was considered safe for the photomultipliers in the TOF system.

### 4.3.5. Cosmic ray tests in the clean room

After the construction of the MDC was finished, cosmic ray tests were performed in the chamber wiring room as a final step of the Q&A procedure. The MDC was rotated to a horizontal position, and two groups of four large scintillating counters covering the full length provide cosmic muon triggers. Approximately 1,800 cells could be tested at once with a fairly high event rate. The MDC was rotated several times until all 6,796 cells were fully tested. No major problems that required replacing wires were found. Small numbers of dead channels, noisy channels and channels with abnormal time or charge distributions due to electrical or electronics problems were found and fixed.

The high voltages of different layers were fine tuned so that measured average charges for minimum ionizing particles agree within 5%. Discrimination thresholds for the inner chamber and the sloped and the stepped sections of the outer chamber were optimized and finalized. Fig. 17 shows the wire efficiency as a function of distance to sense wires in layer 40. The measured average wire efficiency of the MDC is approximately 98%. Fig. 18 shows the residual distribution of the outer layer wires, and the average position resolution of single wires is approximately 120 μm. The fitting procedure was similar to that was described in section 4.2.3.

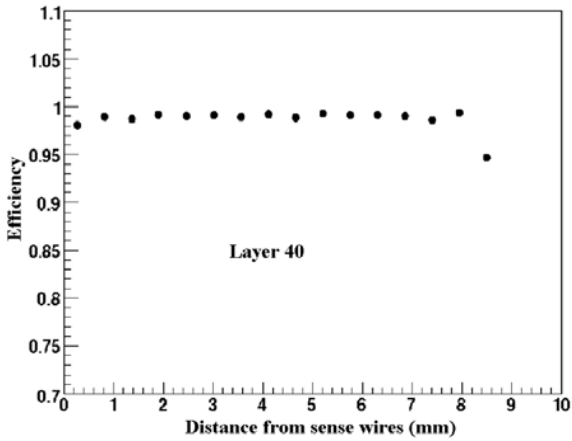

Fig. 17. Efficiency versus distance from the wires in layer 40.

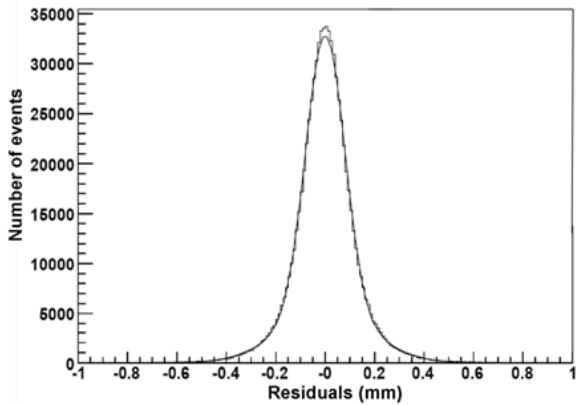

Fig. 18. Residual distribution measured in the clean room using cosmic ray muons.

### 4.4. Electronics

#### 4.4.1. Overview

The MDC electronics system is designed to process the output signals from 6,796 sense wires. The main tasks of the MDC electronics are:
- To measure the drift time of ionization electrons.
- To measure charges collected by sense wires.
- To provide the hit signals to the L1 trigger system.
- To transmit time and charge data to the DAQ system.

The block diagram of the MDC readout electronics system is shown in Fig. 19. It consists of the following main components:
- Preamplifier/HV card.
- Charge and Time measurement Module (MQT).
- Type I and II Fan-out (MF-I and MF-II) Modules.
- Readout Control Module (MROC).
- Calibration and Control Module (MCC).
- Trigger Interface Module (MTI).
- PowerPC (PPC) Controller

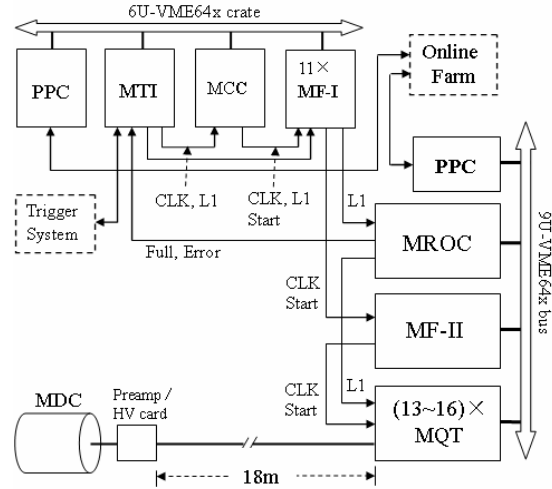

Fig. 19. Block diagram of the MDC readout electronics.

The MDC sense wire signals are first amplified by fast trans-impedance preamplifiers located very close to the wire. Outputs of the preamplifiers are sent to the readout crates via 18 m long shielded twisted pair cables.

Signals are further amplified and split into three branches for timing, charge measurements and the L1 trigger. The discriminated timing signals are digitized by CERN HPTDC chips [27]. The discriminated signals are also sent to the L1 trigger. For charge measurement, the signals are shaped and integrated by charge amplifiers. Signals are then digitized by flash ADC chips. The digital readout and system control logic are implemented in FPGA chips. Valid data are transmitted to the online computer farm via the VME bus and optical links. The MDC readout electronics modules occupy a 6U VME crate and sixteen 9U VME crates.

#### 4.4.2. Performance specifications and test results

Uncertainties of the signal charge and arrival time measurements from electronics must be kept small in order to ensure that their contributions to MDC measurement uncertainties are insignificant. The wire position resolution of 130 μm corresponds to a time resolution of about 3.5 ns. The time resolution design goal for the electronics chain was set to < 0.5 ns so that it has a negligible contribution on the position resolution. In order to cover the ~400 ns maximum electron drift time, the time measurement range of the electronics was set to 500 ns, and the linearity should be better than 0.5%.

The most probable signal charge produced by minimum ionizing particles on a sense wire after amplification is 450 fC determined from the cosmic ray test of the MDC prototype. The required dynamic range of charge measurements is from 15 fC to 1,800 fC to ensure large signals in Landau tails are not saturated. The nonlinearity of the charge measurements should be < 2% over the full range before offline corrections. The maximum linear range of 1,800 fC will result in approximately 5% of signals being saturated, and these will be discarded in track $dE/dx$ calculations.

The expected dE/dx resolution of the MDC is approximately 6% dominated by fluctuations in the signal formation processes. The contribution to charge measurement uncertainties from the electronics should be kept small. The maximum allowed contribution to *dE/dx* measurements from the uncertainty in the single channel charge measurement electronics is required to be ≤ 8 fC. The MDC electronics performance specifications and measured results are summarized in Table 12.

Table 12
MDC electronics performance specifications and measured results

| Items | Required | Test results |
| --- | --- | --- |
| Time resolution | 0.5 ns | 0.1 ns |
| Time dynamic range | 0 - 500 ns | 0 - 500 ns |
| Time integral non-linearity | 0.05% | 0.015% |
| Charge resolution | 8 fC | 6 fC |
| Charge dynamic range | 15 - 1,800 fC | 15 - 1,800 fC |
| Charge integral non-linearity | 2% | 0.5% |

*4.4.3. Front-end preamplifier/HV cards*

In order to operate the MDC with low gas gain and low background, the preamplifier must have high gain and low noise. Controlling the equivalent input noise of the preamplifier can also reduce the electronics contributions to the uncertainties in timing and charge measurements. The preamplifier is a 70 MHz bandwidth trans-impedance amplifier with a 5 ns rise time implemented as a 0.9 cm × 2.6 cm hybrid circuit. Eight such hybrids are mounted on a small printed circuit board that serves 8 channels. High voltages (2,200 V) to the sense wires are distributed through the preamplifier/HV cards. The schematics of the MDC preamplifiers, implemented as hybrid circuits, are shown in Fig. 20. Main parameters of the preamplifiers are summarized in Table 13.

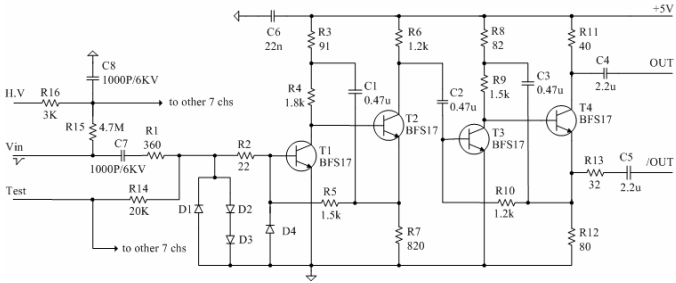

Fig. 20. The schematic of the MDC preamplifier.

Table 13
Parameters of the MDC preamplifier

| Gain | 12 kΩ (±12mV/μA) |
| --- | --- |
| Band width | 70 MHz |
| Rise time | 5 ns |
| Noise | 50 nA |
| Output impedance | 2×50 Ω |
| Power dissipation | 30 mW @ 6V |

The output signals of the preamplifiers are sent to the readout electronics crates over a distance of 18 m via twisted pair cables. In the outer chamber, a piece of short wire connects a sense wire to the input of the preamplifier through the HV decoupling capacitor. Due to space limitations, the preamplifier boards cannot be directly mounted on the end plates of the inner chamber. They are located 0.5 m to 1 m from the end plate, and the wire signals are sent to the preamplifiers by miniature coaxial cables.

The positive high voltages for the 6,796 sense wires in 43 layers are supplied by 43 channels of CAEN HV power supplies (four A1821 modules) and fed to HV distribution circuits as part of the preamplifier/HV cards through 18 m long HV cables. A HV distribution circuit on each card distributes high voltages to 8 channels, each with an RC filter. The test pulse forming circuit, that receives a programmable DC voltage and the start signal from the MQT module, produces standard pulses and sends them to preamplifier inputs for electronics testing and calibration.

*4.4.4. Charge measurement*

The charge and time measurement circuits are implemented in 9U VME modules (MQTs); each MTQ module accommodates 32 channels. The block diagram of the MQT functions is shown in Fig. 21. The MDC readout system requires about 220 MQT modules mounted in sixteen 9U VME crates.

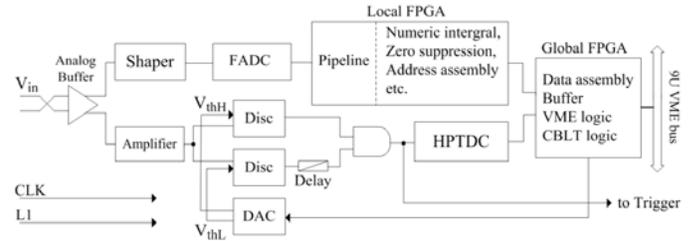

Fig. 21. Block diagram of MQT module.

Buffer amplifiers in MQT modules receive the differential output signals from the preamplifiers. The positive output of the buffer amplifier is sent through a shaper to the FADCs for charge digitization. The negative output is sent to a voltage amplifier, discriminated and sent to the HPTDC chips for time measurement. Digitized charge and time data are processed by the logic circuits in the Global FPGA. The shaper consists of four amplifier stages with a pole zero circuit to cancel the 1/t tails of the wire signals induced due to positive ion drift, and two additional stages of RC shaping circuits. The maximum width of the shaper output signals is set to 800 ns for the FADC waveform digitization.

The detailed block diagram of the charge measurement circuit, including the FADC, the local and global FPGAs that process the data, is shown in Fig. 22. The 41.65 MHz L1 clock of the trigger system is used as the FADC clock. The 10 bit AD9215 FADC chip with a maximum working frequency of 65 MHz was chosen to perform charge integration. The AD9215 with an internal reference voltage source can digitize input signals up to 2 volts at 10 bit precision. The output waveforms of the shaper are sampled by the FADC, and the digital outputs are sent to a FPGA chip (Xilinx Spartan IIE xc2s400e) called "local FPGA" for digital integration to measure the total charge. A local FPGA chip can accommodate 8 channels.

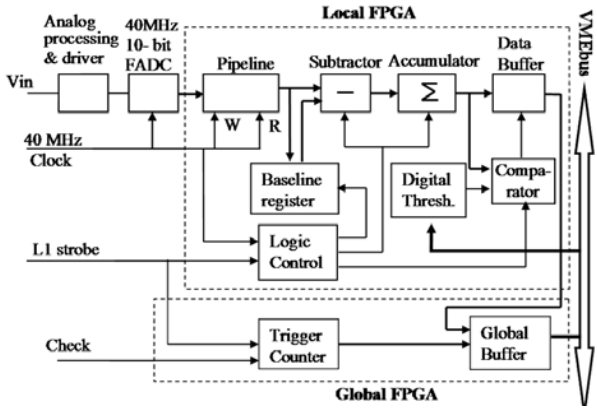

Fig. 22. Block diagram of charge measurement circuit.

Data are placed in the pipelined buffers of the local FPGA. Baseline subtraction and numerical integration are performed with a 1.1 μs charge integration window to ensure later arrival charges are included. The results of the charge integrations are compared to a preset threshold, and valid data are sent to the "Global FPGA" where the zero suppression and data assembly take place. The data stored in the global buffers are transmitted via the VME bus once a L1 trigger arrives. This process then repeats until the next L1 trigger pulse arrives. The pileup rate is expected to be ~4% with the 1.1 μs integration window at a 4 kHz maximum L1 trigger rate.

*4.4.5. Time measurement*

Coordinates of particles that penetrate the MDC are calculated based on the signal arrival times at the MQT module with respect to the time when a collision event occurred. The time measurement circuitry in the MQT is controlled by the 41.65 MHz L1 trigger clock that is synchronized to the accelerator clock. There are three beam crossings within the 24 ns L1 clock cycle. The time measurement hardware cannot determine in which of the 8 ns cycles the collision event has occurred. The time offset caused by this uncertainty is determined in offline analysis.

The time measurement circuitry uses the L1 trigger strobe as the start time. The measured time includes particle flight time before hitting the drift cell, the signal propagation time along the wire and the electronics delay. These time offsets are corrected offline using the electronics calibration database and track reconstruction procedures to obtain the electron drift time.

A dual threshold discriminator scheme, designed to reduce the noise while maintaining the low time walk, is used for timing measurements. The amplified signals first go through a pole-zero tail cancellation circuit and then are sent to two parallel discriminators with "high" and "low" thresholds as shown in Fig. 23. The delayed output of the low threshold discriminator is put in coincidence with the high threshold discriminator output that has very low noise rate. Discriminator threshold voltages are generated locally on each MQT board for the 32 discriminators by a 12 bit DAC chip. The desired threshold settings are obtained via the VME bus.

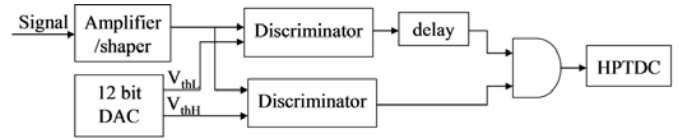

Fig. 23. Block diagram of dual threshold discriminators.

The discriminated signals after the coincidence are sent to HPTDC chips for digitization. In a high rate experiment such as BESIII, the beam bunch spacing is short and collisions occur at a rapid rate. The expected maximum event rate is 4 kHz at BESIII. Conventional TDCs are not suitable for the desired dead time free operation in this environment. The advanced high performance HPTDC chips designed by the CERN microelectronics group provide an ideal platform for the needs of the BESIII experiment. The highly integrated low cost HPTDC has 32 channels per chip and can satisfy the precision requirement for the MDC drift time measurements.

The HPTDC chips have many advanced features, including dead time free operation, programmable time resolution from 250 ps to 20 ps, 5 ns double pulse resolution, digitization of the leading and trailing edges, built-in zero suppression and address assembly. The HPTDC chips used for MDC time readout are synchronized to the 41.65 MHz beam clock, and the signal arrival time can be accurately measured with a 100 ps resolution. In this mode, each HPTDC chip can handle 32 channels. The time walk correction can be performed offline based on the measured charge values of signals as discussed below.

*4.4.6. Operation mode controls and calibration*

As shown in Fig. 22, the Mode Control and Calibration (MCC) is a 6U VME module that controls the working modes of the MDC readout electronics system. In the regular data taking mode, the MCC module receives four signals from the L1 trigger system: the 41.65 MHz system clock, the L1 trigger strobe, the trigger check and reset via optical links. These signals are distributed to the electronics crates to control system operation.

The readout electronics chain described above will be calibrated daily to determine the channel-by-channel variations for certain threshold settings. Standard calibration strobes are produced internally by the trigger system at a rate of about 160 Hz. Variations in time delays and gains among different channels must be corrected in order to obtain optimum system performance.

In the calibration mode, the calibration strobes are produced by the MCC module. Programmable DC voltages are generated by the 12 bit DAC chips on MQT modules and are sent to preamplifiers. The preamplifier cards, triggered by the start pulse they receive from the MQT modules, produce the calibration pulses that have standard amplitudes with the shape mimicking sense wire signals. The calibration precision is better than 1%. The block diagram of the electronics charge calibration circuit is shown in Fig. 24.

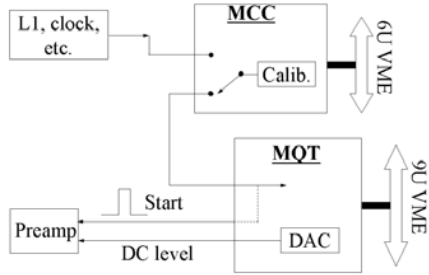

Fig. 24. Electronics charge calibration.

The Trigger Interface Module (MTI) receives control signals from the L1 trigger system via optical links, converts these signals to LVPECL level and distributes them to the MCC module or to the type I Fan-out Module (MF-I) for further distribution. The MTI module also sends the status signals generated by the Readout Control Module (MROC) in the readout crate to the L1 trigger system via optical links. The MTI is a 6U VME module.

The MROC in each 9U VME crate produces interrupts for transmitting the data stored in the temporary buffers to the online computer farm of the L3 event filter. It also provides a trigger fan-out for the 16 MTQ modules in a crate, transmits the status data to the L1 trigger system and controls FPGA programming.

Ten MF-I modules in the 6U VME receive system control signals from the MCC module and fan out (1:16) to the second level fan-out module MF-II in the 9U VME crate. The MF-II then distributes these signals to the MQT modules in the crate via coaxial cables. These two standardized fan-out modules simplify the system design for distributing the large numbers of different control signals among VME crates that are located in different areas.

*4.5. MDC alignment and calibration*

*4.5.1. Introduction*

Alignment and drift time calibration based on reconstructed particle tracks are important for the MDC to achieve optimum performance. Mechanical imperfections, relative alignment uncertainties among different sections of the chamber, time measurement uncertainties caused by electric and magnetic field variations, wire sags in different wire layers and fluctuations of environmental factors can only be corrected through offline calibration procedures. The readout electronic chain described above can also contribute to the uncertainties of time measurements.

Large data samples of cosmic-ray muon tracks were accumulated and used to align and calibrate the MDC. The alignment and calibration were accomplished using both a residual distribution method by fitting plots of residuals as functions of azimuth angles and the Millepede matrix method [28] that can determine a very large number of parameters in a simultaneous linear least squares fit to the data. The initial geometric database and standard parameterizations were used to start the calibration. Based on the reconstructed tracks, a new set of calibration parameters was found and used for the next iteration until the process converged. The alignment and calibration procedure will be refined by using the large number of various types of particle tracks from real collision data, and initial parameters obtained by fitting cosmic ray data will be improved.

*4.5.2. MDC alignment*

The two end plates that hold the MDC sense wires are assembled from 16 independent elements, two end plates of the inner chamber, two end plates of the outer chamber and twelve (2×6) rings of the stepped inner section of the outer chamber. Their coordinate deviations from the designed values were first corrected based on large samples of recorded cosmic ray tracks. Each of the 16 elements requires three alignment constants ($\delta x$, $\delta y$, $\delta z$). Sense wire positions in each end plate element can also deviate from their design positions due to hole drilling tolerances and imperfections of feed-throughs. To define the position of a wire in space, five parameters were chosen for each wire, including the 2D coordinates of the wire ends and the wire tension that determines the wire sag in the middle.

*4.5.3. r-t relation calibration*

The *r-t* relations in drift cells are slightly different in drift cells of different layers since the cell geometry in each layer is slightly different. Electric field configurations in the six boundary layers between the axial and stereo sections are different from the inner layers. Electric field configurations of the inner and outer most layers (layer 1 and 43) are also special. These eight layers must be treated differently in the *r-t* calibration procedure. In addition, the *r-t* relations are different for particles with different entrance angles and the solenoid magnetic field also causes left-to-right asymmetry. We consider 18 entrance angles on either side of the sense wire. By considering all these cases, the total number of parameters needed in the *r-t* relation calibration is 12,384. The *r-t* relations of drift cells in all 43 layers were calibrated using the cosmic ray data. They will be further refined using the collision data.

*4.5.4. dE/dx calibration*

For dE/dx measurements, the gain non-uniformity among sense wires in the 43 layers must be calibrated. Fig. 25 shows the gain distributions of every wire layer before and after the gain calibration using cosmic ray data.

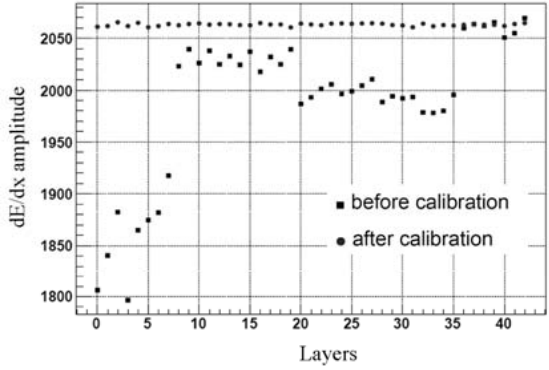

Fig. 25. Gain uniformity in the 43 wire layers before and after calibration.

4.6. Gas system

The MDC holds a gas volume of approximately 4m$^3$ filled with a gas mixture of 60% He and 40% $C_3H_8$ at a pressure 3 mbar above the ambient atmospheric pressure. The flow rate is approximately one volume change per day, and the output gas is discharged to the atmosphere. The gas is mixed onsite from bottled pure gases and the mixing is controlled by a MKS647c mass flow controller. The mixed gas is stored in a buffer tank. The flow of the mixed gas to the MDC goes through another buffer tank and is controlled by another unit of mass flow controller (MKS247c). The pressure and temperature of the buffer tanks and the MDC vessel is monitored by various sensors distributed in the system. To ensure the constant temperature of the MDC, the input gas pipe goes through a water bath that is kept at constant temperature. The over pressure is protected by oil bubblers and pressure relief valves.

## 5. Time-of-flight particle identification system

### 5.1. Overview

Time-of-flight (TOF) detector systems based on plastic scintillation counters have been very powerful tools for particle identification in collider detectors. In addition to identifying the species of charge particles, the TOF system can provide fast trigger signals. Several practical considerations contributed to the decision to use a TOF system based on a double layer barrel plus two single layer end caps for the BESIII particle identification system [29]. Considerations for the available space (TOF counters are placed between the MDC and EMC), the desired solid angle coverage, cost and technical complexities precluded the RICH [30] and DIRC [31] technologies used for particle identification systems by the CLEO-II and Babar experiments.

We do not expect to get the full benefit of the statistical factor $1/\sqrt{2}$ for the system time resolution compared to a single layer counter system in the double layer barrel because of correlated error sources as discussed below. Nonetheless in addition to improving system reliability, two independent measurements of the particle flight time using two layers of scintillation counters provide improvements for the time resolution and significantly extends the physics capability of the system.

The compact design of BESIII limits the inner radius of the first layer TOF counters to be 0.81 m and the second to be 0.86 m. Particles detected by the end cap TOF system have a minimum flight path of about 1.4 m. The short particle flight path makes the TOF system design very challenging.

### 5.1.1 Physical description of TOF system

The cross-sectional view of the TOF system inside the CsI crystal calorimeter is shown in Fig. 1. The TOF system consists of a double layer barrel and two single layer end caps. Each barrel layer has 88 plastic scintillation counters (BC-408), and the two layers are staggered. The reserved radial space for the double layer barrel counters is 11.5 cm between the MDC and the EMC with a 0.5 cm clearance ( 0.81 m < r < 0.93 m). The end cap TOF counters are placed outside of two MDC end plates. The polar angle coverage is $|\cos\theta| < 0.82$ for the barrel and $0.85 < |\cos\theta| < 0.95$ for the end caps. There are small dead gaps between the barrel and two end caps to allow mechanical support structures of the MDC and service lines to go through.

The length of the barrel scintillator bars is 2,300 mm. The bar cross-section is trapezoidal, and the thickness is 50 mm. Two PMTs (Hamamatsu R5924) are attached to the two ends of a barrel counter and coupled by 1 mm thick silicone pads (BC-634A). The length of the PMT is 50 mm, and the total length of PMT assembly is 103 mm, including the base and preamp.

Each end cap TOF station consists of 48 trapezoidally shaped scintillation counters (BC-404) arranged in a single circular layer. The scintillator is 50 mm thick and 480 mm long. Its width at the top is 109 mm and 62 mm at the bottom. The inside end of the scintillator is cut at 45° to reflect light to the photomultiplier. A R5924 PMT is attached on the outside surface of the scintillator facing the 45° cut surface.

### 5.1.2. Factors contributing to time resolution

For the purpose of predicting the time resolution of the BESIII TOF system, the time resolution $\sigma$ of the system can be expressed as

$$\sigma = \sqrt{\sigma_i^2 + \sigma_b^2 + \sigma_l^2 + \sigma_z^2 + \sigma_e^2 + \sigma_t^2 + \sigma_w^2}$$

The definitions and estimated magnitudes of the terms in the above expression are summarized in Table 14.

The "intrinsic" time resolution term $\sigma_i$ is determined by the rise time of the scintillation light, the fluctuations of photon arrival time at the PMT and the transition time spread of the PMT. This term is estimated to be approximately 80 ps averaged over the polar angles. The intrinsic time resolution of the TOF counters depends on the type of particles and their momentum. The time resolution for muons is better than those of kaons and pions. The term $\sigma_b$ is mainly caused by the uncertainty of registering the global timing marker (accelerator RF clock) in the readout electronics. It is estimated to be about 20 ps including jitter due to cables. The term $\sigma_l$ is caused by uncertainties in determining interaction vertices when two beam bunches of 1.5 cm in length (rms) collide. This term is estimated to be about 35 ps. The term $\sigma_z$ is caused by the uncertainty of hit positions along the scintillator bar. The z-position uncertainty of hit points for the barrel counters is estimated to be 0.5 cm corresponding to about 30 ps taking into account the light propagation speed in the scintillator. For end cap counters, the uncertainty of hit positions along the scintillator bars is about 0.8 cm (~50 ps) because charged particles that hit the end caps are only tracked by the inner layers of the MDC and the hit position uncertainty becomes larger. The term $\sigma_e$ is the time resolution of the readout electronics, mainly from the TDCs and is estimated to be 25 ps. The term $\sigma_t$ is the uncertainty of determining the particle flight time based on measuring flight path length and momentum ($\sigma_{p_t}/p_t = 0.5\%$ at 1 GeV/c). This term is estimated to be 30 ps. The last term $\sigma_w$ is the uncertainty caused by the time walk in a fixed threshold discriminator due to signal amplitudes fluctuations. This is the uncertainty remaining after corrections based on signal amplitude measurements.

Table 14
Analysis of TOF time resolution for 1 GeV/c muon

| Item | Barrel (ps) | End cap (ps) |
|---|---|---|
| $\sigma_i$ : counter intrinsic time resolution | 80 ~ 90 | 80 |
| $\sigma_l$ : uncertainty from 15 mm bunch length | 35 | 35 |
| $\sigma_b$ : uncertainty from clock system | ~20 | ~20 |
| $\sigma_\theta$ : uncertainty from θ-angle | 25 | 50 |
| $\sigma_e$ : uncertainty from electronics | 25 | 25 |
| $\sigma_t$ : uncertainty in expected flight time | 30 | 30 |
| $\sigma_w$ : uncertainty from time walk | 10 | 10 |
| $\sigma_1$ : total time resolution, one layer | 100 - 110 | 110 |
| $\sigma_2$ : combined time resolution, two layers | 80 - 90 | - |

Excluding the terms $\sigma_i$ and $\sigma_w$, the total contribution from the other five terms, that are not directly related to the counters, to the uncertainty of the TOF system is about 66 ps for the barrel and 87 ps for the end caps. By using two layers of counters in the barrel, the combined contribution from $\sigma_i$ and $\sigma_w$, that are intrinsic to the TOF counters, can be improved by a factor of about $\sqrt{2}$, whereas the total resolution of the TOF barrel can be improved by approximately 10% compared to the time resolution of the single layer option. The 10% improvement significantly expands the physics reach of the BESIII TOF system.

*5.1.3. Expected particle ID performance*

The acceptance and expected performance parameters of the BESIII TOF system based on Monte Carlo studies and beam tests are summarized in Table 15. Parameters of the BESII TOF are also listed for reference.

Table 15
Parameters and expected TOF performance of BESIII TOF

| Parameters | BESIII | BESII |
|---|---|---|
| Polar angle coverage | | |
| Barrel | \|θ\| >35° | |
| End cap | 18° < \|θ\| < 32° | \|θ\|>50 none |
| Inner radius (m) | 0.81 | 1.15 |
| Time resolution for 1GeV muons | | |
| Barrel | ~90 ps | 180 ps |
| End cap) | ~120 ps | 350 ps |
| 3 σ K/π separation (GeV/c) | < 0.9 | < 0.8 |

Note that in Table 14, the time resolutions given are for 1 GeV/c muons. For pions and kaons the time resolutions in the momentum range of interest are approximately 20% larger based on the experience of similar TOF systems built for other experiments. Fig. 26 shows the simulated K/π separation capability of the BESIII barrel TOF system placed outside of the MDC outer shell at a radius of 810 mm. The thick solid line and thick dashed lines in the figure represent the 3 σ K/π separation capability in the barrel TOF. Conservative time resolutions of 125 ps for the single layer and 105 ps for the double layer TOF system were assumed in the simulation. The K/π separation capability depends on the polar angles of tracks. The 3 σ K/π separation limit is 0.7 GeV/c at 90° where the particle flight time is the shortest, if the time resolution is 105 ps for kaons and pions, and rises to approximately 1 GeV/c near the ends of barrel counters at cosθ = 0.8.

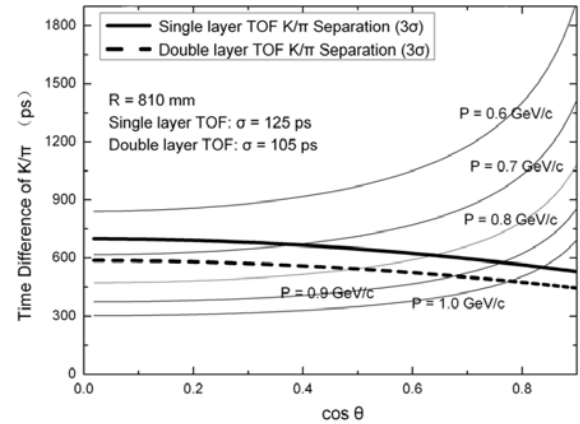

Fig. 26. K/π separation capability as functions of cosθ.

The 3 σ K/π separation limit is expected to be 0.9 to 1.0 GeV/c for the end cap TOF system. The time resolution of the end cap counters is approximately 10% worse than the time resolution of the barrel counters mainly because the end cap has only a single layer and the noise level is higher. The actual performance of the end cap TOF will also depend on how well the inner layers of MDC chamber work.

The particle identification capability was calculated by a likelihood analysis. The results of kaon identification efficiency and probability of kaons misidentified as pions as functions of kaon momentum are shown in Fig. 27. This analysis was based on a time resolution of 90 ps for the double layer barrel and 120 ps for the end caps, combined with the dE/dx of the MDC assuming a 6% resolution. A K/$\pi$ separation efficiency of 95%, and a contamination level of about 5% can be achieved up to 0.9 GeV/c. The proton identification is very good at BESIII.

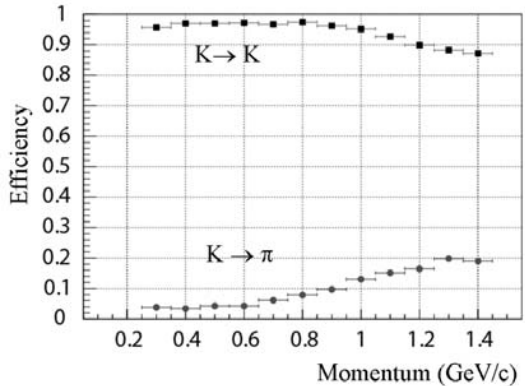

Fig. 27. Expected K/$\pi$ separation efficiency and misidentification rate.

*5.2. Scintillation counter design and construction*

*5.2.1. PMT*

Hamamatsu R5924-70 photomultipliers were chosen to read out the TOF counters. The diameter of the R5924 is 51 mm matching the size of the scintillator bars. The coverage of its 39 mm diameter photocathode is 37% and 39% for the counters of the two barrel layers. The average quantum efficiency is about 23%, the rise time is 2.5 ns and the signal transition time is 9.5 ns with a 0.44 ns rms spread.

The effect of the magnetic field on the gain of PMTs was studied. With 19 stages of fine-mesh dynodes, the R5924 PMT has a sufficiently high gain in 1 T axial magnetic field. Operating at the suggested voltage 2,000 V, the gain reduction factor at 1 T relative to the gain without magnetic field is about 50. The nominal gain of the R5924 without the magnetic field at 2,000 V is $10^7$. A factor of 50 reduction brings the gain down to the $2\times10^5$. The radial component of the magnetic field may further reduce the PMT gain. A fast preamplifier with a gain of 10 is installed in the PMT base to boost signals and extend the lifetime of the PMT.

*5.2.2. Scintillation counters*

The design of scintillator bars, including the scintillator parameters and wrapping, and the performance of the entire counter including the PMT, preamplifier, cable and readout electronics were studied by Monte Carlo and by beam tests [32] conducted at the test beam at IHEP. The beam test setup included a beam line spectrometer that measured the beam momentum and a gas Cherenkov counter that determined the particle species.

Time resolutions of counter samples with 4, 5 and 6 cm thickness and 6 cm wide were tested in the beam. The 5 cm thick counter was found to have the best time resolution. This result was also confirmed by the Monte Carlo simulations.

Five wrapping materials were studied, including aluminum foil, 3M Vikuiti™ ESR ( Enhanced Specular Reflector ) film, Millipore fibrous filter paper, Teflon and Tyvek (DuPont). The aluminum foil wrapping, commonly used by other TOF systems, was found to give the best time resolution for the long barrel scintillator bars, although the attenuation length is shorter and the light yield is lower than some of the other diffusive wrappings such as ESR. For the much shorter end cap counters, the high reflectivity ESR film that produced the highest light output was found to give the best time resolution.

The choice of the scintillators was made based on detailed measurements of full size samples delivered to us by Saint-Gobain (BC-404 and BC-408) [33] and Eljen Technology (EJ200) [34]. Parameters such as the scintillation light yield, rise time and light attenuation length were measured in the electron, pion and proton beams and by using radioactive sources and cosmic rays. The BC-408 scintillator by Saint-Gobain was chosen for the barrel because of its longer attenuation length, and BC404 was chosen for end cap counters. The higher light output and faster rise time of the BC404 result in better time resolution for the much shorter end cap counters. The performance of the EJ-200 scintillator sample was similar to the BC-408. The EJ-200 was not chosen due to cost and engineering issues. Fig. 28 shows the time resolutions of the 5 cm thick EJ-200 and BC-408 barrel counter samples measured in a beam test. Counters were wrapped in aluminum foil. The beam used was a 800 MeV/c electron beam. The readout electronics were the same as used by the final TOF system, and a 18 m long cable was included to transmit the preamplifier output signals to the post amplifier. The time resolution averaged over the length for the BC-408 counters was found to be 86 ps after correcting for the start time jitters of trigger counters and time walk due to pulse height fluctuations.

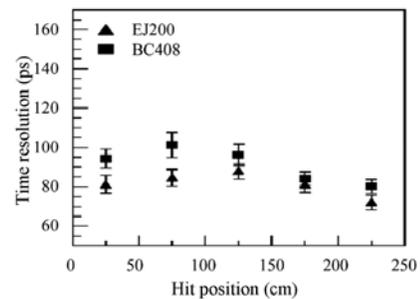

Fig. 28. Time resolution as a function of the hit position.

In Fig. 29, the end cap counter time resolutions as a function of distance from the PMT measured with cosmic rays and an electron beam, and determined from cosmic ray Monte Carlo simulation are given. The time resolutions measured with cosmic ray muons and the electron beam are consistent, and are slightly larger than the cosmic ray Monte Carlo predicted values. The average measured time resolution is approximately 80 ps [35].

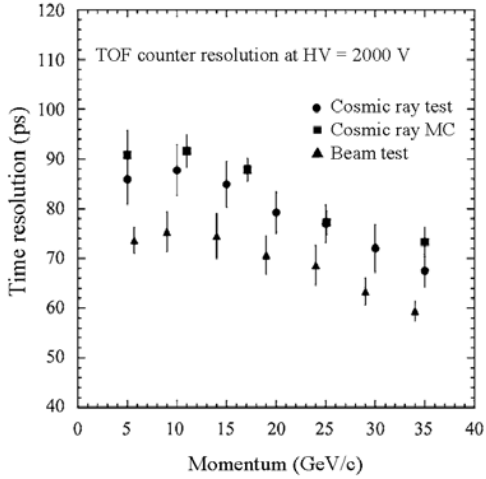

Fig. 29. Simulated and measured time resolution of an end cap counter.

*5.2.3. Counter fabrication and quality control*

The mechanical dimensions of scintillator bars delivered were measured. The bars were then wrapped by a layer of aluminum foil and a layer of black plastic tape. Photomultipliers were attached to the two ends of a scintillator bar using custom made phototube housings, which were attached to the ends of the scintillator bar by screws. The R5924 PMT is forced to make contact to the end of the counter, coupled with optical grease. The end cap counters are trapezoidally shaped with a PMT attached at a 90° angle to the surface of the small end facing a 45° cut surface, that reflects the scintillation light to the PMT.

The gain, quantum efficiency and time response of the 448 Hamamatsu R5924 PMTs for the barrel and end cap TOF counters were tested at Tokyo University in a 1 T magnetic field. The counters were assembled at IHEP and tested using cosmic rays. A group of seven bars were vertically stacked on a shelf. The top and bottom scintillator bars were used to form triggers, while the five counters in the middle were tested. The cosmic ray hit positions in each counter were determined by comparing the time differences of the two ends. Pulse height versus position along the bars was also measured. A cosmic ray test run for a group of counters could be finished in seven hours with sufficient statistics.

Fig. 30(a) shows attenuation lengths of the 180 bars measured by cosmic ray tests. Six bars among 180 bars received had attenuation lengths slightly less than the specified 380 cm. Fig. 30(b) shows the relative light output of the scintillator bars. The measured light output data of the scintillator bars and quantum efficiencies of PMT photocathodes were used to match the bars and PMTs to make the measured numbers of photoelectrons of the final counters more uniform.

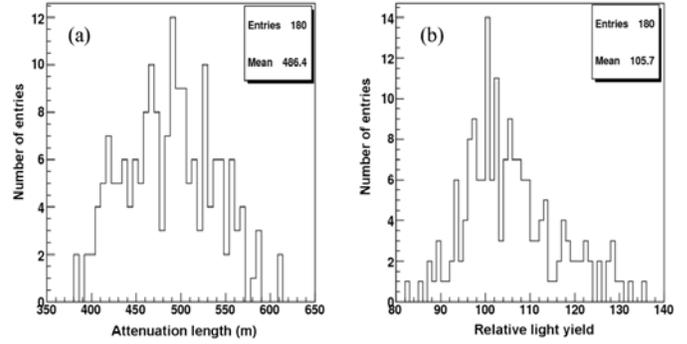

Fig. 30. (a) Attenuation length and (b) Relative light yield.

*5.3. TOF Readout electronics*

*5.3.1. Overview*

The TOF readout system consists of preamplifiers mounted in the PMT bases, cables, signal time and amplitude measurement circuits, the L1 trigger circuits, electronics and a laser calibration system. The output signals of a PMT are amplified by a factor of 10 by a high bandwidth preamplifier. High performance cables transmit fast analog signals over 18 meters with minimum signal rise time deterioration. The 448 signals from PMT preamplifiers are read out with two 9U VME crates, each has 14 Time & Charge measurement modules (FEE), as shown in Fig. 31.

A FEE module receives signals from 16 preamplifiers and splits signals into two branches for time and charge measurements by HPTDC chips. In addition to the FEE modules, each crate has a L1 trigger Fast Control Module that performs L1 fan-out and readout control. Fast signals for triggers are formed by mean timer circuits and sent to the TOF L1 sub-trigger system that is implemented as a 9U VME module via optic links. The digitized time and charge data are transmitted to the DAQ system by the VME bus.

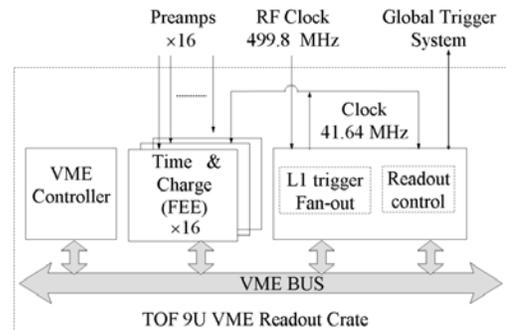

Fig. 31. Architecture of TOF readout electronics crate.

The required dynamic range of the time measurements is 60 ns, and the required time resolution is less than 25 ps. The charge measurement range is from 200 mV to 4 V with a resolution of 10 mV, and the non-linearity of charge measurements should be less than 2%. The expected hit rates of the TOF counters shown in Table 6 are much lower than the specified 20 kHz capability of the TOF electronics.

*5.3.2. Preamplifier and output cable*

A high performance preamplifier was designed for the R5924 PMT using the AD8099 high speed, low noise amplifier chip with a voltage gain of five and a buffer amplifier chip AD8131 with a gain of two. This amplifier has a 550 MHz bandwidth and a slew rate of 1350 V/μs with a gain of 10. The noise level is only 0.95nV/√Hz. The AD8131 [36] with a gain of two is used as a differential driver. The combined rise time is ≤ 2 ns. The preamplifier receives differential signals, which help to eliminate the common mode noise pickup and increases the signal-to-noise ratio.

High performance shielded twisted pair cables bring output signals from preamplifiers to the readout crates. The cables are 7.5 mm in diameter with a high bandwidth to minimize the deterioration of the rising edge of the signal transmitted over the 18 m distance. Its nominal impedance is 100 Ω.

*5.3.3. Buffer, discriminators and TDC*

The basic method used for the time measurements in the TOF system is similar to the one used in the MDC. Buffer amplifiers (THS4500) receive differential signals from the preamplifier outputs, amplify them further and split the signals into three branches: the low and high discriminator branches for time measurements and trigger generation, and the amplitude measurement branch. The discriminators are based on the fast comparator chip MAX9601. Buffer amplifiers, discriminators, TDCs, time-over-threshold circuits that measure amplitudes of signals, and associated logic circuits are implemented as 9U VME modules. Each module accommodates 8 scintillators for the barrel and 16 scintillators for the end cap TOF systems.

Similar to the MDC electronics as shown in Fig. 23, a dual threshold discriminator scheme, designed to reduce noise while maintaining low time walk, is used for time digitization. The outputs from the low threshold (~50 mV) discriminators of the two ends are used to start the precision time-to-digital conversion process and are put into coincidence with the high threshold (~250 mV) discriminator outputs in order to reduce background rates. The outputs of the coincidence circuits are sent to HPTDCs. The high threshold outputs from the two ends are sent to the L1 trigger system after the mean timer circuit for generating fast trigger signals. For precision timing measurements, the time resolution of the CERN HPTDC chips used is set to 25 ps. At this resolution, each HPTDC chip can handle 8 channels. The third branch of the buffer outputs is sent to the time-over-threshold circuits for signal amplitude measurements. The readout electronics system for the end cap TOF counters is similar to the barrel except there is only one preamplifier and no mean timer.

Signal amplitude measurements are necessary for correcting the time walk (slewing) of the discriminators due to signal amplitude variations. Such offline corrections are critical for achieving the 100 ps time resolution since the time walk can be as large as a few ns for fixed threshold discrimination. The signal amplitude measurement circuit is based on the charge-to-time conversion principle. A branch of the buffer amplifier output is used to charge a capacitor. When the measurements start, the capacitor is discharged by a constant current source. The time of the discharge is digitized by the HPTDC. Since the time resolution requirement for the charge measurements is not very high, one HPTDC chip on a VME board can accommodate the 16 channels at low time resolution. The signal charge data from the HPTDC chip are also processed by the data readout and VME interface circuit on the VME board.

All the time and amplitude data are stored in pipelines clocked at 41.65 MHz waiting for the L1 trigger strobe that arrives 6.4 μs after the collision. The data readout circuits and VME interface logic for 16 channels are implemented in an Altera FPGA chip (EP1C12F324). Its functions include providing the HPTDC interface, digital pipelines, non-linearity compensation that is necessary for achieving the optimum time resolution of the HPTDC, event building and the VME interface.

As discussed above, a complete digital approach was taken for the TOF electronics design. Both time and charge measurements are accomplished by the same TDC chips. As a result, the time measurement system is simple, in principle, with high performance and low power consumption.

*5.3.4. Performance of the TOF readout electronics*

The time resolution of the TOF electronics was measured by cable delay tests using two different signal sources [37]. First, test pulses were generated by an Agilent 33250A pulse generator. The test pulses were split into two branches and approximately 8 ns delay was created by adding a short section of cable to one of the branches. The signal arrival time differences of the two branches were measured by two TDC channels for time delays from 1 ns to 30 ns. The maximum rms time spread was 28 ps, corresponding to a time resolution of less than 20 ps for each TDC channel, after appropriate corrections were made.

The above tests were repeated using an end cap TOF counter as a signal source. A cosmic ray telescope was also setup, and the signal output from the PMT mounted on the counter was split into two branches and differences of signal arrival times between the two branches were measured. The results were similar to that obtained by the pulse generator. In the above tests, the contribution from the preamplifiers and the TOF analog signal cables were included.

*5.3.5. Laser monitoring and calibration system*

The laser monitoring system [38], consisting of a powerful laser diode, two optical fiber bundles with a total of 512 optical fibers and associated optical components, delivers fast and stable light pulses to each of the 448 PMTs for monitoring and calibration purposes.

The light source is a PicoQuant PDL 800-B pulsed laser driver with a LDH-P-C-440M laser diode head with a wavelength of 442 nm. Its pulse width is 50 ps FWHM at low power operation and approximately 500 ps FWHM at high power. The maximum peak power of the laser diode is 2.6 W, which is powerful enough for our purpose. The driver can operate at 2.5 MHz to 40 MHz and can be externally triggered. The advantages of this laser diode compared, for example, to a nitrogen-dye laser commonly used for fast timing calibrations are ease of operation, minimum maintenance, stability and long lifetime. The laser head has Peltier cooling, and the power stability over a 12 hour period is 1 % rms.

The laser head emits a collimated elliptical beam of approximately 1.5 mm × 3.5 mm and with divergence angles of 0.32 mrad parallel to and 0.11 mrad perpendicular to the long axis of the beam. The schematic of the laser light delivery system is shown in Fig. 32. The light from the laser diode head is injected to two optical fiber bundles, one for each end of the detector. Each bundle has 256 multimode silica fibers with 100 μm diameter cores and 125 μm outer diameter. The diameter of the input end of a fiber bundle is approximately 2.25 mm.

As shown in Fig. 32, approximately 5% of the laser light is split off for the two reference PMTs by a high transmissivity beam splitter. The majority of the laser beam is further split into two nearly equal intensity beams displaced by 20 mm with respect to each other by a high transmissivity beam splitter (Edmund Optics NT47-189). Two rotary solenoids control which fiber bundle will be illuminated by blocking the undesired beam.

Each laser beam goes through a Thorlabs 11 mm collimator package (F230FC-A) that couples the light into a 550 micron diameter fiber, which mixes the light. The light from the two fiber bundles is directed to 2.5 cm diameter diffusers (Thorlabs ED1-C20) with 20 degree divergence angles that are located 20 mm in front of the input ends of the fiber bundles distributing the laser light uniformly to each fiber in the bundles. The laser light is carried by the fibers to the PMTs.

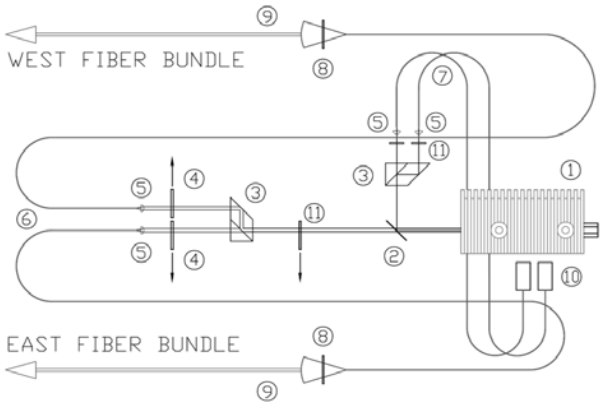

Fig. 32. Schematic of laser box components. (1) Laser diode head, (2) high transmissivity beam splitter, (3) lateral displacement beam splitters, (4) rotary solenoids, (5) Thorlabs 11 mm collimation optics packages, (6 and 7) 550 micron diameter fibers, (8) diffusers, (9) input ends of fiber bundles, (10) reference PMTs, and (11) neutral density filters.

*5.4. Assembly of TOF system*

The barrel scintillator bars were loaded on to the carbon fiber composite shell of the MDC and held temporarily by three circular jigs with screws. Strong glass fiber reinforced adhesive tape was applied to hold the counters in place, after which the jigs were removed. The assembly process was completed by an eight person crew in 4 weeks. The counter placement precision was found to be ±0.8 mm in the ϕ-direction, and the total thickness of the two layers of counters was 114 mm, 1 mm less than the specified 115 mm. The MDC/TOF assembly was successfully inserted into the barrel EMC with only a 6 mm radial clearance. The MDC/TOF barrel is supported by steel brackets through the gap between the barrel and end cap EMC. The brackets are attached to the inner surface of the barrel flux return yoke. The end cap counters have their own separate mechanical supports.

# 6. Electromagnetic calorimeter

## 6.1. CsI(Tl) crystal EMC design

### 6.1.1. Physics requirements

High energy resolution, adequate position resolution and good e/π separation are main design requirements for the EMC. Detecting direct photons and photons from decays of $\pi^0$, η, ρ, etc. produced in ψ′, ψ″, τ and D decays is essential for BESIII physics. In radiative decay processes, such as J/ψ→γππ, J/ψ→γKK, J/ψ→γηη, direct photons must be precisely measured and separated from photons from $\pi^0$ and η decays. As mentioned earlier, the energies of most photons produced in these interactions are quite low. The average multiplicity of photons is approximately four, similar to the multiplicity of charged particles in the final states. Accurately measuring photon energies and their hit coordinates is crucial for reconstructing physics processes and for new discoveries. The desired photon energy detection range for BESIII is from ~20 MeV to the full beam energy of 2.1 GeV for studying the $e^+e^- \rightarrow \gamma\gamma$ process.

An important requirement for the EMC is to accurately reconstruct the invariant mass of $\pi^0$'s. In the laboratory system, the minimum opening angle between the two γ's from $\pi^0$ decay decreases as the $\pi^0$ energy increases. The minimum opening angle between the two γ's from 1.5 GeV $\pi^0$ decays is about $10°$.

Another important consideration for the EMC design is e/π separation. Charged pions can interact in the CsI crystals and generate showers. Energy deposition patterns of some pions, especially at low energies, can be misidentified as electrons. Superb energy resolution is the most powerful tool to discriminate against such misidentification by requiring good agreement between the measured momentum and the measured energy in EMC. Fine segmentation and pattern recognition can also help to improve e/π separation.

Good energy resolution, background free high efficiency detection and accurate photon hit position determination require a well designed, finely segmented crystal electromagnetic calorimeter. Besides high light yield and high density, the calorimeter must have sufficient segmentation for distinguishing electron showers from hadron showers and reducing shower overlap. The calorimeter must be placed inside the 1T SSM. Compactness and the ability to work in the strong magnetic field are essential requirements for the calorimeter design. An high resolution EM calorimeter (EMC) made of sufficiently segmented CsI(Tl) crystals placed inside the superconducting solenoid readout by large area photodiodes was designed and constructed for the BESIII [39]. The design performance specifications of the EMC are:

- Energy range of electron or photon from ~20 MeV to ~2 GeV with energy resolution of 2.5% at 1 GeV and 4 MeV at 100 MeV.
- Photon hit position resolution $\sigma_{x,y} \leq 6\,\mathrm{mm}/\sqrt{E(\mathrm{GeV})}$
- A good $e/\pi$ separation

### 6.1.2. CsI(Tl) scintillating crystals

The CsI(Tl) scintillating crystals are the best choice for the BESIII EMC that must detect low energy photons. The high light yield of CsI(Tl), especially when read out by silicon photodiodes, is critical for high energy resolution of low energy photons. The properties of the CsI(Tl) crystals are listed in Table 16.

### 6.1.3. EMC layout and geometric parameters

Choices made for the CsI(Tl) calorimeter, including crystal size, length, segmentation and geometric parameters, are based on physics requirements described above. Controlling the cost of the entire spectrometer also played an important role in the decision making process. In a well designed CsI(Tl) crystal calorimeter, the energy resolution is mainly determined by the shower energy leakage from backs of crystals, the non-uniformity of light production and by the dead material between crystals.

Table 16
Properties of thallium doped CsI(Tl) crystals

| Parameter | Values |
| --- | --- |
| Radiation length $X_0$ | 1.85 cm |
| Moliere radius | 3.8 cm |
| Density | 4.53 g/cm$^3$ |
| Light yield (photodiode) | 56,000 $\gamma$'s/MeV |
| Peak emission wavelength | 560 nm |
| Signal decay time | 680 ns (64%) |
|  | 3.34 ms (36%) |
| Light yield temp. coefficient | 0.3%/$^\circ$C |
| dE/dx (per mip) | 5.6 MeV/cm |
| Hygroscopic sensitivity | slight |

The length of the crystals was chosen to be 28 cm or 15.1 $X_0$. The cross-sectional view of an EMC can be seen in Fig. 1. The 6,240 crystals are arranged as 56 rings (44 rings in the barrel and 2 × 6 rings in the end caps). Each crystal covers an angle of about 3° in both polar and azimuth directions.

The 5,280 barrel crystals are divided into 44 rings. All crystals point to the interaction region with a small tilt of 1.5° in the $\phi$-directions and 1.5° to 3° in the $\theta$-directions (± 10 cm from the IP in the beam direction) to avoid photons from the interaction point escaping through cracks between crystals.

A barrel ring has 120 crystals. The number of crystals in the six rings of each end cap is 96, 96, 80, 80, 64, 64. Each of the 6 rings are split into two half circles and can be opened for accessing the drift chamber. The end cap crystals have 33 different sizes, and among the 960 end cap crystals, 192 crystals are irregular pentagons.

The barrel and two end caps are separated with 5 cm gaps to allow mechanical support structures and service lines of inner detectors to pass through. The geometric parameters of the EMC are summarized in Table 17.

Table 17
Geometric parameters of EMC

| Parameter | Values |
| --- | --- |
| Crystal length | 28 cm (15.1 $X_0$) |
| Typical front and rear sizes (cm×cm) | 5.2×5.2 - 6.4×6.4 |
| Number of $\phi$-sectors | 120 |
| Barrel |  |
|    Number of $\theta$-rings | 44 |
|    Number of crystals | 5,280 |
|    Inner radius (cm) | 94 |
|    $\theta$-coverage | $\cos\theta < 0.83$ |
|    Total weight (tons) | 21.56 |
| End caps |  |
|    Number of $\theta$-rings | 6 |
|    Number of crystals | 960 |
|    Distance to IP (cm) | ±138 |
|    Ring inner radius (cm) | 88 |
|    Ring outer radius (cm) | 110 |
|    $\theta$-coverage | $0.84 < \cos\theta < 0.93$ |
|    Total weight (tons) | 4.05 |

### 6.1.4. Expected EMC performance

The energy resolution of the EMC calorimeter is affected by many factors, including the crystal quality, dead materials between and in front of crystals, photodiode and amplifier noise, fluctuations of shower energy leakage, calibration errors, etc. At low energies, the photon statistics can also affect the energy resolution. The photon position resolution is mainly determined by the crystal segmentation.

The EMC with 28 cm long CsI crystals can reach the desired energy resolution of $\leq 2.5\%$ at 1 GeV photon energy. A decision to choose 28 cm long crystals, slightly shorter than the 30 cm crystals used in the CLEO-c detector, was made mainly based on cost consideration. The nuclear counter effect due to charged particles leaked from the rear face of crystals hitting the photodiodes is not expected to be significant [40].

The effect of crystal transverse dimensions was also studied by Monte Carlo simulations. The energy and position resolutions as functions of photon energy are shown in Fig. 33. Three different crystal sizes with front face dimensions 4 cm × 4 cm, 4.5 cm × 4.5 cm and 5 cm × 5cm were studied. Energies in 5 × 5 crystal arrays were summed to reconstruct the shower energy. Fig. 33(a) shows the energy resolution as a function of photon energy from 20 MeV to 2 GeV for the three crystal sizes. This study shows that the energy resolution can reach about 2.4% at 1 GeV/c and 3.3 % at 100 MeV with 5 cm × 5 cm crystals. Fig. 33(b) shows that the position resolution of photon hits is from 5.5 mm to 14 mm for the entire photon energy range of BESIII using 5 cm × 5 cm crystals. The mass resolution of $\pi^\circ$'s reconstructed by a pair of photons in $J/\psi \to \rho\pi$ is determined to be 7.3 MeV, as shown in Fig. 34.

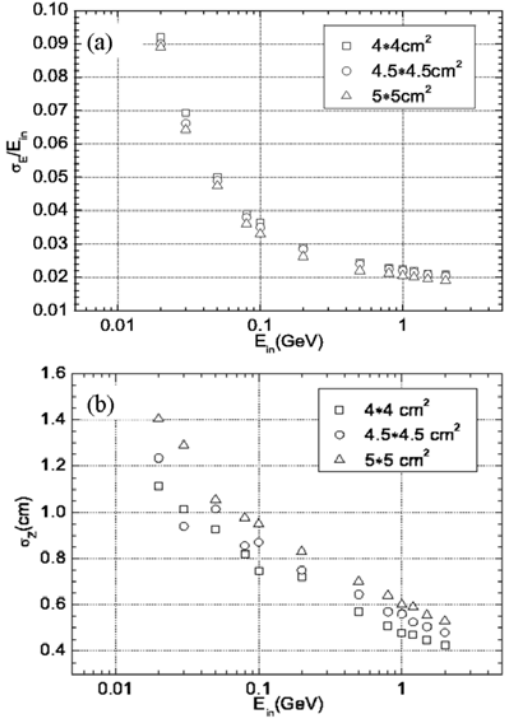

Fig. 33. Results of Monte Carlo simulation: (a) energy and (b) position resolutions as a function of energy

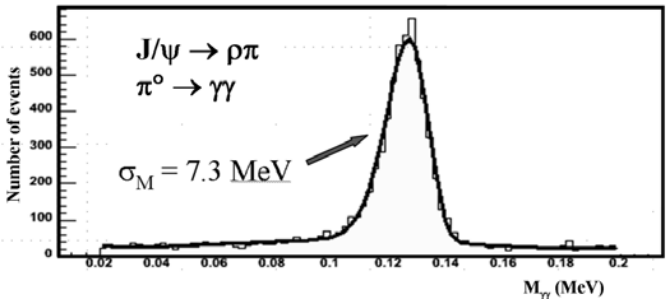

Fig. 34. Results of Monte Carlo simulation: the mass resolution of reconstructed $\pi^\circ$ masses.

In these studies, dead materials such as the beam pipe, the MDC structure and the 10 cm thick plastic scinitllator bars in front of BESIII EMC were not included. A significant fraction of photons will interact in these materials and will make the EMC energy resolution worse. The possibility of making corrections for the material in TOF counters will be discussed below in Section 6.7.3.

The parameters and expected performance of the BESIII EMC based on Monte Carlo simulation and beam tests are summarized in Table 18, along with parameters of the BESII EMC. As reference the parameters and achieved performance of the CLEO-c, Babar and Belle CsI calorimeters are given. The BESIII EMC calorimeter performance is expected to be similar to that of the CLEO-c EMC calorimeter.

Table 18
Parameters and designed performance of the BESIII EMC and parameters and performance achieved by BESII EMC, CLEO-c, Babar and Belle EMC calorimeters.

| Parameter | BESIII | BESII | CLEO-c | BaBar | Belle |
|---|---|---|---|---|---|
| $\Delta\Omega/4\pi$ | 93% | 75% | 93% | 90% | 91% |
| Active media | CsI(Tl) | Gas | CsI(Tl) | CsI(Tl) | CsI(Tl) |
| Depth ($X_0$) | 15 | 12 | 16 | 16 - 17.5 | 16.2 |
| $\sigma_E$ at 1GeV (MeV) | ~25 | 220 | ~20 | ~28 | ~17 |
| $\sigma_E$ at 100 MeV (MeV) | 3.3 | 70 | 4 | 4.5 | 4 |
| Position resolution at 1 GeV/c (mm) | 6 | 30 | 4 | 4 | 6 |

### 6.2. EMC crystal units

#### 6.2.1. Overview

The CsI crystals used in the BESIII EMC have the general shape of a truncated pyramid cut to 28 cm in length. Nominally, the barrel crystals have a 5.2 cm × 5.2 cm front face and a 6.4 cm × 6.4 cm rear face. Dimensions of the end cap crystals that have more complex shapes are similar. The machining tolerances for crystal dimensions were (-0.4 mm/+0.15 mm) for all sides and ±1mm for the length. Crystal sizes were matched during calorimeter assembly to minimize gaps between crystals.

The design of the crystal unit is shown in Fig. 35. Five crystal surfaces are wrapped with reflective material. An aluminum connecting plate is mounted at the rear end by four self tapping screws in 20 mm deep holes drilled in the crystal surface. An aluminum box hosting two preamplifiers is mounted on the connecting plate fixed by four screws. An aluminum angle bracket that is used to attach the crystal to its support beam is mounted on the top cover of the electronics box.

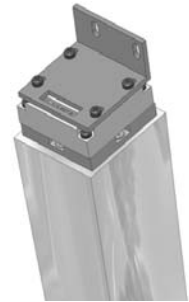

Fig. 35. The EMC crystal unit.

### 6.2.2. Crystal wrapping

Past experience has shown that a CsI(Tl) crystal wrapped by a diffusive white layer backed by an aluminum layer yields the highest light collection efficiency at the rear end of the crystal. The relative light yields of a crystal readout by two photodiodes of the size of 1 cm × 2 cm with 10 different wrapping combinations were measured and are shown in Fig. 36. A layer of laminated sheet of 25 μm thick aluminum and 25 μm thick mylar was wrapped on the outside. The double layer Millipore or Tyvek wrapping and the combined Millipore and Tyvek wrapping produced high reflection efficiencies (10, 5, 7, 8 and 6). Tyvek was chosen as the wrapping material because of its known reliable performance.

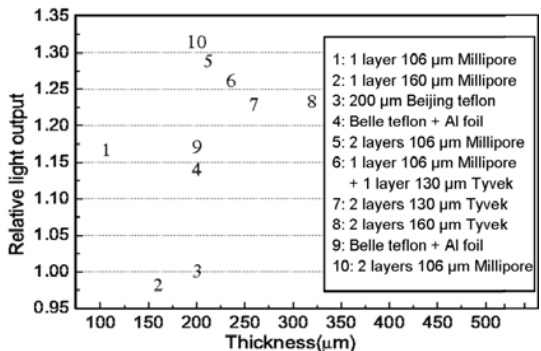

Fig. 36. Comparison of the relative light output (PD readout) of a CsI(Tl) crystal wrapped with different reflection films.

The CsI crystals of the BESIII EMC were wrapped with two layers of 130 μm Tyvek white paper and 25 μm polyester film laminated with 25 μm aluminum. The aluminum laminated polyester film provides further light collection enhancement, hermetic seal, RF shielding, electric isolation between crystals and mechanical protection. The aluminum film is electrically connected directly to the aluminum box of preamplifier box. The total thickness of the wrapping layers is 310 μm. There are no more dead materials between the crystals since they are suspended from their ends without partition walls. Gaps between crystals were kept to a minimum by the design of the mechanical suspension structure with radial adjustments and assembly procedures that will be described later.

### 6.2.3. Photodiodes and preamplifier assembly

The schematic of the crystal unit is shown in Fig. 37. A 50 mm×50 mm aluminum base plate 8 mm thick with a rectangular opening for the photodiodes was fixed to the crystal rear surface by four M4 self-tapping stainless steel screws. Two 1 cm x 2 cm Hamamatsu S2744-08 photodiodes were glued onto a 2 mm thick Lucite light guide, and the light guide was glued directly at the center of the crystal surface in the rectangular opening of the base plate. A two-component optical epoxy (Eccobond 24) was used. The gluing process was done in a dry box with relative humidity less than 10% at room temperature. After the epoxy cured, the rectangular opening in the aluminum base plate was covered with a layer of 500 μm thick Teflon that increased the light yield by about 7%.

Since the total rear surface area of a crystal is approximately 40 $cm^2$; the percentage area covered by the silicon photodiode is about 10%, however ,the fraction of light collected by the photodiodes is more than 10% because of light reflection by reflective material covering the rest of the crystal end surface.

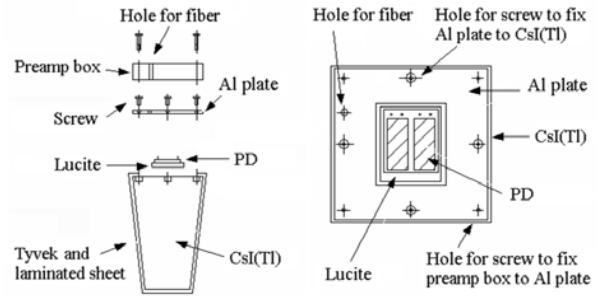

Fig. 37. Mechanical assembly of the crystal unit.

An amplifier box made of aluminum and housing two preamplifiers was mounted onto the aluminum base plate. The two photodiodes were directly coupled to the two preamplifiers via short wires. All metal shields (the aluminum foil wrapping the crystal, the base plate and the amplifier box) were electrically connected together to the ground of the preamplifiers to minimize electronic noise. Output signals of the preamps, power and photodiode bias voltage are supplied by a 20 conductor twisted pair cable. An optical fiber connector is mounted through a hole in the aluminum plate, and laser light can be used to illuminate the rear end of the crystal for system calibration and monitoring.

Quantum efficiencies of photodiodes can vary by about 10%. A database of the quantum efficiency of every photodiodes was established by testing every diode on an optical tester. The measured distribution of the relative photodiode quantum efficiencies is shown in Fig. 38. The database was used to match the quantum efficiencies of the two photodiodes to be attached to a crystal with the light yield of the crystal. A small number of photodiodes with relatively low quantum efficiencies were rejected.

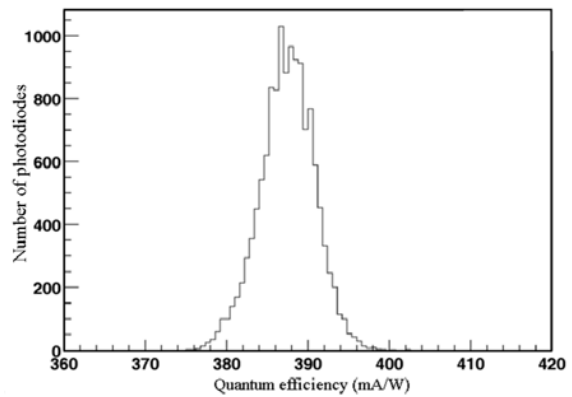

Fig. 38. Distribution of relative quantum efficiencies of photodiodes.

The gain of preamplifiers can also vary by about 10%. The gain of the preamplifiers, the quantum efficiencies of the photodiodes and the light yields of the crystals were taken into consideration during assembly of the units in order to obtain better uniformity among crystals.

*6.3. Crystal fabrication and tests*

*6.3.1. Crystal fabrication and acceptance tests*

CsI(Tl) crystals were fabricated by three manufacturers, Saint-Gobain of France (FR), Shanghai Institute of Ceramics (SH) and Hamamatsu of Beijing (BS), as listed in Table 19. The 33 different types of 960 end cap crystals, including the 192 with irregular shapes, were made by FR. Acceptance tests, including visual inspection and measurements of dimensions, light yields, light uniformity and radiation hardness were performed when crystals were received.

Table 19
Numbers of crystals fabricated by three manufacturers

|  | FR | SH | BS | Total |
|---|---|---|---|---|
| Ordered | 2040 + 960 | 1920 | 1320 | 5280+960 |
| Replaced | 87 (4) | 316 | 79 | 482 (4) |

Light yield is a critical parameter that determines the energy resolution of the calorimeter. The non-uniformity of scintillation light production along the crystal length is also fairly important. Since energy depositions by high energy photons of different energies have different longitudinal distributions, large longitudinal non-uniformity of crystal light yield can affect the photon energy resolution and the energy scale.

Light yield and non-uniformity of crystals were measured by collimated γ-sources moved by a step motor along the crystal length. Eight points at 3 cm steps were tested for every crystal along its length. Three γ energies, 0.662 MeV from $^{37}$Cs, 0.511 MeV and 1.27 MeV from $^{22}$Na, were used. The scintillation light from low energy gamma sources was detected by a standard bialkali photomultiplier tube placed at the end of the crystal. By comparing the uniformity measured in this way with that using cosmic ray muons with photodiodes attached to the end of the crystals, we observed that the different spectral responses between the PMT and photodiode did not affect the uniformity measurements significantly. The full energy peaks of the three γ energies were fitted to a straight line, and the light yield per MeV photon energy was calculated.

The light yields of the EMC crystals were compared to the light yield of a standard crystal cylinder that is 2.5 thick and 2.5 cm in diameter. Crystals with a light yield less than 33% of the light yield of the standard block were rejected.

The non-uniformity of the crystals was required to be better than 7%. Non-uniformity compensation was performed for a small fraction of the barrel crystals. A black line was painted on the crystal surface at the rear end of a crystal, where the light yield was higher, to attenuate the light generated locally. At the front end of a crystal where the light yield was low, a band of high reflectivity ESR film was wrapped around the crystal to enhance the local light reflection. We plan to use the crystal non-uniformity database in the EMC energy calibration procedure to compensate for the effect of different shower maximum depths for different photon energies.

*6.3.2. Quality of accepted crystals*

The measured light yields of the crystals that are used in the EMC relative to the light yield of the standard sized reference crystal are plotted in Fig. 39. The light yields of accepted crystals were higher than 40% of the reference crystal. The distribution of measured non-uniformity of the final crystals is shown in Fig. 40. Only a small fraction of crystals have non-uniformity slightly greater than 7%.

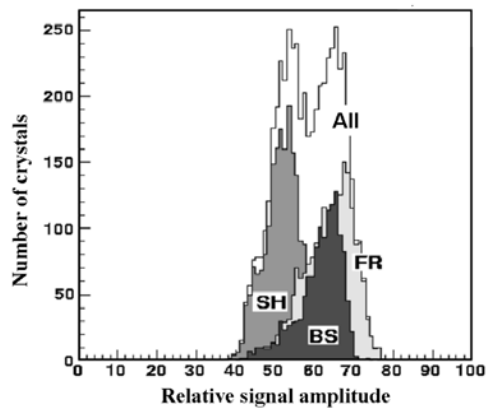

Fig. 39. Light yields relative to a standard reference crystal (%) of crystals delivered by three manufacturers.

The radiation levels at the EMC are not expected to be very high, no more than several hundred rads per year for crystals in the inner rings of the end caps. In order to test the radiation tolerance of the crystals, we required manufacturers to provide several samples of the size $2.5 \times 2.5 \times 12.5$ cm$^3$ for every batch of ingots they produced. These samples were irradiated up to 1000 rads at a rate of 25 rads per hour. The light yield reduction must be less than 20% after irradiation. The results of these tests are shown in Fig. 41. We also irradiated one finished full sized crystal out of every 100 crystals with 100 rads. The light yield reduction of this sample crystal must be less than 9%. A batch of SIC crystals was rejected and replaced by SIC because the radiation hardness was out of specification. This problem was traced to a purity problem of the raw material used. Percentages of rejected and replaced crystals due to various reasons by the other two companies were not very large.

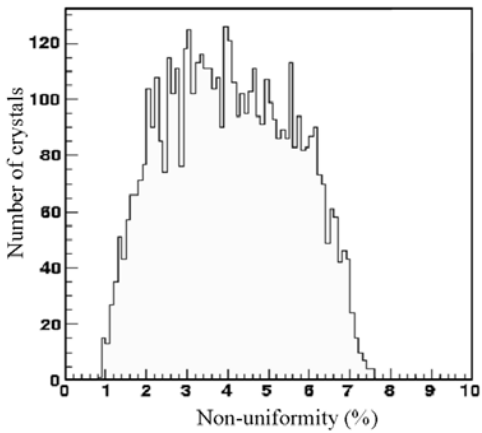

Fig. 40. Non-uniformity in percent along the length of crystals for all crystals.

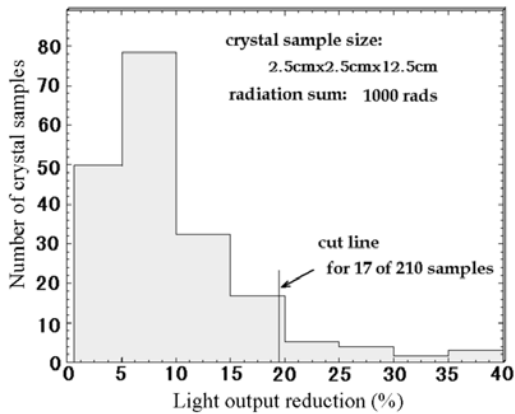

Fig. 41. Results of the radiation test for the representative test samples.

### 6.3.3. Cosmic ray test

Cosmic ray tests [41] were performed for every completed crystal unit. The dE/dx of a minimum ionization particle in the CsI(Tl) crystal is 5.6 MeV/cm, and an average of about 30 MeV is deposited by cosmic ray muons penetrating the 5.2 to 6.4 cm thick crystals. The light yields from muons are much higher than those of γ-sources and can be readout by photodiodes. The purpose of these tests was to check the performance of the crystal units after photodiodes and preamplifiers were mounted and to compare the results to test results using γ-sources and a PMT. The initial database for future EMC calibrations using collision data was established based on these cosmic ray tests.

The cosmic ray test setup consists of two arrays of scintillator strips, oriented 90° with respect to each other, that provide cosmic ray muon triggers and 2D hit position information. Sets of 64 crystal units was tested for 70 hours, and 20k triggers were accumulated. As in the γ source tests, a crystal was divided into 9 sections along its length. Measured deposited energies were plotted and fitted with Gaussian functions. The energy spectra were quite broad due to different dE/dx depositions from cosmic ray muons of different momenta and different entry angles. The typical rms width of a muon energy spectrum is ~20%. With sufficient statistics, peak positions of energy distributions can be determined quite accurately. The uniformity of the nine sectors in a typical crystal measured by the cosmic ray test setup is shown in Fig. 42.

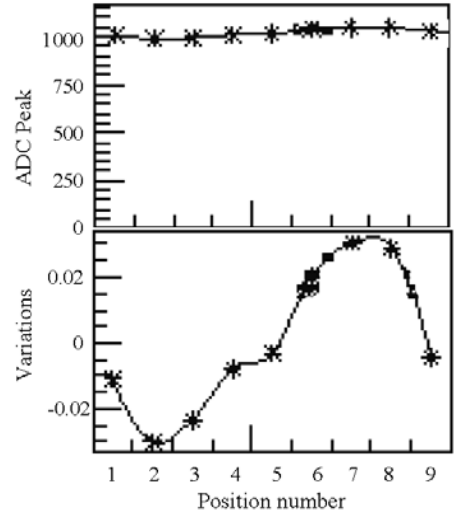

Fig. 42. Top: ADC peak positions; bottom: signal amplitude variation in a typical crystal.

The results of cosmic ray tests agreed well with the γ source tests described earlier. In addition to cross-checking with γ source data, the cosmic ray tests also helped to uncover some defects in the photodiode contacts, loose connections and other problems in the electronics box.

### 6.4. EMC mechanical structure and assembly

#### 6.4.1. Mechanical stress tests

A critical issue for the EMC mechanical design was whether crystals could be hung from a support beam at arbitrary angles by means of brackets attached to the crystals by screws. To test this, a long term stress test of a crystal unit was conducted. A photograph of the test setup is shown in Fig. 43. An electronics box and connecting plate were attached to the rear face of a crystal by four M4 self-tapping screws in 20 mm deep holes drilled in the crystal and the unit was suspended horizontally from a steel stand. With a loading of up to 47 kg of lead bricks placed on top of the crystal, threads of the four screws in the crystal were not stripped although the front-end of the crystal bent by about 80 mm, and the steel angle bracket was deformed. The glue joints between photodiodes and crystal surface also were not affected by the stresses. The CsI(Tl) crystal is very strong mechanically with properties similar to lead. This test proved that crystals have more than enough mechanical strength to be self supporting.

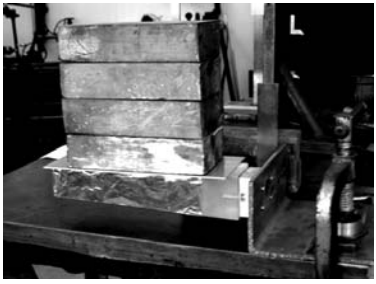

Fig. 43. Loading test of a crystal unit.

Additional mechanical tests were done to a group of crystal units. They were suspended either horizontally or vertically on an actual EMC supporting beam via the aluminum angle brackets. After four days suspension, the light outputs of the crystal modules were measured, and no significant changes were observed. Photodiode glue joints were not broken either.

*6.4.2 Mechanical structure of the EMC*

Based on the mechanical loading test described above, a method to suspend the crystals on support beams was developed in order to minimize the dead material between CsI(Tl) crystals and to simplify the mechanical design of the EMC. Crystal support based on the similar principle was also used successfully by the CMD-2 CsI calorimeter that was a much smaller detector with a different geometry at the VEPP-2M collider [42]. The EMC crystals are individually hung from support beams by suspension brackets screwed on to the rear end of the crystals without partition walls for providing mechanical support to crystals. The 120 sectors of barrel crystals in the azimuth direction are divided into 60 super modules; each has two rows of 44 crystals connected to a common support beam. A section of a prototype super module support is shown in Fig. 44, where five pairs of crystal suspension brackets mounted on a section of the T-shaped stainless steel beam are shown without crystals.

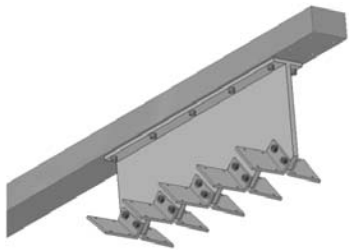

Fig. 44. A section of the prototype support beam with mounting brackets for five pairs of crystals.

The side and the cross-sectional views of the assembly jig for a barrel super module with 44 pair crystals loaded are shown in Fig. 45. Crystal pairs are pointing either ±5 cm from the IP except for the central two pairs of crystals that are pointing at the IP. Such an arrangement avoids photons escaping through cracks between crystals.

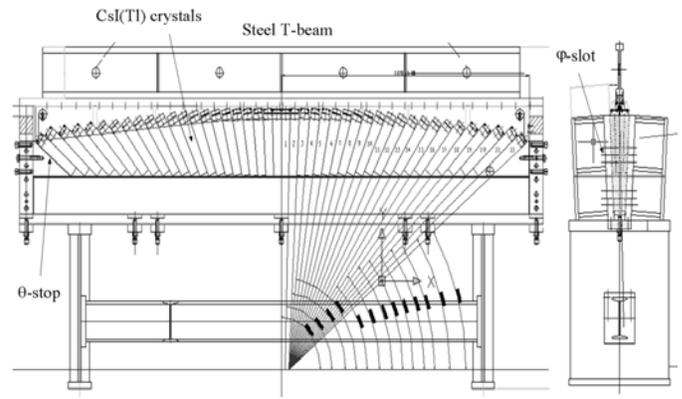

Fig. 45. The side and cross-sectional views of the barrel super module assembly jig.

In the crystal barrel assembly process, a complete super module, hanging vertically, was held by a support beam. The partially assembled barrel below was supported by jigs at the two ends and could be rotated to allow a hanging super module to be dropped down into its final position. Once in place, the support beam of the super module is screwed to the two support rings at the ends of the barrel. The mechanical support structures of the end cap crystals are similar to the barrel in principle. An end cap is divided into two halves that can be opened for inner detector access.

*6.4.3. Crystal size matching*

The two rows of crystals hanging on the support beam were assembled in an steel jig with two 3° side walls. Two crystals with slightly different sizes could be paired to minimize the variations of super module widths. Hanging heights of crystals supported by hanging brackets were individually adjusted to match the widths of crystal pairs at different θ-angles and to minimize gaps between crystals. Another requirement for the crystal matching was to sort crystals so that the light yields of neighboring crystals were similar.

*6.5. EMC readout electronics*

*6.5.1. Introduction*

The purpose of the EMC readout electronics is to measure the energy deposited in every crystal and to provide a fast energy trigger. Specifications for the EMC readout electronics are summarized in Table 20.

The block diagram of the EMC readout electronics is shown in Fig. 46. The system consists of four main components:

- Preamplifiers mounted on the crystals.
- Post amplifier/shaper NIM modules.
- Charge digitization VME modules (Q-module).
- Test and control VME modules (Test-Control).

Signal pulses from photodiode preamplifiers are sent to post amplifier/shaper crates via 18 m twisted pair cables. The post amplifiers/shapers are implemented in NIM modules, each module has 16 channels. The post amplifiers/shapers amplify and shape signals with a 1 μs shaping time.

Table 20
Specifications for the EMC readout electronics

| Parameter | Values |
|---|---|
| Number of channels | 6,240 |
| System clock | 20.8 MHz |
| L1 trigger latency | 6.4 μs |
| Maximum single channel hit rate | ≤ 1 kHz |
| Equivalent noise charge (energy) | 0.16 fC (200 keV) @80 pF |
| Integral nonlinearity | ≤ 1% (before correction) |
| Cross talk | ≤ 0.3% |
| Dynamic range | 15 bits |
| Information to trigger | Analog sum of 16 channels |
| Gain adjustment range for triggers | ≤ 20 % |

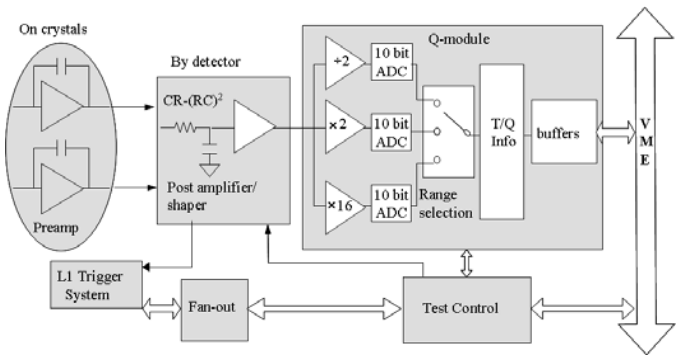

Fig. 46. Block diagram of the EMC readout electronics system.

The EMC readout electronics system has sixteen 9U VME crates and approximately 200 charge digitization modules. Each of these 9U VME modules has 32 channels. Outputs of the post amplifiers are sent via twisted pair cables to Q-modules in VME crates where the charge digitization takes place. Shaped pulses are digitized by FADCs running at 20.8 MHz synchronized with the 41.65 MHz trigger clock. The timing information of the energy signals is extracted. Digitized data are stored in local buffers in each module and then read out via the CBLT (Chained Block Transfer) and virtual global buffers by the VME bus. The digitized signals are stored in digital pipelines in compliance with the 6.4 μs L1 trigger latency without losing data.

The test and control module in the VME crate receives control signals from the fan-out module and distributes them to Q-modules. Fast energy sum signals from every 16 crystals are also formed by the post amplifier modules and sent to the L1 trigger system through the TCBA (Trigger Cell and energy Block Adding) VME modules as explained in Section 8.4.

The fan-out module in the VME crate communicates with the L1 trigger system via optical links. It receives clock, L1 strobe signals from the L1 system and sends them to the fan-out module for distribution. It also sends the fast energy sum trigger signals it receives to the L1 system.

*6.5.2. Preamplifiers*

The output current signals from the two S2744-08 photodiodes are sent to the two charge sensitive preamplifiers mounted on the backs of the photodiodes in the aluminum electronics box via very short wires. The output signals are sent to the post amplifier via 18 m twisted pair cables.

The parameters of the preamplifier are summarized in Table 21. The charge amplifier has a feedback loop consisting of a 1 pF capacitor in parallel with a 50 MΩ resistor. This results in a gain of 1mV/fC for converting the input charge to output voltage and a signal decay time of 50 μs, which does not result in a significant pileup for the expected 1 kHz maximum hit rate in a single crystal.

Table 21
EMC preamplifier parameters

| Gain | 1 mV/fC |
|---|---|
| Equivalent input noise | 700 e |
| Dynamic range | 0.5 to 1,500 fC |
| Linear output range | up to 2.5 V |
| Decay time constant | 50 μs |
| Power consumption | 140 mW |

The equivalent input noise charge (ENC) of the preamplifier is < 700 e (0.12 fC) corresponding to ~150 keV photon energy deposited in a CsI(Tl) crystal. The ENC measurement results of 384 channels, each channel includes two photodiodes and two charge amplifiers, are shown in Fig. 47. The average ENC per channel is 973 e, corresponding to approximately 200 keV with the 80 pF photodiode capacitive loading. The power consumptions of the preamplifiers is 140 mW, and active cooling is needed.

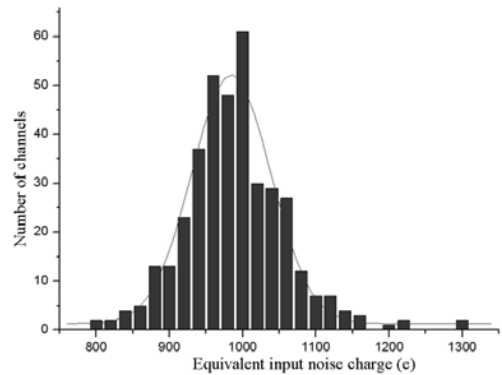

Fig. 47. The measured equivalent input noise charge distribution of 384 preamplifier channels.

A test input is provided at the input of each preamplifier for testing the functionality of the electronics chain and correcting the nonlinearity of each channel. Standard test pulses that mimic the shape of real calorimeter signals are coupled to the input of the charge amplifiers through a 1 pF capacitor via an analog switch. In order to minimize the input noises, test pulses are high impedance current signals that are converted to voltage signals by a 100 Ω resistor at the preamplifier input.

The low voltages of preamplifiers, the bias voltage of photodiodes, and the test pulses are supplied by post amplifier modules. The electronics box is connected to the post amplifier module via a 14 conductor twisted pair cable.

*6.5.3. Post amplification*

Post amplifiers are implemented as single width NIM modules, each accommodates 16 channels. The block diagram of the post amplifier is shown in Fig. 48. The two preamplifier outputs are received by two differential amplifiers and then go through a switching stage with two selections. Normally the selection A + B that sums the two photodiode outputs is used. If one photodiode dies, the selection 2(A + B) that boosts the signals by a factor of two is used.

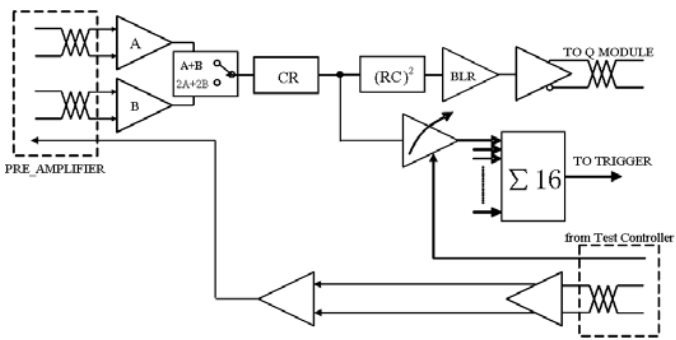

Fig. 48. Block diagram of the post amplifier and analog signal processing.

The post amplifier provides semi-Gaussian shaping to reduce noise and pileup. The time constant of the pole-zero shaping circuit is chosen to be 1 μs, roughly matching the decay time constant of the CsI(Tl) scintillation. This shaping time is found to be optimum for minimizing system noise. The shaped signals go through the baseline restoration circuit (BLR) and are then sent by a differential driver to the charge measurement circuits in the Q-module via twisted pair cables. Shaped signals have a 3 μs peaking time and 10 μs decay time that are sufficiently long for the FADCs at a 20.8 MHz rate. Two or three data points can be sampled at the peak of the signal pulse.

Sixteen analog signals from post amplifier channels are sent to a circuit labeled $\Sigma_{16}$ to generate module analog sums. The gain of each channel is adjusted to compensate for the light yield variations of crystals and response differences of photodiodes. The gain of every channel is calibrated using laser light pulses and gain adjustments are made by the digital potentiometer chip MAX5400. The data for setting the gain adjustments are received by the summing circuit via a serial bus, and gain adjustments take place in the gain adjustment mode operation of the EMC readout electronics.

*6.5.4. Charge digitization*

The photon energy measurement range of the EMC is from 20 MeV to ~2 GeV. The digitization precision is set to 15 bits with the least significant bit corresponding to 0.06 MeV. Three 10 bit FADC channels with different full scale ranges are used to achieve the desired 15 bit dynamic range. The FADC clock speed is one half of the 41.65 MHz of the L1 trigger clock or 20.8 MHz with a sampling cycle of about 50 ns. The charge digitization and peak finding circuits are implemented as single 9U VME modules; each has 32 channels.

The FADCs, which are capable of operating at 20.8 MHz with 15 bit equivalent resolution, are realized by three 10 bit FADC channels with three different gains: ÷2, ×2 and ×16. The output of the unsaturated FADC channel (values < 1,023) that has the highest gain (or lowest range) is selected as valid data by a logic circuit, and a 2-bits range code is added to the data. The 12 bit data are sent to a digital pipeline and the peak finding circuit that must find the peak within 3 μs after the L1 trigger arrives. The peak value is compared with the threshold stored in the threshold register, and signals smaller than the threshold are discarded. Digital values of signals above threshold, including a 10-bit signal amplitude, its 2-bit range code and an additional 6-bit peaking time, are stored in buffers to be read out. The digital logic, pipelines, buffer and VME control circuits are implemented in FPGA chips.

*6.5.5. Test Control*

Each of the VME readout crates in the EMC readout system has a Test Control Module. This module controls the three operation modes of the EMC readout system:
- Collision mode
- Calibration mode
- Gain adjustment mode

Collision mode is the regular data acquisition mode when the collider is in operation. The Test Control Module in each VME crate receives the clock, L1 strobe and reset signals. It also receives the buffer full signal from the Q-module in the same crate and relays it to the trigger system via the fan-out module. Read requests from each Q-module are also sent to the Test Control Module that generates an interrupt to the VME controller for requesting a data transfer. The frequency of the 41.65 MHz trigger clock is divided by two to 20.827 MHz to become the EMC readout system clock. These signals are then distributed to the 32 channels in every Q-module.

In the calibration mode, the test control module generates the 20 MHz clock, trigger strobes, and a series of test pulses corresponding to different amounts of charge to test each electronics channel. An electronics self check is performed, and calibration constants are verified. The energy correction constants can be calculated from the data obtained.

The gain adjustment mode is designed for generating the analog energy sums of crystals for the L1 trigger. The analog sums of signals from blocks of 16 crystals in the barrel and 15 crystals (Trigger Cells or TCs) in the end caps are sent to the L1 trigger system to form EMC triggers. This trigger scheme is designed to reduce the number of cables between the EMC and the trigger system, and simplify the trigger logic. The relative gains among crystals in a cell must be adjusted locally in order to obtain accurate energy information. In the gain adjustment mode, the test control module generates the serial clock and serial data to adjust scale factors for each channel to compensate for light yield differences, photodiode responses and gain variations of electronics channels.

### 6.5.6. Fan-out

The communication between the L1 trigger system and the EMC readout electronics is handled via three optical links. The first link transmits the 41.65 MHz trigger system clock. The second link transmits 16 bit control signals including the L1 trigger strobe and reset. The third link reports the status of the EMC electronics to the trigger system. The fan-out module receives these signals and distributes them to the 16 Test and Control Modules in the 16 VME crates of the EMC readout system.

### 6.6. Beam test of crystal array

A beam test of the crystal array was conducted at the E3 test electron beam line at IHEP. The test array consists of a matrix of $6 \times 6$ crystals. Beam momenta used were 400, 600, 800 and 1,000 MeV/c. Electrons were selected by a gas Cherenkov counter in the beam line.

The measured energy resolution by summing the energy deposits in a $3 \times 3$ crystal matrix was 2.6% at 1 GeV/c, as shown in Fig. 49(a), slightly worse than the result from Monte Carlo simulation due to the rather poor quality of the beam. The fit function used was a Landau distribution with the horizontal axis inverted. The electron test beam as described in section 4.2.1 had relatively large energy, spatial and angular spread determined by the bending magnets and collimators. The electron energies changed within the beam profile and this effect was difficult to completely remove in the data analysis. The measured position resolution was approximately 5.6 mm as shown in Fig. 49(b). Two Gaussian functions were used in the fit.

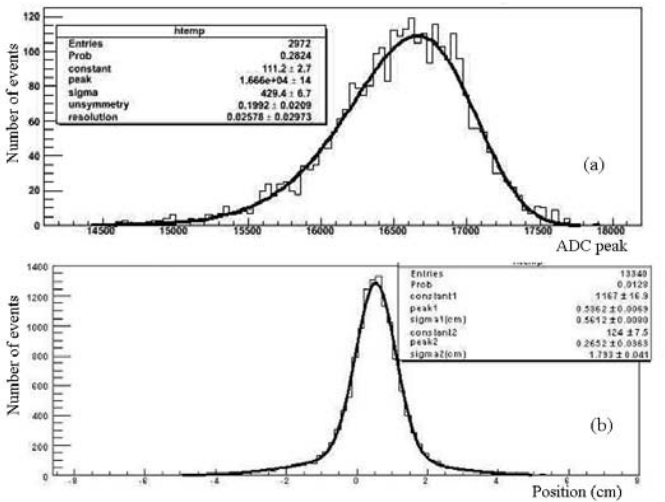

Fig. 49. Measured (a) energy and (b) position distributions.

### 6.7. EMC energy calibration

#### 6.7.1. Light pulsing system

Radiation doses vary with crystal location and crystals can have different degradation rates due to the radiation. Factors such as the photodiode responses, electronics gain variations and environmental factors can also affect the signals. Changes in responses caused by radiation damage to the crystal and other factors are monitored by a light pulsing system as shown in Fig. 50. The stability of the light pulsing system is 0.5%.

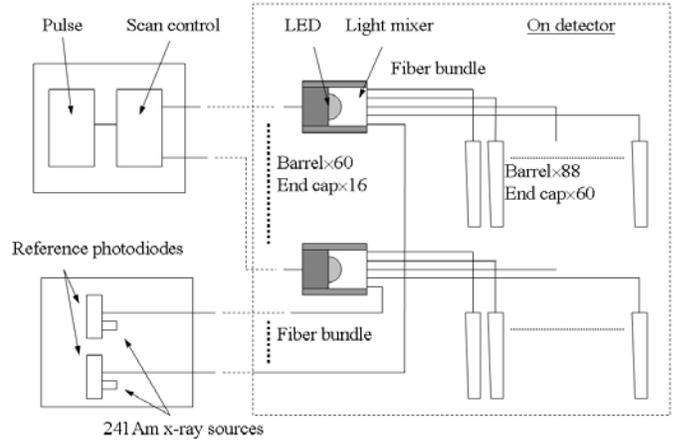

Fig. 50. EMC light pulsing system.

The light pulsing system consists of a pulser, LED light sources and light mixers, optic fiber bundles and a reference system for system stability. The LEDs are LXHL-BL03 made by Luxeon with 1 W output power. The light pulses after a light mixer are distributed by an optical fiber bundle to 88 crystals in the barrel or 60 crystals in the end caps. Quartz fibers with 300 μm diameter were used.

The light mixer is a block of Lucite wrapped with diffusive Teflon and aluminum foil. The light pulses produced are similar in spectrum, rise-time and shape to the scintillation light pulses of CsI(Tl) crystals. The non-uniformity of light delivery by fibers is less than 10%.

The intensity of the LED light pulses is monitored pulse-to-pulse by a reference system consisting of two photodiodes and $^{241}$Am radioactive sources. The photodiodes measure the intensity variations of LED pulses. The stability of the photodiodes are monitored by the 60 keV X-rays from the $^{241}$Am sources. Ten light intensity levels equivalent to photon energies 4 MeV to 2.0 GeV are scanned. The frequency of the LED pulses is 300 Hz. The system stability is 0.5% and can be improved to 0.3% after the offline correction based on temperature compensation.

An important function of the light pulsing system is to monitor the entire readout chain of crystals including the photodiodes. By comparing to the data obtained on the lab bench before the crystal installation, problems occurring during installation and during the operation of the detector can be quickly diagnosed. The optical monitoring results can also monitor the level of crystal radiation damage since the light absorption of the crystal increases with radiation dose received.

*6.7.2. Crystal energy calibration*

The FADC values must be converted to incident photon energies. The formula used to convert the ADC value of crystal $i$ to the photon energy $E_i$ is

$$E_i = \frac{ADC_i - PED_i}{e_i \times c_i}$$

where

- $ADC_i$ is the ADC value
- $PED_i$ is the pedestal value
- $e_i$ is the gain of the electronics chain (mV/C)
- $c_i$ is the energy conversion constant (C/MeV)

The values of $ADC_i$ and $PED_i$ are experimentally measured. The electronics gain $e_i$, which converts the output voltage to the input charge of the preamplifier, is known from the design of the electronics circuits, and is calibrated by sending test pulses to the input of the preamplifier. The energy conversion constant $c_i$ that relates the incident photon energy to the input charge of the preamplifier is the most difficult calibration constant to obtain.

There are several ways to roughly obtain a set of energy conversion constants $\{c_i\}$ for each crystal before the calorimeter is assembled. These include testing every crystal using electron or other charged particle beams, cosmic ray muons and γ-sources with known photon energies. These methods establish preliminary values of $\{c_i\}$ that will be further refined to obtain more precise values by the online calibration procedures.

Every CsI(Tl) crystal was tested in the lab before being installed in the EMC. The scintillation light obtained with a $^{137}$Cs source read out by a PMT was measured. Cosmic ray tests in which the crystals were read out by photodiodes and preamplifiers were performed. The energy conversion constant database established by these tests is used in the BESIII EMC crystal calibrations as initial values. These constants, however, can only be used to determine the relative performance of the crystals.

The final values of energy conversion constants $\{c_i\}$ will be obtained by the calibration procedure using real collision data. Physics processes that will be used include $e^+e^- \to \gamma\gamma$, Bhabha scatters, and radiative Bhabha events. Other physics methods, such optimizing the $\pi^0$ mass reconstructed from the two photons from $\pi^0$ decays can also be used.

The rate of Bhabha scattering events in the BESIII detector at the BEPCII design luminosity of $10^{33}$ cm$^{-2}$s$^{-1}$ will be approximately 1 kHz. Pre-scaled Bhabha triggers will be recorded on tape and be used for high precision EMC crystal calibration. Many effects such as the non-linearity of the crystals, temperature dependence of the light yield, back and side leakages, dead materials between crystals, crystal non-uniformities, dependence on polar angles, radiation damages, etc. can only be compensated for by offline calibration procedures. Cosmic ray runs will be taken to check the crystal calibration data.

*6.7.3. Energy corrections for materials in front of EMC*

The material in front of the barrel EMC can affect the EMC performance. In Table 22, the materials for detector components in front of the EMC are listed. The double layer TOF counters dominate. GEANT Monte Carlo studies show that roughly 30% of photons convert, mostly by $e^+e^-$ pair production in the MDC and TOF before reaching the EMC.

Table 22
Materials in front of EMC at normal incidence

| Parameter | Materials | $X_0$ (%) |
|---|---|---|
| Beam pipe | 1.4 mm Be +14.6 μm gold + 0.8 mm oil | 1.04 |
| DC | 1.32 cm carbon fiber | 6 |
| TOF | 10 cm plastic scintillator | 23.5 |

The simulated energy losses of 1 GeV photons in the barrel TOF counters are plotted in Fig. 51. A small fraction of photons lose a significant amount of energy before they enter the CsI crystals. Energy measurements by the EMC only for such photons will not be very accurate. Monte Carlo simulations show that the energy losses of photons in the TOF counters in principle can be added to the EMC shower energies to partially compensate for the energy losses [43].

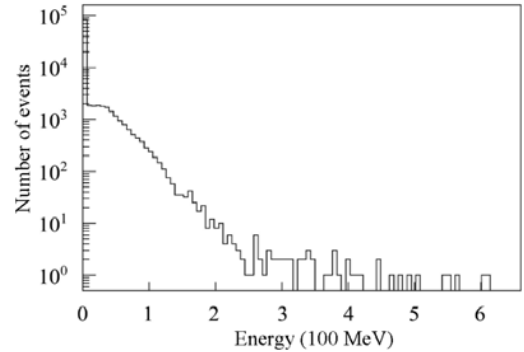

Fig. 51. Simulated energy losses of 1 GeV photons at 90° incident angle in the TOF counters.

Including the energies deposited in 3 neighboring sectors of the two layer TOF counters in front of a 5 × 5 matrix of CsI crystals, the width of the total energy peak for high energy photons becomes narrower, as shown in Fig. 52 for 1 GeV photons. Fig. 53 shows that the energy resolution of the EMC can be improved at high energies by including the energy deposited in the TOF counters. Three sets of MC data were used, 5 × 5 matrix of CsI crystals only, 5 × 5 crystals plus one scintillators in each of the two layers and 5 × 5 crystals plus three scintillators in each of the two layers.

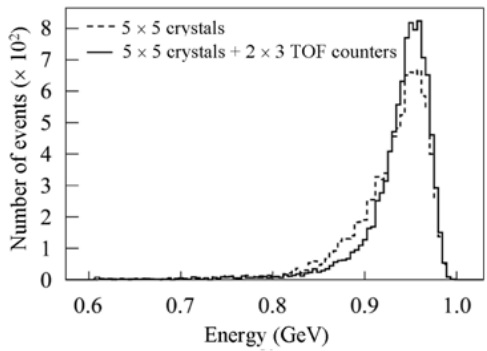

Fig. 52. Simulated energy depositions by 1 GeV photons without and with adding the energies in TOF counters.

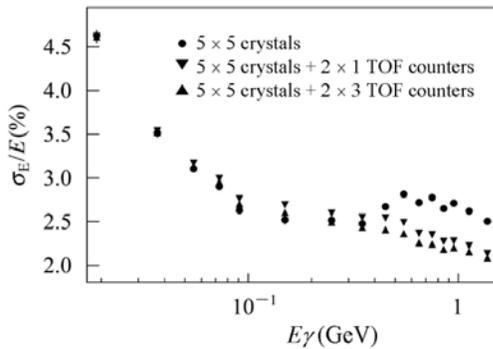

Fig. 53. Simulated energy resolutions as a function of energy with and without the energy depositions in the TOF counters.

### 6.8. Environment control and monitoring

#### 6.8.1. Cooling and temperature monitoring

Active water cooling is required in order to keep the temperature of the CsI(Tl) crystals stable and to remove the heat produced by the preamplifiers that have a power consumption of 0.14 W per channel. Without active cooling, the local temperature at the end of the crystals can be more than 10 °C higher than the ambient temperature. In addition to the heat generated by the EMC preamplifiers, TOF PMTs and bases at the two ends of the EMC barrel and at the inner rings of the EMC end caps are also heat sources. If the heat from these sources are not actively removed, the temperature changes in these areas can cause the EMC energy scale to shift, and the non-uniform light yield will deteriorate energy resolution because the CsI(Tl) crystals have a 0.3%/C° temperature coefficient. Also, the level of noise due to photodiode leakage currents will change rapidly since the leakage current rises exponentially with temperature.

Copper pipes 6 mm in diameter are installed on the aluminum suspension brackets of the crystals to circulate cooling water and remove heat. In the barrel, copper cooling pipes are installed along the direction of the support beams. The temperature of the cooling water is kept to 20 ± 1 °C, and temperature of the crystals is controlled to approximately 25 °C.

Approximately 600 LTM8802 semiconductor temperature sensors are distributed throughout the CsI(Tl) crystal array with a measurement accuracy of ± 0.5 °C. The read out of temperature sensors is done by the BESIII slow control system based on Universal Serial Bus (USB) and the LTM8301 readout module. The recorded temperatures can be used for offline energy corrections.

#### 6.8.2. Humidity control and monitoring

Due to the slight hygroscopic nature of the CsI(Tl) crystal, dry nitrogen gas with a humidity less than 5% is forced to flow through eight nozzles installed at each end of the EMC to maintain a dry atmosphere and to prevent degradation of the crystal surfaces. Approximately 200 humidity sensors, with measurement accuracy of ±3%, are distributed throughout the EMC and are read out by the slow control system.

#### 6.8.3. Radiation monitoring

Radiation damage causes the light yield of CsI(Tl) crystals to be reduced by about 10% per krad of radiation dose received. The EMC crystals are shielded by the focusing magnets and masks. The steel return yoke and collimators around the ring also protect the EMC from beam related radiation. The highest expected radiation dose of the end cap crystals is 300 rads per year. The accumulated radiation doses of the EMC crystals are mostly from beam injection that are not included in the simulated dose rates shown in Table 6. Four PIN diode radiation sensors, identical to the sensors installed on the beam pipe, are installed on the inner surface of EMC end cap, two on each side, to monitor the instantaneous radiation level and record the integrated doses.

In addition, 72 RadFET radiation sensors (NMRC: 400nm IMPL RADFET) used to record the integrated doses are installed on the front faces of the crystals at representative locations. The RadFET is a type of integrated radiation dose sensor that has very small size and low power consumption. The RadFETs are p-type MOSFETs with gate and drain connected together. Their accumulated charges in the RadFET gate region due to radiation causes the output voltage to shift while biased by a constant current source. The RadFET is inherently non-linear. Thus its response curve, $\Delta V_T$ as a function of dose under a given gate bias voltage, must be calibrated usually by using of a γ-ray source with energy around 1 MeV.

## 7. Muon identifier

### 7.1. Overview

Identifications of muons is of great importance in $e^+e^-$ collision physics. Many final states of the BESIII physics processes contain one or two muons. For instance, studies of leptonic decay branching fractions, semileptonic decays of $c\bar{c}$ resonances, rare charmed meson decays such as $D^0 \to \mu^+\mu^-$, and τ decays, all rely on muon identification to separate them from hadrons. The momentum spectra of muons produced in τ and D decays at $\sqrt{s}$ = 3.67 and 3.77 GeV are plotted in Fig. 54.

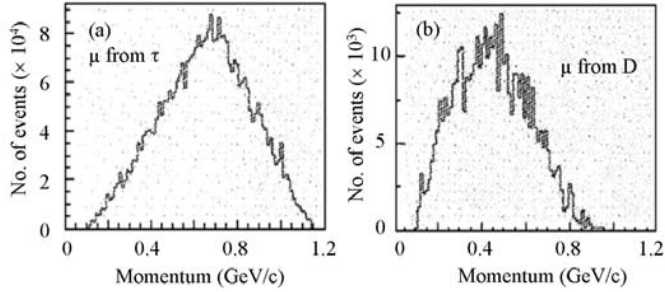

Fig. 54. Momentum distributions of muons produced in τ decays at $\sqrt{s}$ = 3.67 GeV (left) and $D$ decays at $\sqrt{s}$ = 3.77 GeV (right).

The BESIII muon identifier must be able to identify muons with high efficiency well separated from punch through pions and other hadrons with an acceptance as large as possible. Detecting muons with a momentum cut-off as low as possible is also a very important consideration for a muon identifier in the τ-charm energy region.

The BESIII muon identifier is constructed of resistive plate counters (RPCs) inserted in the steel plates of the magnetic flux return. The steel barrel of the flux return is divided into nine layers and a total of nine layers of RPCs are installed. Due to space limitations in the end caps, eight layers of RPCs are installed in the gaps between the nine layers of steel plates.

By associating hits in muon counters with tracks reconstructed in the MDC and energy measured in the EMC, muons can be identified with low cut-off momentum. The RPC modules have readout strips in both the θ and φ directions for reconstructing charged particle tracks. The requirement for the position resolution of the RPCs is modest because of the multiple scattering of low momentum muons in the EMC, magnet coil and the steel layers. The typical width of the readout strips in the RPCs is about 4 cm. The momentum of identified muons can be precisely measured by the MDC.

In addition to performance considerations, the active detectors for the muon identifier system must be highly reliable with low cost since the total area to be covered is quite large and almost inaccessible once the detectors are installed. A safe working gas must be used for the RPCs because of their large area and gas volume. The design performance parameters of the BESIII muon identifier are given in Table 23. Parameters of the BESII muon identifiers are given as reference. The details of the detector construction are give in Ref. [44].

Table 23
Design performance parameters of BESIII muon identifier

| Parameter | BESIII | BESII |
|---|---|---|
| $\Delta\Omega/4\pi$ | 89% | 65% |
| Number of layers (Barrel/End caps) | 9/8 | 3 |
| Technology | RPC | Proportional tubes |
| Cut-off momentum (MeV/c) | 400 | 500 |

A 3D solid model of the BESIII muon identifier is illustrated in Fig. 55. The overall shape of the instrumented flux return is octagonal. Layers of steel plates are joined together by connecting plates to form octants. Rectangular shaped RPC barrel modules are installed in gaps between steel plates. The end cap steel and RPC modules are constructed as quadrants. Two quadrants are joined together to become movable end doors. The mechanical parameters of muon identifier are listed in Table 24.

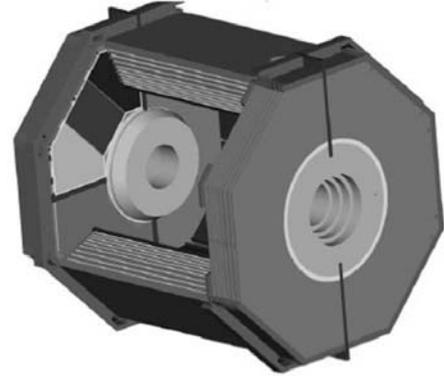

Fig. 55. The 3D models of the BESIII muon identifier.

Table 24
Mechanical parameters of the muon identifier

| Barrel | Parameters |
|---|---|
| Inner radius (m) | 1.700 |
| Outer radius (m) | 2.620 |
| Length (m) | 3.94 |
| Weight (tons) | 300 |
| Steel plate thicknesses (cm) | 3, 3, 3, 4, 4, 8, 8, 8, 15 |
| Gap between plates (cm) | 4 |
| No. of RPC layers | 9 |
| Polar angle coverage | $\cos\theta \leq 0.75$ |
| End cap | Parameters |
| Inner distance to IP (m) | 2.050 |
| Outer distance to IP (m) | 2.800 |
| Weight (tons) | $4 \times 52$ |
| Steel plate thicknesses (cm) | 4, 4, 3, 3, 3, 5, 8, 8, 5 |
| Gap between plates (cm) | 4 |
| No. of RPC layers | 8 |
| Polar angle coverage | $0.75 \leq \cos\theta \leq 0.89$ |

The first layer of barrel RPC modules is placed in front of the first layer of steel plates. Eight subsequent layers of RPC modules are inserted into the gaps between the 9 layers of steel plates. The thicknesses of the steel plates, especially the first five layers, are made intentionally thin. This design ensures the cut-off momentum for muon detection is as low as possible. The end cap muon identifier has 8 layers of RPC modules inserted in gaps of the 9 layers of steel plates. It is desirable to place the end cap steel as close to the superconducting coil as possible to improve the magnetic field uniformity of the solenoid. The total area covered by the RPC modules is approximately 700 m$^2$.

*7.2. Resistive plate counter development*

*7.2.1. New resistive plate electrodes*

Based on experience with the BaBar RPC system and the problems associated with the linseed oil coating, we decided to investigate new approaches for constructing bakelite RPCs when we started to consider the design of the muon identifier for BESIII. A new type of phenolic paper laminate has been developed to construct the RPCs. The surface quality of these laminates is superior to bakelite plates that have been used to construct RPCs elsewhere. We also developed a method for adjusting the resistivity of these laminates and the surface resistivity of the coating. Tests have shown that prototype RPCs made using our resistive plates without the linseed oil treatment can achieve a level of performance comparable to RPCs with linseed oil treated bakelite or resistive glass electrodes in streamer mode.

The bakelite plates used as electrodes for RPCs are made by a high pressure lamination process in which paper layers, after going through a resin bath and a roller system, are placed in a large press on a polished stainless steel plate. Another steel plate is placed on top of the paper layers. The stack of paper layers and steel plates are then held in a high pressure press for up to several hours at elevated temperature. The hardened laminates are then removed from the press and cut to size. The resin used is usually phenolic or a mixture of phenolic and melamine. In order to improve the surface quality of the bakelite plates, the paper layers used on the outside of the laminates are substituted with more refined paper layers and specially formulated melamine based resin was used [45] in order to improve the hardness and smoothness of the surface of the bakelite plates. The surface quality and the cleanliness of the stainless steel plates, which are repeatedly used, also play a critical role in determining the surface quality of the finished bakelite plates. The stainless plates used to produce bakelite plates for the BESIII RPCs were hand selected.

A number of prototype counters were manufactured using the special phenolic paper laminates that we developed with improved surface quality and without the linseed oil treatment. These prototypes adopted a typical single gap RPC design. The RPC prototype was working in streamer mode with 4 cm wide readout strips placed outside of the RPC gas gap. Extensive tests using cosmic rays were conducted, and the gas mixture $Ar/C_2F_4H_2/C_4H_{10}$ 50:42:8 was selected for the BESIII RPC system. After a short period of training at a working voltage of 8 kV, the single gas gap efficiency reached approximately 96%, the singles rate was approximately 0.04 Hz/cm$^2$, and the dark current was less than 1 $\mu$A/m$^2$. [46].

*7.2.2. RPC single gas gaps*

Details of the RPC single gas gaps used in the BES muon identifier are shown in Fig. 56 The bakelite thickness is 2 mm and the gas gaps are also 2 mm. Spacers of 12 mm in diameter, arranged as a matrix with 10 cm spacing, are glued to the inner surface of the bakelite electrodes. The gas volume is enclosed by gluing a 10 mm wide frame made of ABS plastic around the edges. Gas inlet and outlet tubes penetrate the 2 mm thick frame. The efficiency of typical single gas gap is better than 95%.

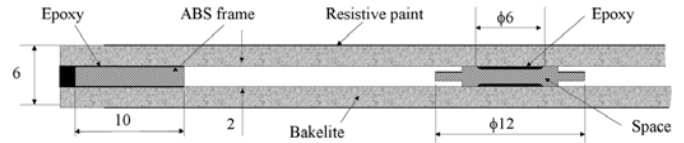

Fig. 56. The cross-sectional view of the RPC gas gap.

The water molecules absorbed when the bakelite plates are exposed to room air play a critical role in electrical conduction. In a dry environment, it is believed that the conductivity of the bakelite plates comes from the migration of positive ions of small organic molecules remaining after the phenolic resin polymerization process. The conductivity of bakelite can be controlled by the manufacturing parameters such as resin formulation, temperature, pressure, and time in the press. The RPC modules were produced at the Beijing Gaonengkedi Science and Technology Co. Ltd. (BGST) in close collaboration with IHEP [47]. Some initial testing and training of the produced RPC gas gaps were done at the company. The resistive plate production process, including the chemical formulation and production parameters, were strictly controlled at the bakelite plate factory and supervised by BGST engineers and physicists of the BESIII muon group. The surface quality and resistivity of the plates were checked before they were used. The bulk resistivity of the bakelike plates used was in the range of $2\times10^{11}$ $\Omega\cdot$cm to $2\times10^{13}$ $\Omega\cdot$cm. The resistivity of bakelite plates for end cap RPCs is lower than the resistivity of bakelite used in the barrel counters, as shown in Fig. 57. Polycarbonate spacers 2 mm thick and 12 mm in diameter were glued every 10 cm to the surfaces of the bakelite plates to maintain the 2 mm gas gap. The areas of RPC gas gaps range from 0.7 m$^2$ to approximately 1.7 m$^2$ with an average of approximately 1.2 m$^2$. A resistive paint consisting of a mixture of colloidal carbon and epoxy was sprayed on the outer surfaces of the bakelite plates to form electrodes for supplying high voltages. The finished conductive layer is 0.1 mm thick with a surface resistivity of 0.2 to 1.0 M$\Omega$/square.

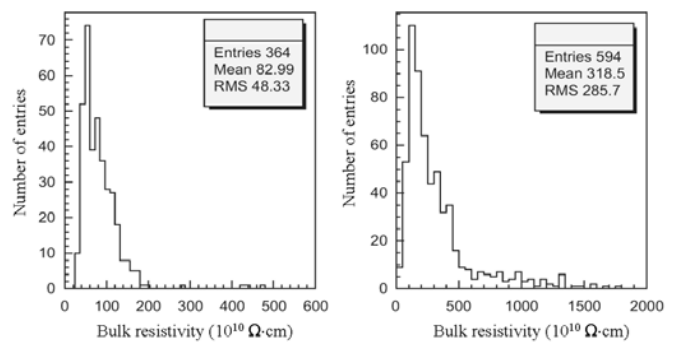

Fig. 57. Distributions of bakelite bulk resistivity used in end cap (left) and barrel (right) counters.

The quality assurance procedures at BGST included a gas leak test, spacer glue joint check, and initial performance test and training. The gas leak test was done at 10 cm $H_2O$ over-pressure. Three main parameters: cosmic ray efficiency, dark current, and singles rate, were tested at a time for a group of RPC gas gaps in the test setup. The RPC gas gaps that passed the Q&A were delivered to IHEP for further testing.

As with most gas detectors, the RPC gas gaps were subjected to a training procedure. The method used for the end cap counter training was conventional. The regular working gas was used, and the counters underwent cosmic ray tests while the training took place. The high voltage was cycled from 6 kV to 9 kV while the counter efficiency plateau was recorded. It took quite a long time, as long as many days in some cases, for the counter performance to reach an acceptable level.

In order to speed up the training process, a very aggressive training method was later used at IHEP for the barrel counters. The gas gaps were trained in pure argon gas under high voltage at 10 kV. The initial current of a counter can be as large as several hundred $\mu A/m^2$. Dark currents usually dropped quickly and after about 24 hours, the counter currents were often reduced to below 100 $\mu A/m^2$. Counters with current still higher than this level were rejected after an additional 24 hour training. Performance of the accepted gas gaps was tested with the RPC operating gas mixture at 8 kV.

The performance parameters of the single gas gap counters are summarized in Table 25. The data were taken at 8 kV. The average efficiency measured using cosmic rays with temperature in the range of 20°C to 22 °C is 96% for the barrel and 95% for the end cap counters. Note that the percentage of surface area occupied by the spacers is about 1.6%.

Table 25
Performance of barrel and end cap RPC single gaps at 8 kV

| Barrel | Barrel | End cap |
|---|---|---|
| Average efficiency | 96% | 95% |
| Dark current ($\mu A/m^2$) | 1.1 | 1.8 |
| Dark current ($Hz/m^2$) | 0.1 | 0.15 |

A total of 384 end cap and 594 barrel single gap counters were produced. The total counter area is 1272 $m^2$. Among all single gas gaps produced, five were rejected because they had either more than one unglued spacer or their dark currents were too high during the initial test. Another nine were rejected because they failed the cosmic ray test. Most of the rejected modules were end cap counters that were produced before the final quality assurance procedures were put into place.

*7.2.3. Design and construction of BESIII RPC modules*

The accepted single gas gaps were assembled into 136 super modules, 72 in the barrel and 64 in the end cap. The RPC modules adopts a standard double-gap design to improve the muon detection efficiency. Two single gas gaps are stacked together to form a double-gap with the readout strips sandwiched in between as shown in Fig. 58. The muon tracking efficiency can reach approximate 98% using the double-gap design. The double-gaps were placed in a 32 mm thick aluminum box to become a super module. Two layers of 6 mm thick polycarbonate honeycomb panels were used for mechanical protection.

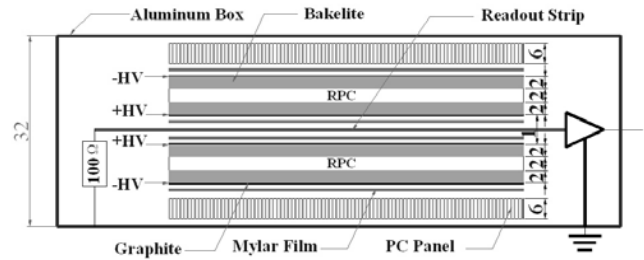

Fig. 58. Cross-sectional view of the double gap RPC.

A typical end cap super module, which covers a quadrant of an octagonally shaped end cap, consists of six trapezoidal shaped single gas gaps, and a typical barrel super module consists of eight rectangular single gaps. Single gas gaps are not equal in size and they were staggered in a super module so that there are no dead regions except regions around the box edges where the gas gap frame, electronics and service lines are placed. The sizes of the barrel super modules vary according to layer number. The fractions of dead area range from 5.4% to 7.4% for the barrel and 6.9% to 7.8% for the end cap. The high voltage power supplies are common for all double gap modules in a super module and the gas flow is connected in series.

Each of the RPC superlayers can measure only one coordinate and the orientation of signal strips alternate by layer. In the barrel, odd numbered layers having strips in the Z-direction measure azimuthal coordinates, and even numbered layers having strips in the Φ direction measure the longitudinal coordinates. The readout strips in the end caps are arranged alternately in the horizontal (X) and vertical (Y) directions with odd numbered layers measuring the Y-coordinates and even numbered layers measuring the X-coordinates as indicated in Fig. 59.

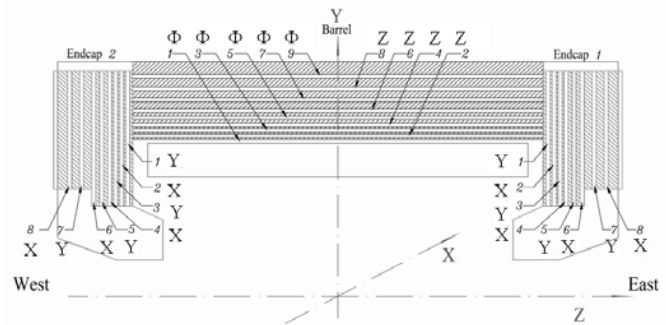

Fig. 59. The RPC layer arrangements and strip orientations.

The barrel contains 72 rectangular super modules that span the full length of the 3.94 m long barrel. Spaces are reserved at each end for electronics and services. The two end caps have a total of 64 super modules. The design parameters of the RPC super modules are listed in Table 26.

Table 26
Super module parameters

| Barrel | Parameters |
|---|---|
| No. of super modules | 72 |
| Total No. of channels | 5,056 |
| Module length (m) | 3.8 |
| Module width (m) | 1.050 to 1.848 |
| Layers measuring Φ-coordinates | 1, 3, 5, 7, 9 |
| Φ-strip pitch in each layer (mm) | 20, 27, 30, 33, 37 |
| No. of Φ-strips per SM | 48 |
| Layers measuring Z-coordinates | 2, 4, 6, 8 |
| Z-strip length (mm) | 1120, 1187, 1261, 1382 |
| Z-strip pitch (mm) | 39 |
| No. of Z-strips per SM | 96 (octant 1 has 112) |
| End cap | |
| No. of super modules | 64 |
| No. of strips per SM | 64 |
| Total number of channels | 4096 |
| Layers measuring Y-coordinates | 1, 3, 5, 7 |
| Layers measuring X-coordinates | 2, 4, 6, 8 |
| Strip pitch (mm) | First strip: 49 |
| | Last strip: 32 - 54 |
| | Other strips: 35 |

The muon identification efficiency and purity are the main considerations in determining the design of the muon identifier. Based on the results of Monte Carlo simulations, the pitch of the signal strips are chosen to be in the range of 20 mm to 54 mm. The exact choices of strip pitches are not crucial because of the multiple scattering.

### 7.3. Gas system

Gas flows to the 136 super modules are parallel and metered individually. Flows of pure argon, freon 134a and iso-butane gases from bottles are individually controlled by a MKS247B mass flow controller to form the working gas mixture $Ar/C_2F_4H_2/C_4H_{10}$ 50:42:8. The total gas volume of the RPC system is approximately 4 m$^3$, and the flow rate is approximately one volume change per day. The gas mixture from the mass flow controller fills the buffer tank, and the gas flow is then divided into 136 branches. The flow rate of each branch is controlled by a meter valve and a flow indicator. The output gas from the super modules goes through bubblers and is vented to the atmosphere. The maximum differential pressure allowed at the inlets of the RPCs is 4 cm H$_2$O limited by bubblers.

### 7.4. RPC readout electronics

#### 7.4.1. System overview

The RPC system has approximately 10,000 channels that must be read out at a maximum trigger rate of 4 kHz. Front-end cards (FECs) consisting of discriminator chips, pipelined data buffers and data transmission controllers are mounted near the edges of the RPC super modules in the gaps of the flux return steel plates. Signals from the RPC super modules are brought to the FECs by short twisted pair cables. Sixteen FEC cards are daisy-chained to form a "data chain". A total of 40 data chains, each having up to 16 FECs (256 channels) are connected to 10 readout I/O modules in the readout crate via 30 m long shielded twisted-pair cables. Control commands, clocks and serial data are also transmitted. The system clock is 20.8 MHz and synchronized with the accelerator clock. The Fast OR signals of the first four layer barrel RPCs are used as inputs to the global L1 trigger. The muon data stored in the FECs are transmitted to the VME crate after an L1 trigger is received.

#### 7.4.2. The FEC

A FEC card is 30 cm × 5 cm in size, and each FEC has 16 channels. Working in the streamer mode, the signal sizes produced by RPC gas gaps are quite large. For the double-gap counters working at 8 kV, the most probable signal amplitude is about 500 mV across a 50 Ω load, with a rise time of a few ns. Note the signal amplitudes depends on the bakelite resistivity, which was affected by the relative humidity and temperature at the time when tests took place.

A block diagram of the FEC is given in Fig. 60. The front-end electronics boards host 16 channels of discriminators and a data controller that receives the L1 trigger. The four channel CMP401GS comparator chips (Analog Devices Inc.) are used to directly discriminate the signals. Thresholds of the discriminators are set by an 8-bit DAC chip (TLC7528CDW, Texas Instruments) on the card, and its reference voltage is 2.5 V. Thresholds can be remotely set, and the corresponding minimum threshold value is about 10 mV. In normal RPC operation, the threshold is set to 100 mV. A test pulse is provided to the input of every discriminator channel. The 16 channels of discriminated signals are sent to a FPGA chip (Altera ACEX 1K series EP1K30TC144-3) on the FEC.

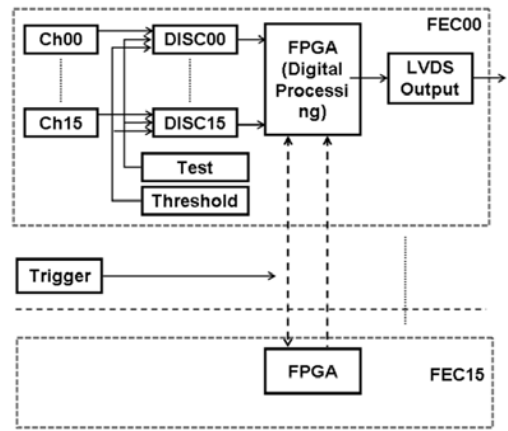

Fig. 60. Block diagram of the front end card (FEC) of the RPC system.

Sixteen buffers operating at 20.8 MHz are implemented in the FPGA chip: eight pipelined buffers in front of the trigger for the input data and eight buffers acting as derandomizers after the trigger in order to achieve a dead time less operation. Data control logic is also implemented in the FPGA chip. Within 16 clock cycles or 800 ns, with respect to the event time, data in the de-randomizers are transmitted serially through the daisy chain to the data readout modules. The wide trigger gate is enough to account for jitter in the trigger latency.

The output data from the FPGA are converted into LVDS format and transmitted to the readout module through 30 m long shielded twisted-pair cables. The FPGAs on the FECs can be remotely configured, which is useful for adjusting the parameters and also important if the FPGA program is lost due to bursts of radiation resulting from unusual beam losses.

*7.4.3. RPC readout crate*

The RPC readout system is hosted in a VME crate that contains a master controller, 10 readout modules and a control module. The system reads the compressed muon data of 600 bytes per event, and the total data rate is 2.4 MB/sec at a maximum trigger rate of 4 kHz.

(a) Master Controller

The master control module is a Motorola MVME5100 based on the 450 MHz PowerPC750 with 512 MB of memory and an Ethernet interface running the real-time VxWorks operating system and user programs.

(b) Readout module

Each readout module can handle four data chains. The functions of readout modules include initializing the FECs, transmitting control commands, clock and trigger signals to all FEC data chains, receiving data from FECs, compressing and storing the data in buffers. These modules also monitor the status of the system operation and add status information to the data chain for the system problem diagnostics.

(c) Fan-out module

The fan-out module receives the trigger signals (L1, Clock, Check, and Reset) from the trigger system via optical links. It decodes and converts the optical signals into TTL signals and makes the signals available on the VME bus. In self-testing mode, the fan-out module generates a system clock and triggers to test the muon readout system.

*7.5. RPC super module cosmic ray tests*

The RPC super modules were tested in a large cosmic ray stand in order to obtain 2D efficiency maps and measure the position resolution [48]. Eight super modules could be tested at once and readout strips in four super modules were perpendicular to the readout strips in the other four super modules. They were read out by the 16-channel FECs. The fast OR outputs from FEC cards mounted on the trigger stations were used to generate cosmic ray triggers. The FECs were read out to a computer by specially designed readout electronics modules with USB interface. Clean single muon events were selected in the data analysis by excluding multi-muon events, cosmic air showers and other irregular events.

Normally about 600k events were recorded for each test. Two dimensional efficiency maps were obtained for every module. A super module was divided into small cells with the length and width of a cell roughly equal to the width of RPC signal strips. The average efficiency is determined for every cell and an efficiency map was generated.

Results of 72 barrel super modules are shown in Fig. 61. Fig. 61(a) shows the efficiency distribution of all cells. Cell efficiency is over 99% except for areas near the boundaries. The overall efficiency of the double gap super modules is 98%. Fig. 61(b) shows the distribution of the dark current of the barrel super modules and their mean is approximately 1.1 $\mu A/m^2$.

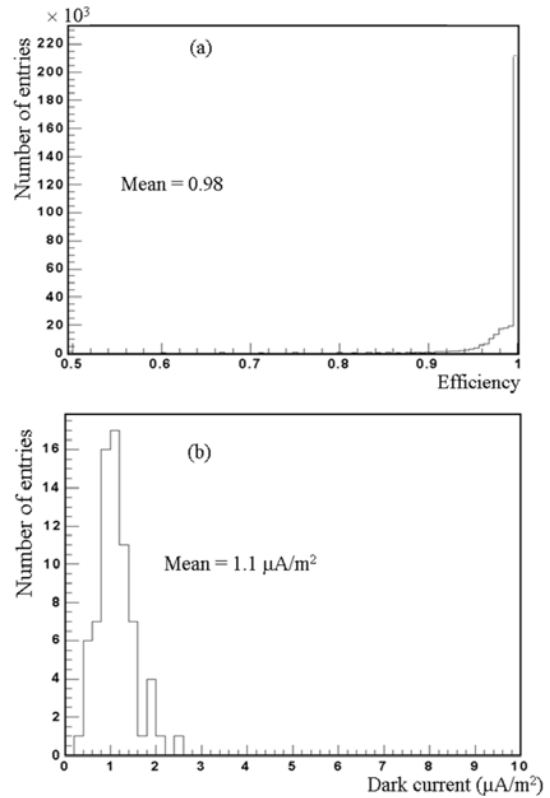

Fig. 61. (a) Cell efficiencies and (b) dark current of barrel super modules.

The end cap RPCs were tested using cosmic rays after they were installed in the end cap yokes. The end cap flux return yokes were placed flat on the ground during the test. Fig. 62(a) shows measured efficiencies and Fig. 62(b) shows the dark currents.

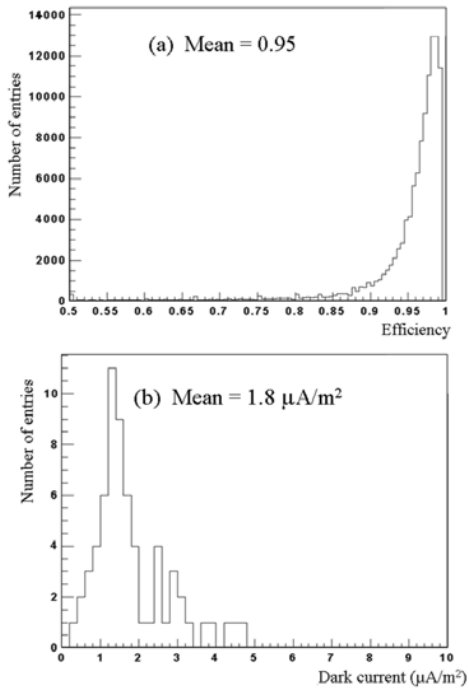

Fig. 62. (a) Cell efficiencies and (b) dark current of end cap super modules.

*7.6. Expected system performance*

The total solid angle coverage of the BESIII muon identifier is $\Delta\Omega/4\pi = 0.89$. The single layer position resolution measured by cosmic ray muons is ~1.2 cm, significantly smaller than the position uncertainties caused by multiple scattering. To reduce the electronics cost, each RPC layer only measures one coordinate. The $\theta$ and $\phi$ positions of muon tracks are measured in alternate layers. Monte Carlo simulation was performed to study the performance of BESIII muon identifier. The minimum momentum of muons that can be identified is determined to be approximately 0.4 GeV. For muons with momentum over 0.5 GeV, the identification efficiency is higher than 90%, and the pion contamination is less than 10% at various incident angles, as shown in Fig. 63.

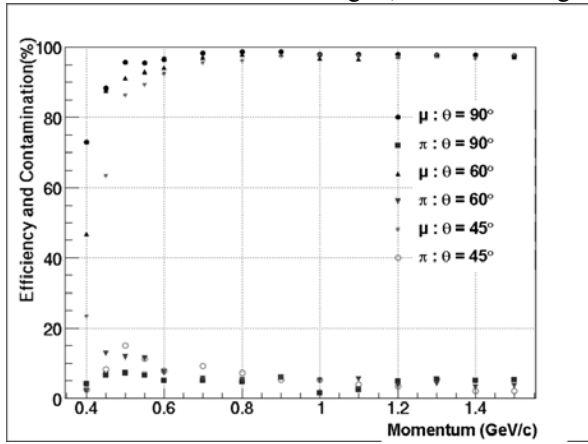

Fig. 63. The simulated muon identification efficiency and pion contamination rate.

## 8. Level 1 Trigger

*8.1. Overview*

*8.1.1. Data flow*

The BESIII trigger system [49] is a two level system consisting of the Level-1 hardware trigger and the Level-3 software trigger (event filter), performed in the online computer farm. The data flow diagram in the BESIII is shown in Fig. 64. The L1 trigger is designed to select good physics events with high efficiency and to reduce the cosmic ray and beam related backgrounds to a level smaller than the 2 kHz physics event rate expected at the J/$\psi$ peak. The maximum L1 trigger rate is designed to be 4 kHz.

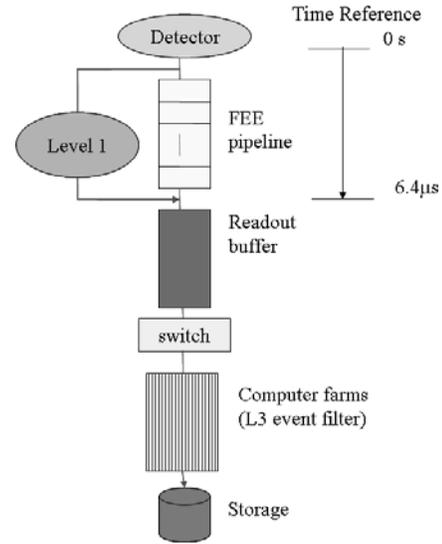

Fig. 64. Data flow diagram of the L1 trigger.

The BESIII frond-end electronics and DAQ system are completely pipelined. Data in the front-end buffers are read out once an L1 accept signal is received, 6.4 μs after the collision. Data are transmitted to an online computer farm where event building takes place. The event stream goes through an event filter based on IBM eServerBlade computers. Background events are further suppressed by the L3 trigger, and data are eventually written to the permanent storage device.

*8.1.2. Event rates*

At the peak luminosity of $10^{33}$ cm$^{-2}$s$^{-1}$, the physics event rates are expected to be 2 kHz and 600 Hz at the J/$\psi$ and $\psi'$ resonance, respectively. Physics events are detected with high efficiency and logged in permanent storage for off-line analysis. The rate of Bhabha events in the detector coverage ($|\cos\theta| < 0.95$) is expected to be about 800 Hz. Pre-scaled Bhabha events will be recorded for luminosity measurements and detector calibration purposes.

At ground level in Beijing, the cosmic-ray event rate in the BESIII is estimated to be about 170 m$^{-2}$ s$^{-1}$ × 3 m × 3 m = 1.5 kHz. Cosmic ray backgrounds are suppressed mainly by the MDC and TOF triggers at L1, to approximately a level of 200 Hz.

Because of the high beam currents, beam related backgrounds, dominated by the lost electrons and positrons are expected to be high. Since beam background rates are very sensitive to actual accelerator conditions, it is difficult to make reliable estimates. The beam current of BEPCII will be 40 times higher than that of BEPC and the beam lifetime is expected to be only half that of BEPC. Therefore the electron/positron loss rate, and corresponding beam backgrounds could be 80 times the BEPC background rate. Assuming the lost particles are uniformly distributed around the storage ring, the rate of lost electrons/position hitting the spectrometer can be $1.3 \times 10^7$ Hz. This is a conservative estimate since the particle losses will be larger around the final focusing magnets near the IP. A Monte Carlo study based on a modified DECAY TURTLE gives similar rate estimates [15]. Extensive collimators and masks are installed around the storage ring to reduce the possibility that lost particles hit the spectrometer. The L1 trigger system must thus reduce the background trigger rate to less than 2 kHz. The event rates of the BESIII trigger system at the J/$\psi$ peak are summarized in Table 27.

Table 27
Expected trigger rates at the J/$\psi$ peak

| Processes | Event rate (kHz) | After L1 (kHz) | After L3 (kHz) |
|---|---|---|---|
| Physics | 2 | 2 | 2 |
| Bhabha | 0.8 | Pre-scaled | Pre-scaled |
| Cosmic ray | < 2 | ~0.2 | ~0.1 |
| Beam background | > $10^4$ | < 2 | < 1 |
| Total | > $10^4$ | 4 | 3 |

*8.1.3. L1 trigger architecture*

The block diagram of the BESIII L1 trigger system is shown in Fig. 65. The main features of the L1 system are:

- All signals from the sub-detector front-end electronics are pipelined to the trigger system with a clock frequency of 41.65 MHz, a twelfth of the 499.8 MHz RF frequency. A L1 decision is made every clock cycle (24 ns).
- Optical link transmission links are used between the trigger system and all the front-end and readout electronics crates to reduce common-ground noises. RocketIO Multi-Gigabit Transceivers, implemented in a Xilinx Virtex-II Pro FPGA, are used to reduce the number of optical fibers and connections.
- Data in the sub-detector readout electronics pipelines are read out for event building and L3 trigger assessment after a L1 accept signal is received, 6.4 μs after the collision.
- Advanced FPGAs with online re-configuration capability are used to provide great flexibility and reliability.

In Fig. 65, the components on the left side of the vertical dashed line are sub-detector front-end electronics (FEE) and readout electronics crates located near the spectrometer, while the components on the right side are L1 trigger crates, located in the counting room. Data and control signals are transmitted between the two sides via 45 m high speed serial optical links to avoid ground loops and to simplify the cable connections [50].

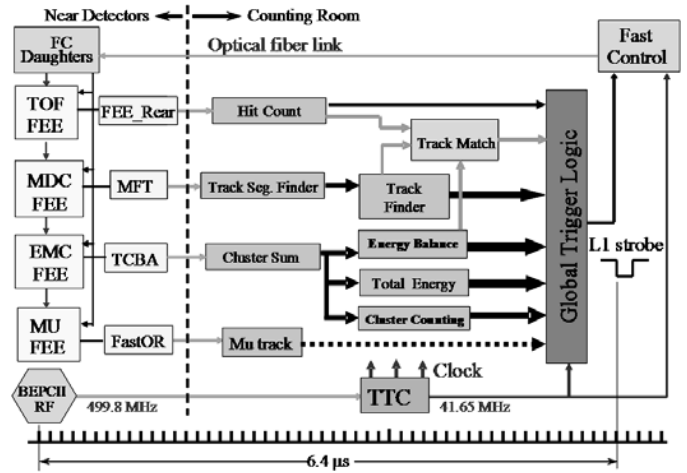

Fig. 65. Block diagram of the trigger system.

Three sub-trigger systems based on the MDC, TOF and EMC provide inputs for the L1 trigger decisions. Fast OR signals from the first four layers of muon RPCs, after simple track finding, are also sent to the L1 global trigger system, although they are not yet used in the current trigger scheme.

The sub-detector signals needed for the L1 triggers are initially processed in the four sub-system modules: FEE_Rear for the TOF, MFT (MDC Fiber Transmitter), TCBA (Trigger Cell and energy Block Adding) for EMC and FastOR for the muon identifier. The FEE_Rear, MFT, TCBA and FastOR are electronics modules located in the readout electronics crates of the corresponding sub-systems. Signal stretching and parallel-to-serial conversion are the main function of FEE_Rear and MFT modules. Trigger Cell formation, energy summation and parallel-to-serial conversion are the main functions of TCBA module. Online test mode operation is also implemented in each module.

Output signals of these modules are transmitted to the L1 trigger crates via optical links. The serial optical signals received by the trigger crates are first converted to parallel electrical signals by optic receivers on the trigger interface modules. The TOF hit patterns; track segment finding and final tracking in the MDC; isolated cluster energies and patterns and total energy of EMC crystals are generated and processed by the appropriate logic units to yield basic trigger conditions. These calculations include TOF hit counts and back-to-back conditions; short and long tracks and their distributions in MDC; energies of isolated clusters and blocks in the EMC. These trigger conditions are sent to the Global Trigger Logic (GLT) for making final trigger decisions.

The TOF, MDC and EMC sub-trigger conditions are also sent to the Track Matching Logic (TML) unit in which TOF hit patterns are matched with the tracks found in the MDC and energy clusters in the EMC. Outputs from the track matching logic are also sent to the GLT where the final L1 accept signal is produced.

L1 accept signals, along with other control commands and the system clock, are sent to the Fast Control system to be distributed to the various readout electronics crates via optical links. These fiber-optic signals are received by the fast control daughter cards in the readout electronics crates of sub-detectors.

The 6.4 μs total L1 decision latency is mainly determined by the time needed to find peaks from the slow EMC signal shaping time that have a 1 μs peaking and 3 μs decay time. The time signals for MDC and EMC triggers are extracted by the L1 trigger processors. The event time for charged particles is determined by the TOF trigger, which can determine the event time to ~30 ns.

The trigger information of the sub-triggers and global triggers, including trigger or event number, trigger conditions, system status and error messages are stored in pipelined buffers to be passed to the DAQ system for offline analysis.

*8.1.4. Trigger hardware implementation*

The L1 trigger system is a compact system with one NIM crate, one 6U and two 9U VME crates. The NIM crate is used to host test modules. The 6U VME crate hosts the system clock distribution modules, the fast control distribution module, the readout control module (TROC), a PowerPC crate control module and 2 TDC modules. One of the 9U VME crate hosts ten MDC track finder modules (8 TKFs for outer chamber and two ITKF for inner chamber), two track counter (TKC) modules, as well as a readout control module (TROC) and a PowerPC crate control module. The other 9U VME crate hosts the Trigger Cell (TC) and energy Block Adding (TCBA) modules and Energy Adding and Cluster Counter (EACC) module for the EMC subsystem, the TOF trigger processor (TOFT), signal alignment and fanout (SAF) and GTL modules for the global trigger, and also a readout control module (TROC) and a PowerPC module. Each of these modules will be described later in this section.

*8.2. TOF sub-trigger system*

*8.2.1. TOF L1 trigger signal flow*

The TOF signals with a timing uncertainty of 30 ns are the fastest among the L1 trigger signals. The TOF signals are used to determine the event time in the L1 trigger system based on charged particle hits. The block diagram of the TOF trigger signal flow is shown in Fig. 66.

The 176 mean-time outputs of the two layer barrel counters and the 96 signals from the end caps are generated by the TOF FEE modules in the TOF readout VME crates. These 272 bits of counter hit information are converted into 22 branches of serial data by the optical drivers of the rear plug-in cards (FEE_rear) mounted on the back of the crates. The data are sent to the trigger master modules in the trigger crate via 45 m optical cables. TLK1501 (Texas Instruments) serial-to-parallel transceiver chips and HFBR-5921L (Agilent) optical transceivers are used for the TOF signal optical transmission.

The TOF trigger master is a 9U VME module that receives the serial optical data, converts them back to parallel electrical signals, re-synchronizes them with local trigger clock and places them in a FIFO buffer. Digital buffers and trigger logic are all FPGA-based and can be remotely programmed via the VME bus and controller.

The TOF trigger processor (TOFT) generates six TOF trigger conditions (7 bit each) that are sent to the global trigger logic and 136-bit hit pattern signals that are sent to the track matching logic for matching with MDC tracks and EMC clusters.

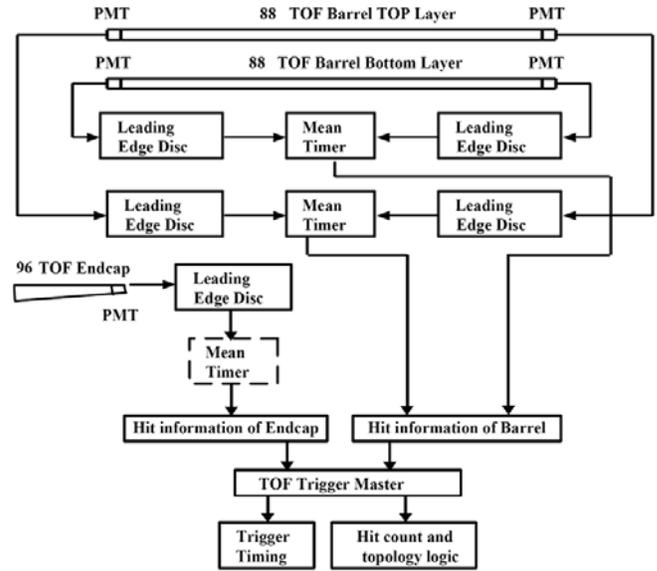

Fig. 66. Block diagram of the TOF L1 trigger data flow.

*8.2.2. TOF L1 trigger conditions*

The TOF sub-trigger provides 6 TOF trigger conditions to the global trigger logic. They are:
- NBTOF ≥ 1 : Number of hits in the barrel TOF ≥ 1.
- NBTOF ≥ 2 : Number of hits in the barrel TOF ≥ 2.
- NETOF ≥ 1 : Number of hits in the end cap TOF ≥ 1.
- NETOF ≥ 2 : Number of hits in the end cap TOF ≥ 2.
- TBB : Barrel TOF back to back.
- ETBB : End cap TOF back to back.

These 7-bit TOF trigger data are transmitted to the global trigger crate via a 3 m long twisted-pair cable. The hits in counters of two barrel layers are first put into coincidence. The back-to-back condition is satisfied if a hit exists in the section of 13 counters opposite to another hit. The back-to-back condition for the two end caps is such that a hit in one end cap must have a corresponding hit in the other end cap within a range of 9 counters.

*8.3. MDC sub-trigger system*

*8.3.1. Overview*

The drift cell hit patterns projected onto the r-φ plane are used for identifying tracks originating from the IP with momentum higher than a $p_t$ threshold. The MDC sub-trigger can discriminate background tracks associated with beam losses, synchrotron radiation and cosmic rays. It checks hits for a track segment with a transverse momentum greater than $p_t$, determines "long" or the "short" tracks and counts the number of the tracks. Trigger conditions are based on the number of tracks, the angles and positions of tracks to form global triggers, and the track position information is also available to the track matching logic.

The signal flow of the MDC L1 trigger system is shown in Fig. 67. Signals are transmitted from the MDC electronics crates located in the collision hall via optical links to the L1 trigger crates that are located in the counting room. The MDC sub-trigger system consists of MDC signal fiber transmitters (MFTs), track finders for the axial layers of outer chamber (TFKs), track finders for inner chamber signals (ITKFs) and track counter (TKC) modules. Tracks classified as long tracks, short tracks and inner-chamber tracks are counted in the TKC and MDC trigger conditions are determined.

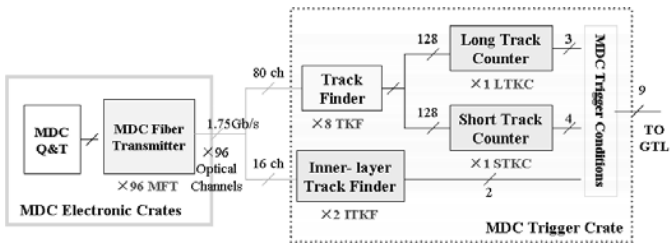

Fig. 67. Signal flow chart of the MDC sub-trigger.

The current MDC L1 trigger [51] is based on tracks using hits in the four axial superlayers SL3, SL4, SL5 and SL10, as shown in Fig. 68. Trigger logic using the two stereo inner layers is implemented although there is no immediate plan to use the inner chamber in the L1 trigger. The number of sense wire signals used in the L1 trigger is 2,796, including the 484 sense wires of the inner chamber. Backgrounds due to beam losses and cosmic rays may be suppressed by including inner chamber tracks, but the disadvantage of including the inner chamber in track finding requirements is that the efficiency of triggering is low for neutral particles decaying in the tracking volume.

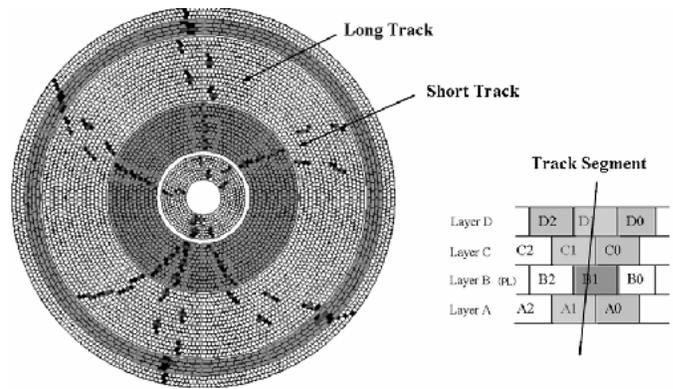

Fig. 68. Graphic display of track segment, long and short tracks in MDC.

*8.3.2. MFT and TKF communication*

The MFTs are 6U VME modules, and each can accommodate signals from 32 sense wires. There are a total of 96 MFTs in the system for the 2,796 sense wire signals that are transmitted to the L1 trigger crates. The TKF and ITKF are 9U VME modules. There are a total of eight TKF modules, each module covers an MDC octant with necessary overlaps on either side.

Optical fibers are used between the MFT and TKF (ITKF) to eliminate ground-loop current interference between the trigger and FEE electronics systems. Xilinx RocketIO transceivers implemented in the Virtex-II Pro FPGA chips are used for parallel-to-serial and serial-to-parallel conversion. The 32-bit wide signal words, plus some control words, are packed into a serial format in MFT and transmitted to ITK (ITKF) in the MDC L1 trigger crates via 45 m long optical fiber cables at a rate of 1.75 Gb/s, under the control of the 41.65 MHz system clock. The serial optical data received are converted back to parallel electrical signals and re-synchronized with the 41.65MHz L1 system clock by the TKFs and ITKF.

*8.3.3. Track finder*

Track segments are defined in the superlayers based on a 3 out of 4 (3/4) logic that requires 3 hit layers in a 4 layer superlayer. The 3/4 logic improves the efficiency of finding track segments. Cell hit patterns are examined according to a memory look-up tables (LUT) to test for the presence of a candidate track segment. The LUT is generated based on Monte Carlo simulation of superlayer responses to charged tracks in the MDC. Originating from the IP, charged tracks that are able to reach SL10 must have at least a $p_t$ of 110 MeV/c. Anchored by one cell in the track segment finding process, the number of possible cell combinations is fairly large, in order to achieve the desired low track transverse momentum trigger threshold.

Track segments are checked further in LUTs for long and short tracks. Tracks that contain track segments in superlayers 3, 4, 5 and 10 and tracks that have valid track segments in superlayers 3, 4 and 5 are defined as long tracks and short tracks respectively, and counted separately. Tracks that can reach superlayer 10 should have a minimum transverse momentum of 110 MeV/c in a solenoid field of 1 Tesla and they must have 70 MeV/c transverse momentum to reach superlayer 5. To reject cosmic and beam-related backgrounds, transverse momentum cuts of 120 MeV/c and 90 MeV/c, respectively, are used for Long Track and Short Track decisions. The simulated efficiencies of finding long and short tracks as a function of $p_t$ are plotted in Fig. 69. Here the wire efficiency was assumed to be 100%. For two tracks, back-to-back conditions are also defined. Because of the desired low $p_t$ threshold, the back-to-back requirement is broadened to a 40° cone.

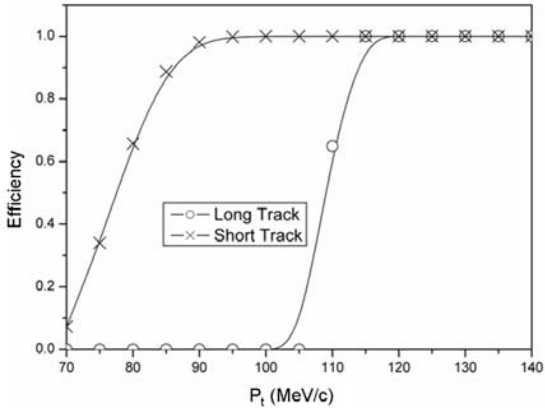

Fig. 69. Simulated efficiencies of finding long and short tracks as functions of $p_t$.

The MDC trigger conditions are categorized and counted. Based on Monte Carlo simulation, trigger conditions that can be used to maintain high trigger efficiency for good physics events and suppress the backgrounds are established. The MDC L1 trigger defines nine basic trigger conditions:

- NLtrk ≥ 1: Number of long tracks ≥ 1.
- NLtrk ≥ 2: Number of long tracks ≥ 2.
- NLtrk ≥ N: Number of long tracks ≥ N (N is programmable).
- NStrk ≥ 1: Number of short tracks ≥ 1.
- NStrk ≥ 2: Number of short tracks ≥ 2.
- NStrk ≥ N: Number of long tracks ≥ N (N is programmable).
- STrk-BB: Short track back-to-back.
- NItrk ≥ 1: Number of inner tracks ≥ 1.
- NItrk ≥ 2: Number of inner tracks ≥ 2.

The conditions that require the number of tracks ≥ N are intended to identify events in which an entire region triggers because of beam background or noise. The MDC trigger information is transmitted to the track matching module to be combined with the TOF and EMC trigger information, and also directly to the global trigger logic.

### 8.4. EMC sub-trigger system

#### 8.4.1. Overview

The EMC sub-trigger generates the L1 trigger [52] based on the total energy, the energy balance in different regions and the number of isolated energy clusters. Basic trigger cells consist of 16 (4 ×4) crystals in the barrel and 15 crystals in the end-caps, as shown in Fig. 70.

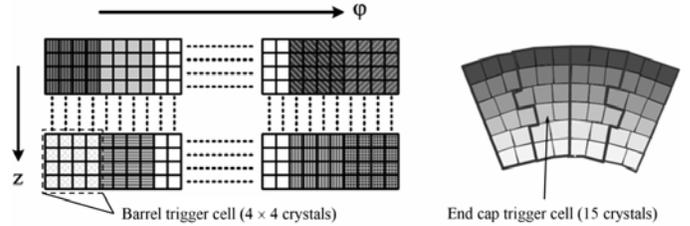

Fig. 70. EMC trigger cell arrangements in the barrel and end caps.

Fig. 71 shows the block diagram of the EMC sub-trigger data flow. Analog energy sums of trigger cells (TCs) are generated by the post amplifier modules in EMC readout NIM crates and transmitted to the TCBA modules by short twisted-pair cables. The TCBA modules are in a 9U VME crate next to the NIM crates of the EMC readout electronics system. TCBA modules process the received analog sums of the TC energies.

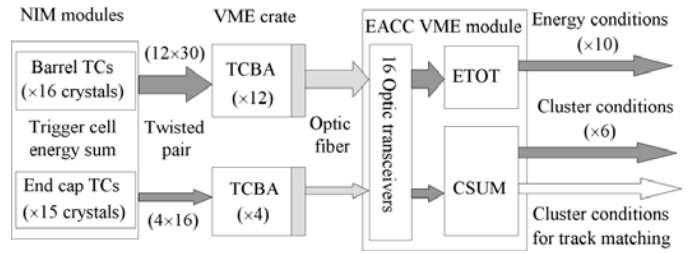

Fig. 71. Trigger cells of EMC barrel (left) and 1/8 end cap (right).

Trigger cell energies are compared to a predetermined threshold in a discriminator for each cluster. The energies of TCs within a cluster are summed together and digitized by a flash ADC. Also, isolated energy clusters are found by joining adjacent TCs.

The total energy in a module and the numbers and distributions of the isolated clusters are sent to the EACC module via an optical link. The digital signals of trigger cells and trigger blocks are processed by the two function blocks in the EACC module. The L1 trigger conditions of the EMC, including the energy of trigger blocks, the counts and positions of isolated energy clusters, are generated. Isolated TC positions are sent to the track matching logic to be combined with the TOF and MDC trigger information. The EMC trigger conditions are generated and sent to the global trigger logic for final trigger decision.

*8.4.2. TCBA module*

A block diagram of the TCBA module is shown in Fig. 72. The TCBA module receives a maximum of 30 analog sums from the TCs (16 or 15 crystals) via AD8130 differential amplifier chips, and performs signal shaping and discrimination (MAX912) using a dual threshold scheme, to extract the time and correct for time walk. Since crystal signals have a 1 μs rise time and 3 μs decay time, the time is important in order to know to which of the 24 ns trigger cycles the event belongs. Energy and timing information of TCs is passed to pipeline memory in the FPGA chips at each clock cycle, to be transmitted to L1 trigger crates.

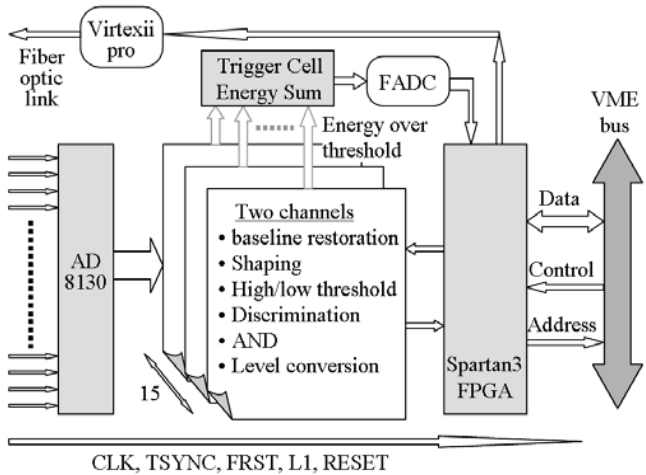

Fig. 72. Block diagram of the TCBA modules.

For the trigger system, the barrel EMC is divided into 12 energy blocks (regions) and each end-cap is divided into two energy blocks. A TCBA module covers a trigger energy block. The energies of valid TCs are added together in a TCBA module to form an analog sum of a trigger block. The analog sum signals are shaped and digitized by a FADC chip (Texas Instruments, THS1030, 10 bits) running at 20.825 MHz, half of the system clock. The optical data links utilize the Xilinx RocketIO transceivers in the Virtex-II Pro FPGA chips operating at a rate of 1.75 Gbps.

*8.4.3. EMC trigger based on isolated clusters*

A high energy shower may deposit energies in several adjacent crystal cells forming energy clusters. Isolated clusters are identified in the CSUM function block by connecting these cells. The number of isolated clusters is counted and their positions are recorded. Positions of the isolated EMC TCs are used to match the TOF hits and MDC tracks in the track matching module.

For physics events, the lower the threshold of the TCs, the higher the trigger efficiency. Too low a threshold, however, will result in more background hits. From trigger simulations, it is determined that an adjustable threshold in the range of 80-100 MeV for the TCs can be used to effectively discriminate beam backgrounds while maintaining high trigger efficiency for minimum ionizing particles.

The following six EMC trigger conditions are defined based on the numbers of isolated clusters.

- NClus ≥ 1 : Number of clusters ≥ 1.
- NClus ≥ 2 : Number of clusters ≥ 2.
- BClus_BB : Barrel clusters back to back.
- EClus_BB : End-cap clusters back to back.
- Clus_PhiB : Barrel clusters balance in φ direction.
- Clus_PhiE : End-cap clusters balance in φ direction.
- Clus_Z : One cluster in the east half, one cluster in the west half.

*8.4.4. Energy balance and total energy triggers*

Energy blocks, twelve from the barrel and four from the two end caps, are used to define the energy balance conditions and to calculate the total energy. Nine trigger conditions are generated based on energy spatial balances and thresholds.

BEtotH: Total energy of barrel exceeding the high threshold.
EEtotH: Total energy of end cap exceeding the high threshold.
EtotL:  Total energy of all EMC exceeding the low threshold.
EtotM:  Total energy of all EMC exceeding the middle threshold.
BL_Z:   Z direction energy balance (include barrel and end-cap).
DiffB:  Energy difference balance between each half barrel.
DiffE:  Energy difference balance between each half end cap.
BL_BLK: Energy balance of barrel blocks.
BL_EEMC: Energy balance between the east and west end-caps.

The Energy TOTal (ETOT) modules calculate the total energies of events and compare the total energies with three digital thresholds for different physics processes. The total energy measured in the EMC is a powerful parameter that can be used to select different types of physics events and discriminate again various types of backgrounds. Three energy thresholds, Etot_L, Etot_M and Etot_H are set for the following purposes.

- Etot_L ≥ 200 MeV, to reject beam-related backgrounds.
- Etot_M ≥ 700 MeV, for neutral physical events.
- BEtot_H or EEtot_H ≥ 2.3 GeV, for Bhabha events.

Pure neutral events refer to events that have only photons and neutral hadrons in their final states.

*8.5. Global trigger logic and fast control*

*8.5.1. Overview*

The trigger condition signals from TOF, MDC and EMC sub-triggers are used in the Global Trigger Logic (GTL) for making the final L1 trigger decision. The GTL generates the L1 trigger signals and fans them out to all the readout electronics systems via optical fiber cables. The track matching logic that combines the TOF, MDC and EMC sub-trigger position information has not yet been implemented. The muon trigger based on the FastOR signals from the barrel muon system is also not implemented yet.

As mentioned before, the task of the L1 trigger system is to select physics events with high efficiency and reject backgrounds from various sources. The simulated trigger efficiencies for representative physics channels and background rejections are given in Table 28. The simulated event rate of the beam-related backgrounds, by far the most serious background source, is about 40 MHz. A L1 trigger rejection factor of $4.6 \times 10^{-5}$, if achieved, will bring beam background down to 1.84 kHz.

Table 28
Simulated trigger efficiencies for some representative physics channels and background rejection

| Processes | Passing fraction (%) |
|---|---|
| $J/\Psi \to$ any thing | 97.66 |
| $\Psi' \to$ any thing | 99.50 |
| $J/\Psi \to D\bar{D} \to$ anything | 99.90 |
| $J/\Psi \to \omega\eta \to 5\gamma$ | 97.85 |
| $J/\Psi \to \gamma\eta \to 3\gamma$ | 97.75 |
| $J/\Psi \to K^+K^-\pi^0$ | 97.39 |
| $J/\Psi \to p\bar{p}\pi^0$ | 97.94 |
| $J/\Psi \to p\bar{p}$ | 95.82 |
| $e^+e^- \to e^+e^-$ | 100 |
| $e^+e^- \to e^+e^-\gamma$ | 100 |
| Beam related backgrounds | $4.6 \times 10^{-3}$ |
| Cosmic ray backgrounds | 9.4 |

*8.5.2. Global trigger*

The block diagram of the global trigger and fast control system is shown in Fig. 73. The trigger signals from different sub-systems arrive at the GTL at different times because of various delays occurred in detector electronics. After level transformation, these signals are adjusted separately in signal alignment and fan-out (SAF) modules to be aligned and checked for time matches. The original and delayed signals are monitored by a TDC and a multi-channel scalar for online monitoring and adjustment.

A trigger table, defining trigger types based on trigger conditions is realized in LUT using FPGA chips and can be modified online. The trigger table includes charged type (CHARGE1, CHARGE2) for events with charged particles, neutral type (NEUTA, NEUTB) for neutral events, common type (COMMON) for crosscheck, Bhabha type (BHABHA BEMC, BHABHA EEMC) for Bhabha event tests, and random type (RAND, 1/s) for test and background study.

The L1 signal output timing is determined by timing logic, which checks for the presence of timing signals by order of priority: TTOF, TMDC, TEMC and TIMEOUT, to ensure the best L1 timing.

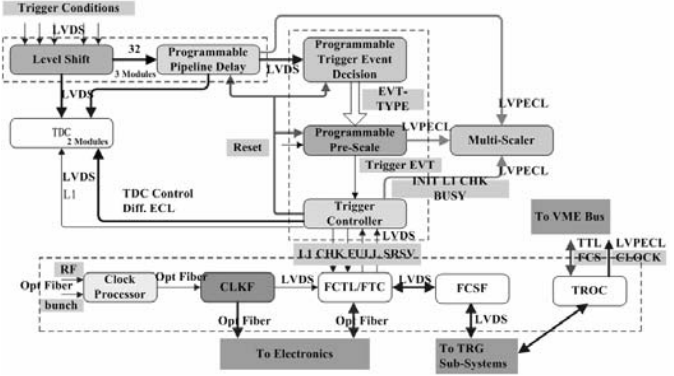

Fig. 73. Block diagram of global trigger and fast control system.

*8.5.3. System clock and Fast control*

The trigger system is pipelined and controlled by the BESIII system clock of 41.65 MHz which is derived from the 499.8 MHz accelerator RF clock after a division by 12 in the clock processor. These clock signals are sent directly to the TOF electronics, and also distributed to trigger and other detector readout electronics by a fan-out module via optical fiber clock links. The clock fiber link goes together with the fast control link in order to have a reliable synchronization.

The fast control fiber links are used to transmit L1 signals to readout electronics modules. The three most important signals from trigger system to the readout electronics are:

1. RST: An initialization signal to be used for system initialization and initialization of RocketIO and optical transceivers.
2. L1: When a L1 signal is received, the L1 counter is increased by one, and the number is used as the event number for the data in the readout buffer.
3. CHK: A signal 500 ns after the 256$^{th}$ L1 signal used for checking and resetting of the event number counter.

The three most important status signals from readout electronics modules sent back to trigger system:

1. RERR：When an error is found in the readout electronics, the RERR signal is issued to inform the online computer via the trigger system. If an event number mismatch occurs when the CHK signal comes, the RERR signal is also issued.
2. FULL: When accumulated data in the buffer reaches a pre-defined percentage (~80%), a FULL signal is issued to the trigger system to hold the generation of L1 signals.

3. EMPT: When the data stored in the readout buffer become less than a predefined percentage (10%), an EMPT signal is issued by the readout electronics system to the trigger system and the latter releases the hold on L1 generation.

## 9. DAQ and event filter

### 9.1. Overview

The architecture of the BESIII DAQ system [53] is shown in Fig. 74. The system consisting of high performance computers and can be divided into two subsystems: the front-end system consists of approximately 40 readout VME crates controlled by PowerPC CPUs and the online computer farm, based on 42 nodes of IBM eServerBlade HS20. Communications between the two systems and among computer nodes are handled by high speed optical links, high speed Ethernet ports and switches.

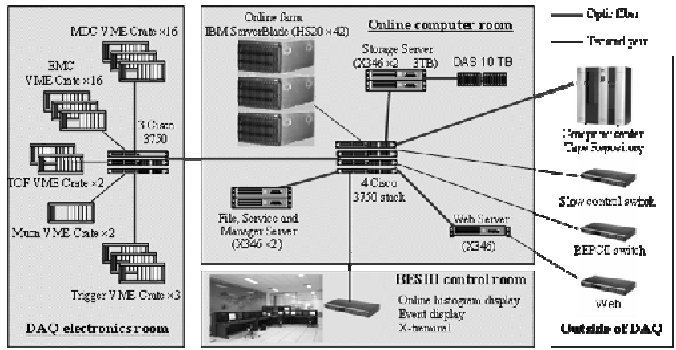

Fig. 74. Architecture of the data acquisition system.

The data rate that the DAQ system must be able to handle is quite high. In Table 29, the BESIII data rates at different stages are compared to several major HEP collider experiments. The data rate written to tape in the BESIII experiment is between the rates of D0 and CMS. The BESIII data acquisition system is designed with a bandwidth of 80 MB/s for reading out data from the VME crates. The online computer farm has a throughput of 50 MB/s, and the capability of writing data to tape exceeds 40 MB/s, after the event filtering. This capacity can be easily expanded if necessary.

Table 29
Comparison between BESIII and other modern HEP experiments

|  | BESIII | Belle | D0 | CMS | ATLAS |
|---|---|---|---|---|---|
| L1 trigger output rate (kHz) | 4 | 0.3 | 1 | 100 | 100 |
| L3 trigger output rate (kHz) | 3 | 0.25 | 0.05 | 0.10 | 0.2 |
| Ratio of L1/L3 | 1.33 | 1.2 | 200 | 1,000 | 500 |
| Average event size (kB) | 14 | 40 | 250 | 1,000 | 1,000 |
| Data rate to tape (MB/s) | 42 | 15 | 12.5 | 100 | 200 |

Multi-level buffering, parallel processing, high-speed VME readout, high-performance computing and network transmission techniques are utilized in the BESIII DAQ design.

The goal of the DAQ design is to achieve high reliability, stability, scalability, portability and maintainability. The components of the online computer system are summarized in Table 30. The system is highly scalable and can be expanded easily when needs arise.

Table 30
The BESIII online computer system

| Functions | Model | Quantity | Software |
|---|---|---|---|
| VME controller | Motorola MVME5100 PowerPC 750 CPU | 39 | VxWorks 5.4 |
| Online computer farm | IBM eServerBlade HS20 (dual 3.0 G Xeon CPU) | 42 | SLC3 |
| File server | IBM 346 2U server (dual 3.0 G Xeon CPU) | 5 | SLC3 |
| Monitoring PCs | Lenovo dual CPU 2.4G | 4 | SL5 |
| Temporary storage | Raid 5 | 10 TB | |

The online computer farm receives event fragments from the VME crate controllers via the Ethernet switches. It assembles event fragments into full events and runs event filtering software to further reduce the data rate. Accepted events are written to tapes. The online event filtering is also called the L3 trigger.

There are five additional IBM xSeries eServer x346 computers, each with dual 3.0 GHz Xeon CPUs and 1.5 TB RAID5 hard disks. Among these five 2U server nodes, two are used as file servers to provide system management and other services, two are used as data servers to control the data file transfer to the computer center for writing to tapes, and one is used as a web server for remote system access. Additional PCs are used to perform various control and monitoring tasks such as run control, event display, histogram generation and process monitoring.

The DAQ software used in BESIII was developed based on the framework of the ATLAS TDAQ software [54]. It accomplishes the data collection, event building, filtering and recording of event data. It also provides additional control and test functions.

### 9.2. VME based DAQ front-end

The required performance of the DAQ front-end is determined by the trigger rate and event size. The data rate of the BESIII sub-detectors at 4 kHz L1 trigger rate are summarized in Table 31.

Table 31
BESIII DAQ front-end parameters

| Sub-detector | No. of VME Crates | No. of Channels | Occup. | Data (kB) | L1 rate (MB/s) |
|---|---|---|---|---|---|
| MDC (T + Q) | 17 | 6796+6796 | 0.10 | 6.4 | 25.6 |
| EMC | 16 | 6240 | 0.17 | 5.6 | 22.4 |
| TOF (T + Q) | 2 | 448+448 | 0.10 | 0.4 | 1.6 |
| Muon | 1 | 9152 | 0.01 | 0.2 | 0.8 |
| L1 trigger | 3 | 400 | 1 | 1.6 | 5.6 |
| Total | 39 | ~30K | - | 14.2 | ~56.8 |

The total number of VME crates, readout channels, expected occupancies and event data sizes are given. The total number of readout channels is approximately 30 k and the average event data size is approximately 14 kB. The DAQ electronics consists of 39 VME crates located in the collision hall and in the trigger electronics room. The crate controllers are based on embedded PowerPC CPUs. These embedded PowerPCs perform four main tasks:

- Data readout: respond to the interrupts issued by VME modules in crates and read out the data.
- Event building: assemble data received from the VME modules into event fragments with headers.
- Event buffering.
- Upload event segments: transfer event segments to online farm via Ethernet ports.

The VME crate controllers (Motorola MVME5100) are single board computers based on PowerPC 750 CPUs running the VxWorks 5.4 real time operating system with a 100 Mb/s Ethernet port for data communication. The CBLT (Chained Block Transfer) DMA readout method is used to readout event fragments from different front end modules in the VME crate and to transfer the data to the VME controller in order to reduce the number of interrupts required.

Signals from sub-detector front-end electronics are processed by various VME modules and transmitted to the crate controllers. The first level of event building takes place in the crate controllers. A system transfers event fragments stored in the VME controllers to the online computer farm via 100 Mb/s Ethernet ports. The event fragments are then assembled into full events by the online computer farm.

### 9.3. Event building and filtering

The IBM eServerBlade HS20 servers are based on Intel Xeon dual CPUs running the SLC (Scientific Linux CERN) operating system. Among the 42 nodes, 11 are used to receive and process the event fragments from the VME controllers, one node is used as a data flow controller and the remaining 30 are used for full event building and event filtering.

Fragments of events stored in the pipelined buffers of the VME controllers are transferred via the 100 Mb/s Cisco 3750 Ethernet switches to the 11 dedicated nodes of the online computer farm. The fragments that belong to the same L1 trigger from various crates are assembled into events. The farm supervisor controls the event building process. There are two queues in the supervisor. One is the queue of the event to be built, which comes from the L1 trigger. The other one is the queue of free nodes in the computer farm. According to these two queues, the supervisor informs one of free nodes to read out the proper sub-events from the nodes in the readout system. All the subsequent event building and data processing tasks are done on this assigned node.

The primary function of the event filter is to fetch full events from the event builder computers, analyze them with physics event selection algorithms, and send the accepted events to the event storage subsystem. The framework of the BESIII event filter is based on Atlas Event Filter framework.

Some new features are added in order to increase system reliability.

Algorithms that are more flexible and more sophisticated than used by the L1 trigger are adopted for the L3 event filtering. For BESIII, the required reduction factor for the event filter is rather modest. The BESIII event filter algorithms [55] are designed to suppress the background rate by about one half from ~2 KHz, or about 50% of the expected maximum L1 rate of 4 kHz, to less than 1 kHz, in order to reduce the data rate written to tape.

In addition to further background event rate suppression, the event filtering software also categorizes the types of events into hadronic, Bhabha, $\mu\mu$, cosmic ray and beam related backgrounds, and performs online control and monitoring tasks. For events that require further classifications, fast reconstruction algorithms for MDC tracks and fast EMC energy cluster finding can be invoked. To keep the DAQ system running smoothly, the average latency of the event filter is designed to be less than 5 ms.

### 9.4. Inter process communications

Data communications of the online computer farm are controlled by Cisco Gigbit Ethernet switches via high speed Ethernet cable or optical links. Custom software is necessary for the communication process control, synchronization and data transfers among the large number of computer nodes. A custom $C^{++}$-based software package, IPC (Inter Process Communication) [56], was used as the framework for developing the communication software. IPC is based on the Naming Service of CORBA (Common Object Request Broker Architecture) utilizing omniORB [57].

## 10. Detector control system

### 10.1. System requirements

The principal tasks of the detector control system (DCS) are to supervise and control the status of the BESIII experiment, to guarantee its safe operation and to record data that are critical for online/offline detector calibrations and data corrections. The DCS must provide a uniform, coherent, simple and user friendly interface to the operators. The BESIII DCS [58] is divided into six subsystems: temperature and humidity (TH), low voltage power supplies (LV), high voltage power supplies (HV), electronics VME crates (VME), gas control (GAS), safety interlocking and global control system (SI&GCS). Each system has its own online computer to control the data acquisition and perform functions such as data processing, recording, analysis and alarm generations. The SI&GCS system provides the interlock between the BESIII and BEPCII to guarantee the safe operation.

There are over 3,000 sensors of various types and thousands of monitoring outputs from devices such as power supplies, gas mass flow controllers, VME crates, etc. The entire DCS has approximately 100 components, including computers, programmable logic controllers (PLCs) and network communication switches. A more detailed breakdown of the DCS system is given in Table 32. Data from the approximately 9,000 points are recorded once every 10 seconds on average. A portion of the data is made available to the online computer farm for calibration and monitoring purposes.

Table 32
The detector control system

| Subsystem | Parameters | No. of devices | Data points |
|---|---|---|---|
| Detector environment | Temperature and humidity | 1,580 sensors | 1,770 |
| Low voltage power supplies | Voltage, current, temperature humidity | 1,188 sensors | 1,215 |
| CEAN H.V. power supplies modules | Status, V_mon, V_set, I_mon, I_set, Trip, Ramp_up, Ramp_down and temperature | 654 H.V. ch. 59 Temp. sensors | 5,232 59 |
| VME crates | On/Off, voltage, current, fan speed, temp. | 860 | 860 |
| Gas system | Pressure, mass flow, on/off, Status, temp., humidity, flammable gas alarm, oxygen, helium | 85 Sensors and I/Os | 85 |
| SI&GCS | Luminosity, SC solenoid, beam pipe, gas, TOF, FEE, door control and radiation | 283(I/Os | 283 |
| Total | | 4,709 | 9,504 |

*10.2. DCS hierarchy*

The DCS is hierarchically organized into three layers: a Front End Layer (FEL), a Local Control Layer (LCL) and a Global Control Layer (GCL). In the FEL, devices range from simple sensors up to complex computer-based devices like embedded CPUs and Programmable Logical Controllers (PLC). A LabVIEW-based software framework has been developed for the LCL and GCL. Network communication and web-based technologies have been used for the GCL. The architecture of the DCS is shown graphically in Fig. 75.

Sensors and data points to be monitored are distributed throughout the spectrometer and experimental complex. They are read out by the FEL devices that include Programmable Logical Controllers (PLC) and network interface devices under the control of the local control layer. Data are transmitted to the six sub-system computers from the front layer devices, and all data acquired are inserted into a MySQL database by the DB server. The data are accessible by the local control system and also through the Web server.

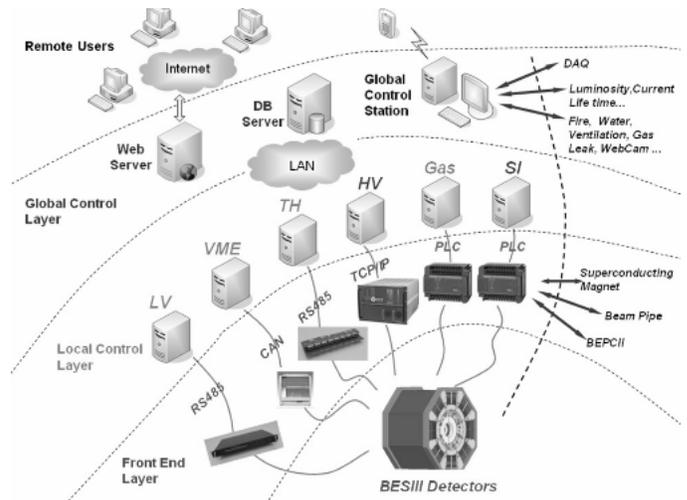

Fig. 75. Architecture of detector control system.

*11. The BESIII offline software*

The BESIII Offline Software System (BOSS) [59] has been developed using the C++ language and object-oriented techniques within the Scientific Linux CERN (SLC4) operating system [60]. Software production and configuration management is facilitated using the Configuration Management Tool, CMT [61]. The BOSS framework is based on Gaudi [62], which provides standard interfaces for the common software components necessary for data processing and analysis. The overall architecture is shown in Fig. 76.

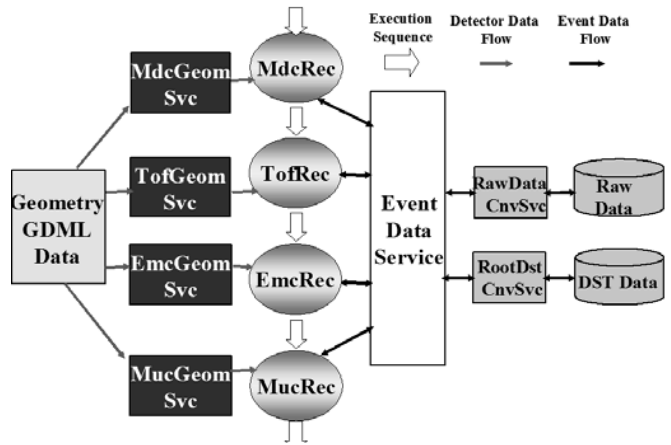

Fig. 76. The overall BESIII software architecture.

The framework employs Gaudi's Event Data Service as the data manager. Reconstruction algorithms can access the raw event data from the Transient Data Store (TDS) via the Event Data Service. However, it is the Raw Data Conversion Service that is responsible for conversions between persistent raw data and transient raw data objects. The detector's material and geometry information are stored in Geometry Design Markup Language, GDML [63], files. Algorithms can retrieve this information by using the corresponding services. Through the DST Conversion Service, the reconstruction results can be written into ROOT [64] files for subsequent physics analyses.

Furthermore, the BOSS framework also provides abundant services and utilities to satisfy the requirements from different BESIII algorithms. For instance, the Magnetic Field Service provides the value of magnetic field for each space point within the detector. The Navigation Service helps users trace reconstructed tracks back to Monte Carlo particles. Using the Particle Property Service, a particle's properties can be accessed by various software components

The BESIII Object Oriented Simulation Tool (BOOST) [65] is based on Geant4 [66]. The material and geometry needed for the detector simulation are read from GDML data files. The standard shapes in Geant4 are not general enough to describe all BESIII sub-detector modules. For example, the G4TwistedTube [67] shape is used to describe MDC stereo cells and G4IrregBox shape is used for EMC end-cap crystals. After importing these two external packages, the BESIII detector can be successfully described.

The calibration software, which is based on GLAST's scheme [68], consists of a calibration framework and calibration algorithms. The framework provides reconstruction algorithms and a standard way to obtain the calibration data objects. The calibration constants for each sub-detector, produced by the associated calibration algorithm, are stored in a ROOT file and subsequently into the database. When a reconstruction algorithm requests calibration constants, it will trigger the conversion of calibration data objects from the ROOT file. During this process, the framework will first search for the appropriate database record, then fetch only one particular calibration dataset, and finally return transient objects to the reconstruction algorithm.

Data reconstruction is the central task for the offline data processing. For BESIII, a complete chain of reconstruction algorithms has been developed and fit into the BOSS framework. The reconstruction system contains the MDC tracking algorithm, dE/dx and TOF reconstruction algorithms, EMC clustering and shower maker, a muon track finder, etc.

The MDC tracking algorithm [69] starts with a formation of track segments from hits using pre-calculated patterns. Then, it links the found axial segments to circular tracks and performs circle fits using a least-squares method. After that, it adds stereo segments to track candidates and performs an iterative helix fit. Finally, track refitting is performed after including additional hits in the MDC that might possibly belong to these tracks.

The dE/dx reconstruction algorithm applies many corrections to the ADC values of MDC channels, and determines the energy loss of each charged particle in the MDC, which is important for particle identification.

A Geant4-based algorithm [70] was developed to extrapolate an MDC track to outer sub-detectors. Considering the magnetic deflection and ionization loss of the charged particle in the detector, the algorithm provides the precise position and the momentum of the particle in all the outer sub-detectors. The associated error matrix at a given space point is calculated with multiple scattering effects taken into account. By comparing the results of track extrapolations with results from a full simulation, it is found that they are consistent with each other. This track extrapolation is widely used by other reconstruction and analysis algorithms.

For TOF reconstruction [71], MDC tracks are first extrapolated to the TOF and matched with TOF hits. Then, the TOF reconstruction algorithm applies various corrections, such as the effective velocity correction, attenuation correction, etc., and calculates the time that a charged particle takes to travel from the interaction point to the TOF layers.

Data reconstruction in the EMC consists of three concatenated steps. The first step is an energy calibration. The ADC value of each crystal is converted to real energy using the EMC calibration constants. Then a clustering algorithm is used to form EMC clusters for both barrel and end-caps. Finally, reconstructed showers, with calculated positions and energies, are determined. In this step, seeds are selected using the same criteria in each cluster. The seed is the local maximum, which has the highest energy among its neighbours. If a cluster contains a unique seed, the shower is defined. Otherwise, a splitting function is invoked to split this cluster into multiple showers.

The MUC reconstruction algorithm [72] starts two searches for hit matches among axial hit wires and transverse hit wires, respectively. Then it combines the two sets of hits into track candidates and matches them with tracks reconstructed by the MDC. Low momentum muons may pass through too few layers in the MUC to form sets of hits. In order to improve the tracking efficiency, a second search for tracks is performed for the remaining hit wires. In this step, MDC tracks will be used as seeds. The seed track is extrapolated to the Muon Counter, and hit wires located inside a hit window will be assigned to the muon track.

The analysis program fetches the reconstruction results from TDS through the Event Data Service, then builds data objects that are more suitable for physics analysis and puts them back to the TDS. In this process BESIII physics tools, including Vertex Finder [73] and Particle Identification [74], are used. Particle Identification will combine TOF information, dE/dx measurements, energy deposits in the EMC, and MUC information to provide physics analysis particle IDs.

**Acknowledgements**

The authors are grateful for the tremendous effort of the engineering team at IHEP who contributed to the success of the BESIII construction over the past 6 years. They also would like to thank support received from their home institutions and the supporting staff. The authors would like to express special thanks to Prof. T.D. Lee for his strong support of the BEPCII/BESIII project. The authors would also like to thank Alexander Bondar for his useful suggestions on the design of the EMC.

This work was supported by BEPCII project, CAS Knowledge Innovation Program U602 and U-34 (IHEP), National Science Foundation of China (10491305,10225524) and the U.S. Department of Energy under Contract No. DE-FG02-04ER41291 (U. Hawaii).

**References:**

[1] The BESIII detector, IHEP-BEPCII-SB-13 (2004) Beijing.

[2] C. Zhang for BEPC & BEPCII Teams, Performance of the BEPC and Progress of the BEPCII, Proceedings of APAC (2004) 15-19, Gyeongju, Korea.

[3] Minghan Ye and Zhipeng Zheng, BEPC, the Beijing electron-positron collider, International Journal of Modern Physics A, v 2, n 6, (1987) 1707-25; Fang Shouxian and Senyu Chen, The Beijing electron positron collider, Particle Accelerators, 26 (1990) 51-61.

[4] J.Z. Bai et al., The BES detector, Nucl. Instr. and Meth. A 344 (1994) 319-334;

[5] J.Z. Bai et al., The BES upgrade [Beijing Spectrometer], Nucl. Instr. and Meth. A 458 (2001) 627-637.

[6] BESIII Collaboration, The construction of the BESIII experiment, Nucl. Instr. and Meth 598 (2009) 7-11; Frederick A. Harris for the BES Collaboration, BEPCII and BESIII, Nuclear Physics B (Proc. Suppl.) 162 (2006) 345–350.

[7] Yukiko Yamada et al, P-wave charmed-strange mesons, Phys. Rev. C 72, 065202 (2005); J. Vijandea,, A. Valcarce and F. Fernández, Multiquark structures in heavy-light meson
systems, Nucl. Phys. A, Vol. 790 (2007) 506c-509.

[8] Y. Z. Wu et al, BEPCII Interaction Region Design and Construction Status, Proceedings of 2005 Particle Accelerator Conference, Knoxville, Tennessee; Q. L. Peng, BEPCII Interaction Region Superconducting Magnet System, IEEE Trans. on Appl. Supercond. Vol. 14 (2004) 539-41.

[9] Xunfeng Li et al., Design and cooling of BESIII beryllium beam pipe, Nucl. Instr. and Meth. A 585 (2008) 40–47.

[10] http://www.deetee.com/, Detection Technology, Inc., Finland,

[11] YSI Inc. http://www.YSI.com, USA.

[12] Wang Yong-gang et al., Real time bunch-by-bunch luminosity monitor for BEPCII, IEEE Instr. and Meas. Tech. Conf. (2008) 1155-8,

[13] ZHOU Neng-Feng et al., Monte Carlo Simulation of BEPCII Beam-Repated Barckground (I)- Synchrotron Radiation, High Energy Phys. Nucl. Phys. 28 (2004) 227-231 (in Chinese).

[14] JIN Da-Peng et al. Monte Carlo Simulation of the BEPC II/BESIII Backgrounds (II)-Beam-Gas-Interactions, High Energy Phys. Nucl. Phys. 28 (2004)1197-1202 (in Chinese); JIN Da-Peng et al, Monte Carlo Simulation of the BEPC II/BESIII Backgrounds (III)-Touschek Effect, High Energy Phys. Nucl. Phys. 30 (2006) 449-53 (in Chinese).

[15] K.L. Brown and Ch. Iselin, CERN 74-2, Feb. 1974.

[16] http://wwwasd.web.cern.ch/wwwasd/geant/

[17] The BESIII detector magnet, Zhu Z., etc., Proceedings of the Twentieth Internationa Cryogenic Engineering Conference (ICEC20) (2005) 593-596, Beijing, China; B. Wang et al., Design, Development and Fabrication for BESIII, Superconducting Muon Detector Solenoid, IEEE Trans. Appl. Supercond., Vol. 15 (2005), 1263-1266.

[18] Wang NMR Incorporated, Livermore, CA 94551 USA.

[19] Shike Huang et al., Full power test of superconductor of BESIII magnet, Cryogenic 47 (2007) 406-408.

[20] Q.J. Xu et al., Numerical study on quench process and protection of BESIII superconducting detector magnet, Cryogenic 47 (2007) 292-299.

[21] C. Chen et al., The BESIII drift chamber, IEEE Nuclear Science Symposium Conference Record, 2007, 1844-6.

[22] D.G. Cassel. et al., Design and construction of the CLEO II drift chamber. Nucl. Instr. and Meth. A252 (1086) 325; D. Peterson, Construction of the CLEO III tracking system: silicon vertex detector and drift chamber, Nucl. Instr. and Meth. A 409 (1998) 204-209.

[23] Wu, Lin-Hui et al., Design and Optimization of Cell Structure and Layer Arrangement of the BESIII Drift Chamber, High Energy Physics and Nuclear Physics, Vol. 29, No. 5 (2005) 476-480 (in Chinese).

[24] http://consult.cern.ch/writeup/garfield/files/

[25] Jin Da-Peng, Study of the compensation voltage on the boundary field layer of the BESIII drift chamber, High Energy Physics and Nuclear Physics, v 30 (2006), 665-9 (in Chinese).

[26] J.B. Liu at al., A beam test of a prototype of the BESIII drift chamber in magnetic field, Nucl. Instru. and meth. 557 (2006) 436-444.

[27] The CERN High-performance time-to-digital Converter, http://micdigital.web.cern.ch/micdigital/hptdc.html.

[28] V. Blobel, Software alignment for racking detectors. Nucl. Instr. and Meth. A 566, 2006.

[29] Yuekun, Heng, The two scintillator detectors on BESIII, , IEEE Nuclear Science Symposium Conference Record, 2007, 53-56.

[30] M. Artuso et al., The CLEO III ring imaging Cherenkov detector, Nucl. Instr. and Meth. A 461 (2001) 545-548.

[31] David W.G.S. Leith, DIRC—the particle identification system for BaBar, Nucl. Instr. and Meth. A 494 (2002) 389-401.

[32] Chong Wu, The timing properties of a plastic time-of-flight scintillator from a beam test, Nucl. Instr. and Meth. A 555 (2005) 142-147; Sun Zhi-Jia et al., Beam Test for a 1:1 Module of Time of Flight Counter of BESIII, High Energy Physics and Nuclear Physics, Vol. 29, No. 10 (2006) 933-937 (in Chinese); AN Shao-Hui, et al., Beam Test of the Time Resolution of the BESIII End Cap TOF, High Energy Physics and Nuclear Physics, Vol. 29, No. 8 (2005) 775-780 (in Chinese).

[33] http://www.bicron.com.

[34] http://www.eljentechnology.com/ej-200.html.

[35] S H An et al., Testing the time resolution of the BESIII end-cap TOF detectors, Meas. Sci. Technol. 17 (2006) 2650-2654.

[36] http://www.analog.com.

[37] GUO Jian-Hua et al., Time Measurement System Prototype for BESIII TOF, High Energy Physics and

Nuclear Physics, Vol. 30, No. 8 (2006) 761-766 (in Chinese).

[38] F.A. Harris, et al., BESIII time of flight monitoring system, Nucl. Instr. and Meth. A 593 (2008) 255-262.
[39] Mingyi Dong, Development of the BESIII Electromagnetic Calorimeter, Ph.D. Thesis, Institute of High Energy Physics, April 20, 2008 (in Chinese); Dong Ming-Yi et al, Design and assembly of the CsI(Tl) crystal module of the BESIII electro-magnetic calorimeter, Chinese Physics C, v 32, n 11, 908-11, Nov. 2008.
[40] Wang Zhi-Gang et al., Nuclear Counter Effect of Silicon Photodiode Used in CsI(Tl)Crystal Calorimeter, High Energy Physics and Nuclear Physics, Vol. 30, No. 9 (2006) (in Chinese).
[41] SHANG Lei, Cosmic Ray Test on CsI(Tl) Modules of BESIII Electromagnetic Calorimeter, High Energy Physics and Nuclear Physics, Vol. 31, No. 2 (2007) 177-182 (in Chinese).
[42] V.M. Aulchenko et al., CMD-2 barrel calorimeter, Nucl. Instr. and Meth. A 336 (1993) 53.
[43] HE Miao et al., Energy loss correction for a crystal calorimeter, Chinese Physics C (HEP & NP), Vol. 32, No. 4 (2008) 269-274.
[44] Xie Yuguang, Development and Offline Calibration of the BESIII Muon Detector, Ph.D. Thesis, Institute of High Energy Physics, April, 2007 (in Chinese).
[45] Chinese patent application No. ZL 200820117783.1.
[46] Jiawen Zhang et al., A new surface treatment for the prototype RPCs of the BESIII spectrometer, Nucl. Instr. and Meth., A 540 (2005)102-112.
[47] Jiawen Zhang, The design and mass production on Resistive Plate Chambers for the BESIII experiment, Nucl. Instr. and Meth. A 580 (2007) 1250-1256.
[48] Jifeng Han, Cosmic ray test results on resistive plate chamber for the BESIII experiments, Nucl. Instr. and Meth. A 577 (2007) 552-557.
[49] Zhen'An Liu et al., Trigger system of BES III, 5th IEEE-NPSS Real Time Conference (2007) 701-704
[50] Hao Xu et al., FPGA based high speed data transmission with optical fiber in trigger system, IEEE Nuclear Science Symposium Conference Record (2007) 818-821.
[51] Xu Hao, Physics Design of Main Drift Chamber Trigger for BESIII Experiment, High Energy Physics and Nuclear Physics, Vol. 29, No. 4 (2005) 376-382 (in Chinese).
[52] Shu-Jun Wei, Introduction to BESIII EMC sub-trigger system, Nucl. Instr. and Meth. A 598 (2009) 323–327
[53] Tao Ning et al, A prototype of the read-out subsystem of the BESIII DAQ based on PowerPC, Plasma Science and Technology, v 7, n 5, Oct. 2005, 3065-8.
[54] ATLAS Trigger and Data Acquisition, Available at https://twiki.cern.ch/twiki/bin/view/Atlas/TriggerDAQ.
[55] FU Cheng-Dong et al., Study of the online event filtering algorithm for BESIII, Chinese Physics C (HEP & NP), Vol. 32. No. 5 (2008) 329-337; ATLAS Event Filter, available at: https://twiki.cern.ch/twiki/bin/view/Atlas/EventFilter.
[56] Li Fei et al., Distributed inter process communication framework of BESIII DAQ online software, Nuclear Electronics& Detection Technology, Vol. 26 (2006) 605-608 (in Chinese).
[57] CORBA homepage http://www.omg.org/corba/.
[58] Xihui Chen et al., Design and Implementation of the BESIII Detector Control System, Nucl. Instr. and Meth., A 592 (2008) 428-433.
[59] W. Li et al., "The Offline Software for the BESIII Experiment", Proceedings of the International Conference on Computing in High Energy and Nuclear Physics, 225 (2006).
[60] See http://linux.web.cern.ch/linux/scientific4/.
[61] C. Arnault, "CMT: a Software Configuration Management Tool", Proceeding of CHEP2000, Padova, Italy (2000); http://www.cmtsite.org.
[62] Barrand, G. et al., "GAUDI - A software architecture and framework for building HEP data processing applications", Comput. Phys. Commun., 140: 45 (2001); http://proj-gaudi.web.cern.ch/proj-gaudi.
[63] Yutie Liang et al., "The Uniform Geometry Description for Simulation, Reconstruction and Visualization in BESIII Experiment", Nuc. Inst. & Meths. A, Vol. 603, 325 (2009); http::/www.cern.ch/gdml.
[64] ROOT: http://root.cern.ch/drupal.
[65] DENG Zi-Yan et al. "Object-Oriented BESIII Detector Simulation System" (in Chinese), High Energy Physics and Nuclear Physics, Vol. 30(5), 371 (2006).
[66] S. Agostinelli et al., Nucl. Instr. and Meth. In Phys. Res. A, Vol. 506, Issue 3, 250 (2003); J. Allison et al., IEEE Transactions on Nucl. Science, Vol. 53, No. 1, 270 (2006); http://www.geant4.org/geant4.
[67] K. Hoshina et. al.,"Development of Geant4 Solid for Stereo Minijet Cells in a Cylindrical Drift Chamber", Comput. Phys. Commun. 153, 373 (2003).
[68] J. R. Bogart, Calibration infrastructure for the GLAST LAT, SLAC-PUB-9890, CHEP-2003-MOKT001 (2003); http://www.glast.slac.edu.
[69] ZHANG Yao et al., "The Reconstruction and Calibration of the BESIII Main Drift Chamber", CHEP2007, Victoria, B.C., Canada (2007).
[70] WANG Liang-Liang et al.. "BESIII Track Extrapolation and Matching" (in Chinese), High Energy Physics and Nuclear Physics, Vol. 31(02), 183 (2007).
[71] Jifeng Hu et al., "Correlation Analysis in Time-of-Flight Calibration of BESIII", Vol. 31(10), 893 (2007).
[72] LIANG Yu-Tie et al., "Study on BESIII MUC Offline Software with Cosmic-ray Data", Chinese Physics C, 33(7), 562 (2009).
[73] XU Min et al., "Decay vertex reconstruction and 3-dimensional lifetime determination at BESII", Chinese Physics C, Vol33(06), 428 (2009).
[74] QIN Gang et al., "Particle Identification Using Artificial Neural Networks at BESIII", Chinese Physics C, Vol. 32(01), 1 (2008).